\documentclass[a4paper]{article}
\usepackage[margin=25mm]{geometry}
\usepackage{amsmath}
\usepackage{amsfonts}
\usepackage{amssymb}
\usepackage{amsthm}
\usepackage{graphicx}
\usepackage{subfig}
\usepackage{authblk}
\usepackage{cite}

\newcommand{\be}{\begin{equation}}
\newcommand{\ee}{\end{equation}}

\numberwithin{equation}{section}

\begin{document}

\title{Impact of the non-canonical approach to the exact solution of the ideal one-dimensional electron gas confined with an anisotropic quantum wire of oscillator-shaped profile}%

\author[1,2]{E.I. Jafarov\thanks{Corresponding author: ejafarov@physics.science.az}}
\author[1]{S.M. Nagiyev\thanks{sh.nagiyev@physics.science.az}}
\author[3]{J.~Van~der~Jeugt\thanks{Joris.VanderJeugt@UGent.be}}

\affil[1]{Institute of Physics, Ministry of Science and Education, Javid av.\ 131, 1143, Baku, Azerbaijan}
\affil[2]{Baku State University, Z. Khalilov str.\ 33, 1148, Baku, Azerbaijan}
\affil[3]{Department of Mathematics, Computer Science and Statistics, Ghent University, Krijgslaan 281-S9, 9000 Gent, Belgium}

\date{} 

\maketitle

\begin{abstract}
We study an exactly solvable model that can be interpreted as an ideal one-dimensional electron gas confined with an anisotropic quantum wire potential of oscillator-shaped profile. 
The homogeneous nature of the quantum wire is broken by the introduction of the effective electron mass, which changes with radial distance. 
We solve the problem described both within the canonical and the non-canonical approach. 
Analytical expressions of the wavefunctions of the stationary states for both cases in terms of the Laguerre polynomials are obtained, as well as the discrete energy spectrum related to these wavefunctions. 
Additionally, an exact solution to the angular position part of the position-dependent mass Schr\"odinger equation within the non-canonical approach leads to the angular-part wavefunctions of the even and odd states expressed through the Gegenbauer polynomials. 
Possible limit relations and special cases are studied too.
\end{abstract}

\section{Introduction}

The ideal electron gas consisting of electrons moving freely strictly in one direction with confinement applied to them in the other two directions greatly impacts the development of various branches of condensed matter physics and related areas~\cite{barnham2001}. 
The classical problem neglecting any quantum effects is treated in detail in~\cite{kunz1974}. 
The same problem can also be considered within the quantum mechanical context by just applying two-dimensional confinement in the form of a quantum wire. Namely, one needs to consider a physical system, the sizes of which can be compared with the de Broglie wavelength of the electrons. Then, one will observe stronger and more unpredictable quantum effects. The ideal electron gas being confined in two directions comparable with the electron de Broglie wavelength forms a quantum wire and moves across it. The best experimental examples of such quantum structures are the discovery of the quantum point contact~\cite{vanwees1988,wharam1988} and carbon nanotubes~\cite{dresselhaus2001}.
Fabrication and both theoretical and experimental investigations of quantum wire structures are of considerable interest for a lot of researchers of various areas, because these structures are able to exhibit considerable higher nonlinear optical characteristics, higher sensitivity, electron localization, transport, tunneling, and a number of other advanced properties due to ``doubled" confinement application~\cite{kapon1989,tserkovnyak2002,havu2004,steinberg2006,pugnetti2009,guclu2009,owen2016}. 
These advantages make them one of the best candidates for the potential application in quantum computing and nanoelectronics~\cite{isailovic2004,fitzsimons2006,arora2008}. 
The simplest approach for obtaining such a structure is the application of the additional confinement by potential steps to the two-dimensional electron gas already confined via an infinite potential well. 
Its exact solution in three dimensions is well known. 
However, the weakness of such a model is the strict rectangular profile of the two-dimensional quantum wire applied to the moving electrons. 
We revisit this problem and present here an exact solution to the ideal one-dimensional electron gas confined with an anisotropic quantum wire of non-rectangular profile. 
Non-rectangularity of the profile and non-uniform nature of the quantum wire are achieved thanks to the replacement of the constant effective mass with a mass that varies with radial distance $\rho$ in both the kinetic energy operator and harmonic oscillator potential of the quantum system under consideration. 
One-dimensional exactly solvable versions of the similar problem, leading to the semi- or full confinement of the potential of the non-rectangular profile is well-known~\cite{carinena2004,yu2004,schmidt2007,ovando2019,jafarov2020a,jafarov2020b,elnabulsi2020a,elnabulsi2021a,jafarov2021,elnabulsi2021c,elnabulsi2022,sari2022,jafarov2022,lima2023,dossantos2023,quesne2023,nagiyev2024}. 
Our exact main goal here is to construct an analytically solvable ideal model under the two-dimensional confinement, which will behave in the form of a generalized version of both triangular-shaped potential and of an infinitely deep quantum well non-rectangular profile in two dimensions. 
Both these potentials have a great impact and applications in experimental physics and connected topics~\cite{nagamiya1940a,nagamiya1940b,feinberg1964,card1972,vlaev1994,benavides2007,benavides2018,al2022,al2026}. Additionally, \cite{ezaki1997} considered a very similar problem and achieved limited results due to the available numerical computations for the electronic structures under consideration. Despite the numerical nature and lack of exact solutions, this already leads to a huge number of successful applications in low-dimensional systems fabrication, equipment instrumentation, and radiation physics~\cite{yoffe2001,reimann2003,rontani2006,huseynov2015,borgonovi2016,zhou2024
}.

In this paper, we decided to perform all our computations within both the canonical and the non-canonical approach. 
We discussed this question thoroughly in~\cite{jafarov2008,jafarov2025}.
Here, we briefly highlight that the question was opened for discussion by Wigner in his seminal paper~\cite{wigner1950}, where it is shown that the exact solution of the Heisenberg--Lie equations of motion for the non-relativistic quantum harmonic oscillator problem, as well as for any Hamiltonian of quadratic behavior, does not uniquely imply the canonical quantum mechanical commutation relations between momentum and position operators. Then the discussion immediately led to the development of new general commutation rules and generalized superalgebras with new statistics different from the known boson and fermion statistics~\cite{pais1950,lang1951,green1953,greenberg1955}. It also has a huge number of attractive applications in different branches of modern physics. For example, \cite{henneaux1982,ercolessi2010} analyze the approach described in~\cite{wigner1950} together with the inverse problem in the calculus of variations and show how these approaches jointly can lead to the possible discovery of alternative structures in quantum mechanics as well as direct equivalence of Lagrangians of some simple classes of forces with a certain set of exactly-solvable second order differential equations. Dehghani et al.~\cite{dehghani2015} studied new cat-states, which are built on the basis of the non-canonical commutation rules introduced in \cite{wigner1950} and discusses possible limit cases, for which the known Schr\"odinger cat-states are retrieved. Fakhri and Sayyah-Fard~\cite{fakhri2021} studied thoroughly the Jaynes--Cummings model of a two-level atom in a single-mode para-Bose cavity field via the interaction between the light field of the para-Bose Glauber coherent states and matter. Hamil and L\"utf\"uoglu~\cite{hamil2022} investigated the thermal quantities of graphene under a constant magnetic field with the Dunkl formalism and observes that parity parameters appearing as a result of the non-canonical commutation rules modify the known results at low temperatures. Merabtine et al.~\cite{merabtine2023} studied the ideal Bose gas and the blackbody radiation within this non-canonical approach and showed that it increases the value of the entropy and changes all related heat functions. Benarous et al.~\cite{benarous2025} analyzed the direct influence of the Wigner parameter, which appears as a result of the proposed non-canonical commutation rules on the internal energy and the heat capacity functions of a harmonically trapped ideal Bose gas without change of the possible phase transitions. Recently, \cite{huerta2025} reported the successfully performed experiment based on the quantum simulation of para-particle oscillators satisfying non-canonical commutation relations and demonstrated how such a system accurately reproduces the dynamics of para-bosons and para-fermions of even order.
Now, we think that it could have a great impact in the field of low-dimensional systems, if one tries to solve exactly the ideal electron gas problem confined by a quantum wire of the specific profile, just combining together two important foundations of modern quantum physics -- non-canonical commutation relations and a mass that varies by radial distance~\cite{giaever1960a,giaever1960b,harrison1961,bendaniel1966}. 

We structured our paper as follows. 
In Sect.~2, general properties of the time-independent Schr\"odinger equation of the ideal one-dimensional electron gas confined with an anisotropic quantum wire are discussed, and basic mathematical tricks are presented for the mass that is central-symmetric in the cylindrical coordinates and depends only on the radial distance. 
Then, an analytical expression of such a central symmetric mass is introduced in Sect.~3. 
Such an introduction allows one to solve the radial Schr\"odinger equation exactly. 
Section 3 includes further details of this exact solution in terms of the generalized Laguerre polynomials. 
In Sect.~4, we generalize all obtained results to the non-canonical case. 
We show that if we are dealing with the non-canonical case, then one needs to solve the hypergeometric equation for the angular position part of the wavefunction additionally. 
Section 5 discusses the general properties of the obtained results and possible limit relations, for which the well-known circular oscillator potential is retrieved.

\section{The time-independent Schr\"odinger equation of the ideal one-dimensional electron gas confined with an anisotropic quantum wire}

Our starting point is the following time-independent Schr\"odinger equation of the ideal one-dimensional electron gas in a field in which the confinement potential only depends on the Cartesian coordinates $\left(x;y\right)$ due to the fact that the electrons move freely in one direction (formally, this is the $Z$-direction in our calculations):
\be
\label{scheq-01}
\left[ {\frac{{\hat {\vec p}}^2}{{2m_0 }}  + V\left( {x,y} \right)} \right]\psi \left( {x,y,z} \right) = E\psi \left( {x,y,z} \right).
\ee
Here, ${\hat {\vec p}}^2  = \hat p_x ^2  + \hat p_y ^2  + \hat p_z ^2$. 
$m_0$ acts here as a constant electron effective mass and can be interpreted as the level of the electron mobility in each compound. 
In other words, this value of the mass defines the electrons as lighter (heavier) particles during their motion inside a certain compound. 
For example, it behaves as $0.067 \times m_e$ for the conducting band of $GaAs$ binary compound
(where $m_e$ is the constant electron mass), $0.077 \times m_e$ for $InP$ binary compound, $0.096 \times m_e$ for $CdTe$ binary compound, etc. 
However, we will keep in our computations of its parametric value without application to a certain compound.  
The field here being central-symmetric can be taken as a two-dimensional (or circular) harmonic oscillator of the following form:
\be
\label{p-uc-osc}
V\left( {x,y} \right) = \frac{{m_0 \omega^2 }}{2}\left( {x^2  + y^2 } \right).
\ee
Then, the Schr\"odinger equation (\ref{scheq-01}) of such a circular oscillator (\ref{p-uc-osc}) yields:
\be
\label{scheq-02}
\left[ {\frac{{\hat p_x ^2  + \hat p_y ^2  + \hat p_z ^2 }}{{2m_0 }} + \frac{{m_0 \omega^2 }}{2}\left( {x^2  + y^2 } \right)} \right]\psi \left( {x,y,z} \right) = E\psi \left( {x,y,z} \right).
\ee

An exact solution to this equation is well known if one assumes that the momentum operator in the position representation is defined as a first-order ordinary derivative with respect to position satisfying the canonical commutation relation~\cite{flugge2012}. 
The generalization of the momentum operator to the non-canonical case preserves similarity of the solution of the corresponding radial equation but drastically changes the solution of the Schr\"odinger equation with respect to the angular position. 
One observes that it is necessary to solve now a second order differential equation very similar to the special case of the seminal P\"oschl-Teller potential problem~\cite{flugge2012}. 
Some details of such exact solutions are already known~\cite{tempesta2001,tremblay2009,post2012,genest2013}. 
We will demonstrate in Sect.~4 how such a solution appears in our case and then solve exactly the arisen second order differential equation.  

We are going to apply the two-dimensional confinement effect of a specific behavior to this quantum system. 
Such a specific behavior of the confinement can be achieved through the introduction of a position-dependent effective mass $M\left( {x,y} \right)$ in the $\left(x;y\right)$ coordinates instead of a homogeneous effective mass $m_0$, but assuming that this effective mass preserves its constant nature in the $Z$-direction. 
We introduce the analytical expression of varying mass $M\left( {x,y} \right)$ in Sect.~3 and show mathematically as well as via the graphical visualizations how this changes the oscillator potential in such a way that the potential exhibits both triangular- and well-shaped behavior. The best example of the successful application of the varying effective electron mass is a seminal experiment by Giaever on the discovery of the electron tunneling phenomena in superconductors~\cite{giaever1960a,giaever1960b} and its perfect theoretical explanation in~\cite{harrison1961}. Furthermore, at present there is a well known advanced method for the experimental fabrication of the confined multi-quantum semiconductor structures with layer-to-layer thickness variation profiles, which have dozens of successful applications~\cite{weisbuch1981,miller1984,herman2012,mccormack2022}.

However, direct replacement $m_0 \to M\left( {x,y} \right)$ in Eq.~(\ref{scheq-02}) violates the hermiticity property of the Hamiltonian under consideration, because then we would have a non-Hermitian kinetic energy operator $\left(\hat p_x ^2  + \hat p_y ^2 \right)/2M\left( {x,y} \right)$. The case is discussed in~\cite{barnham2001}. 
Therefore, we follow the rule suggested in~\cite{jafarov2025,mustafa2007}, which states that a replacement $m_0 \to M\left( {x,y} \right)$ in the definition of the momentum operator will slightly deform, but, preserve the nature of the oscillator Heisenberg-Weyl algebra only if von Roos parameters $\alpha$, $\beta$ and $\gamma$ (with an additional condition $\alpha+\beta+\gamma=-1$) introduced in~\cite{vonroos1983} will equal to $\alpha=\gamma=-1/4$ and $\beta=-1/2$. 
Application of this rule leads to the following modified Schr\"odinger equation with a Hermitian free Hamiltonian:
\be
\label{scheq-03}
\left[ {\frac{1}{2}\frac{1}{{M^{\frac{1}{4}} }}\hat p_x \frac{1}{{M^{\frac{1}{2}} }}\hat p_x \frac{1}{{M^{\frac{1}{4}} }} + \frac{1}{2}\frac{1}{{M^{\frac{1}{4}} }}\hat p_y \frac{1}{{M^{\frac{1}{2}} }}\hat p_y \frac{1}{{M^{\frac{1}{4}} }} + \frac{{\hat p_z ^2 }}{{2m_0 }} + \frac{{M \omega ^2 }}{2}\left( {x^2  + y^2 } \right)} \right]\psi \left( {x,y,z} \right) = E\psi \left( {x,y,z} \right).
\ee
We substitute here the analytical expression of the momentum operator in the position representation within the canonical approach
\be
\label{mom-can}
\hat p_x  =  - i\hbar \frac{\partial }{{\partial x}},\quad\hat p_y  =  - i\hbar \frac{\partial }{{\partial y}},\quad \hat p_z  =  - i\hbar \frac{\partial }{{\partial z}},
\ee
and solve this equation in a cylindrical coordinate system $\left( {\rho ,\varphi ,z} \right)$, where $\rho  = \sqrt {x^2  + y^2 }$ ($\rho \geq 0$) is the radial distance, $\varphi  = \arctan \frac{y}{x}$ ($0 < \varphi  < 2\pi$) is the angular position and axial coordinate $z$ of the cylinder is preserved as $-\infty <z<+\infty$. 
It is clear that the two-dimensional oscillator defined via (\ref{p-uc-osc}) is a central symmetric potential. 

In general, the application of cylindrical coordinate transformations $x = \rho\cos \varphi$ and $y = \rho\sin \varphi$ requires $M\left( {x,y} \right) \to M\left( {\rho,\varphi} \right)$. 
However, in order to preserve the simpler exact solubility property of eq.~(\ref{scheq-03}), one needs to assume that the mass $M\left( {\rho,\varphi} \right)$ is of central symmetric nature and depends only on the radial distance $\rho$. 
Hence, $M\left( {\rho,\varphi} \right) \to M\left( {\rho} \right)$ should be assumed. 
Straightforward computations yield:
\be
\label{scheq-04}
\begin{split}
-\frac{1}{2}\hbar^2 \frac{1}{{M^{\frac{1}{4}} }}&\left[ {\frac{\partial }{{\partial x}} \frac{1}{{M^{\frac{1}{2}} }}\frac{\partial }{{\partial x}} + \frac{\partial }{{\partial y}} \frac{1}{{M^{\frac{1}{2}} }}\frac{\partial }{{\partial y}} } \right]\frac{1}{{M^{\frac{1}{4}} }}\\
&=- \frac{{\hbar ^2 }}{{2M}}\left[ {\frac{{\partial ^2 }}{{\partial \rho ^2 }} + \left( {\frac{1}{\rho } - \frac{{M'}}{M}} \right)\frac{\partial }{{\partial \rho }} - \frac{1}{4}\frac{{M''}}{M} + \frac{7}{{16}}\left( {\frac{{M'}}{M}} \right)^2  - \frac{1}{4}\frac{1}{\rho }\frac{{M'}}{M} + \frac{1}{{\rho ^2 }}\frac{{\partial ^2 }}{{\partial \varphi ^2 }}} \right].
\end{split}
\ee
Hence, its substitution in Eq.~(\ref{scheq-03}) yields the following second order differential equation for eigenfunction $\psi \left( {x,y,z} \right) \equiv \Psi \left( {\rho ,\varphi } \right)Z\left( z \right)$ and eigenvalues $E\equiv E_{\rho ,\varphi }  + E_z$:
\be
\label{scheq-05}
\begin{split}
&\left\{ { - \frac{{\hbar ^2 }}{{2M}}\left[ {\frac{{\partial ^2 }}{{\partial \rho ^2 }} + \left( {\frac{1}{\rho } - \frac{{M'}}{M}} \right)\frac{\partial }{{\partial \rho }} - \frac{1}{4}\frac{{M''}}{M} + \frac{7}{{16}}\left( {\frac{{M'}}{M}} \right)^2  - \frac{1}{4}\frac{1}{\rho }\frac{{M'}}{M} + \frac{{\partial ^2 }}{{\partial \varphi ^2 }}\frac{1}{{\rho ^2 }}} \right] + \frac{{M\omega ^2 }}{2}\rho ^2 } \right\}\Psi \left( {\rho ,\varphi } \right)Z\left( z \right) \\
&- \frac{{\hbar ^2 }}{{2m_0 }}\frac{{\partial ^2 }}{{\partial z^2 }}\Psi \left( {\rho ,\varphi } \right)Z\left( z \right) = \left( {E_{\rho ,\varphi }  + E_z } \right)\Psi \left( {\rho ,\varphi } \right)Z\left( z \right).
\end{split}
\ee
Now, this equation can be split in $\left( {\rho ,\varphi} \right)$- and $z$-dependent parts as follows:
\begin{align}
\label{scheq-05-01}
\begin{split}
&\left\{ { - \frac{{\hbar ^2 }}{{2M}}\left[ {\frac{{\partial ^2 }}{{\partial \rho ^2 }} + \left( {\frac{1}{\rho } - \frac{{M'}}{M}} \right)\frac{\partial }{{\partial \rho }} - \frac{1}{4}\frac{{M''}}{M} + \frac{7}{{16}}\left( {\frac{{M'}}{M}} \right)^2  - \frac{1}{4}\frac{1}{\rho }\frac{{M'}}{M} + \frac{{\partial ^2 }}{{\partial \varphi ^2 }}\frac{1}{{\rho ^2 }}} \right] + \frac{{M\omega ^2 }}{2}\rho ^2 } \right\}\Psi \left( {\rho ,\varphi } \right) \\
&= E_{\rho ,\varphi } \Psi \left( {\rho ,\varphi } \right), 
\end{split}
\\ 
\label{scheq-05-02}
&  - \frac{{\hbar ^2 }}{{2m_0 }}\frac{{\partial ^2 }}{{\partial z^2 }}Z\left( z \right) = E_z Z\left( z \right).
\end{align}
The solution to Eq.~(\ref{scheq-05-02}) is well-known:
\be
\label{z-wf}
Z\left( z \right) \equiv Z_{\kappa _z}\left( z \right) =\frac{1}{\sqrt{2\pi}} e^{i\kappa _z z} ,\quad \kappa _z  = \sqrt {\frac{{2m_0 E_z }}{{\hbar ^2 }}} .
\ee

Next, Eq.~(\ref{scheq-05-01}) can be rewritten in terms of eigenfunctions $\Psi \left( {\rho ,\varphi } \right) \equiv {\rm P}\left( \rho  \right)\Phi \left( \varphi  \right)$ and eigenvalues $E_{\rho ,\varphi }$ as follows:
\be
\label{scheq-05-03}
\begin{split}
&\left[ {\rho ^2 \frac{{\partial ^2 }}{{\partial \rho ^2 }} + \left( {\rho  - \rho ^2 \frac{{M'}}{M}} \right)\frac{\partial }{{\partial \rho }} - \frac{1}{4}\frac{{M''}}{M}\rho ^2  + \frac{7}{{16}}\left( {\frac{{M'}}{M}} \right)^2 \rho ^2  - \frac{1}{4}\frac{{M'}}{M}\rho  + \frac{{2M\rho ^2 E_{\rho ,\varphi } }}{{\hbar ^2 }} - \frac{{M^2 \omega ^2 }}{{\hbar ^2 }}\rho ^4 } \right]{\rm P}\left( \rho  \right)\Phi \left( \varphi  \right) \\ 
 &+ \frac{{\partial ^2 }}{{\partial \varphi ^2 }}{\rm P}\left( \rho  \right)\Phi \left( \varphi  \right) = 0.
\end{split}
\ee
The above equation can also be split into the following two second-order differential equations:
\begin{align}
\label{scheq-rho-01}
& \left[ {\rho ^2 \frac{{\partial ^2 }}{{\partial \rho ^2 }} + \left( {\rho  - \frac{{\rho ^2 M'}}{M}} \right)\frac{\partial }{{\partial \rho }} + \frac{{7\rho ^2 }}{{16}}\left( {\frac{{M'}}{M}} \right)^2  - \frac{{\rho ^2 M''}}{{4M}} - \frac{{\rho M'}}{{4M}} + \frac{{2\rho ^2 ME_{\rho ,\varphi } }}{{\hbar ^2 }} - m^2  - \frac{{M^2 \omega ^2 }}{{\hbar ^2 }}\rho ^4 } \right]{\rm P}\left( \rho  \right) = 0, \\ 
\label{scheq-phi-01}
&  - \frac{{\partial ^2 }}{{\partial \varphi ^2 }}\Phi \left( \varphi  \right) = m^2 \Phi \left( \varphi  \right).
\end{align}
The solution to Eq.~(\ref{scheq-phi-01}) is also well-known. 
The wavefunctions $\Phi \left( \varphi  \right)$ have the following orthonormal analytical expression:
\be
\label{wf-phi-01}
\Phi \left( \varphi  \right) = \frac{1}{{\sqrt {2\pi } }}e^{im\varphi } ,\quad m = 0, \pm 1, \pm 2, \pm 3, \ldots .
\ee

One observes that the solution of the Schr\"odinger equation (\ref{scheq-05}) in coordinates $\varphi$ and $z$ is almost trivial. 
In the next section, we are going to demonstrate how its $\rho$-dependent radial part~(\ref{scheq-rho-01}) can be exactly solved.

\section{Analytical expression of the mass $M\left( {x,y} \right)$ and exact solution to the radial Schr\"odinger equation}

In the preceding section, we discussed that Eq.~(\ref{scheq-rho-01}) contains a mass $M \equiv M\left( {x,y} \right)$ that varies with position in $x$ and $y$ directions, but exhibits homogeneous behavior in $z$ direction. 
In general, the application of the known cylindrical coordinate transformations leads to $M\left( {x,y} \right) \to M\left( {\rho,\varphi} \right)$. 
But, one can assume that the mass $M\left( {\rho,\varphi} \right)$ is of central symmetric nature and depends only on the radial distance $\rho$. 
Hence, the simplification $M\left( {\rho,\varphi} \right) \to M\left( {\rho} \right)$ is adopted for further computations.

Now, we assume the following additional conditions for the analytical expression of the varying mass $M\left( {\rho} \right)$:
\begin{itemize}

\item The analytical expression should preserve the exact solubility property of the radial Schr\"odinger equation (\ref{scheq-rho-01});

\item Its substitution in the circular harmonic oscillator potential (\ref{p-uc-osc}) allows potentials of triangular shape and of infinite well-shape behavior;

\item In the limit when $M\left( {\rho} \right)$ reduces to a constant effective mass $m_0$, the obtained expression tends to its non-relativistic analogue.

\end{itemize}

An expression of the varying mass satisfying the above-listed conditions, inspired by \cite{jafarov2025,mustafa2007,alhaidari2002,keshavarz2013}, is the following:
\be
\label{m-rho}
 M \equiv M\left( \rho \right) = m_0 \lambda _0 ^{2a} \left(\rho^2\right)^a,  \qquad -1<a<+\infty, \qquad \lambda _0  = \sqrt {\frac{{m_0 \omega }}{\hbar }}.
\ee

\begin{figure}[b]
\begin{center}
\resizebox{0.32\textwidth}{!}{%
  \includegraphics{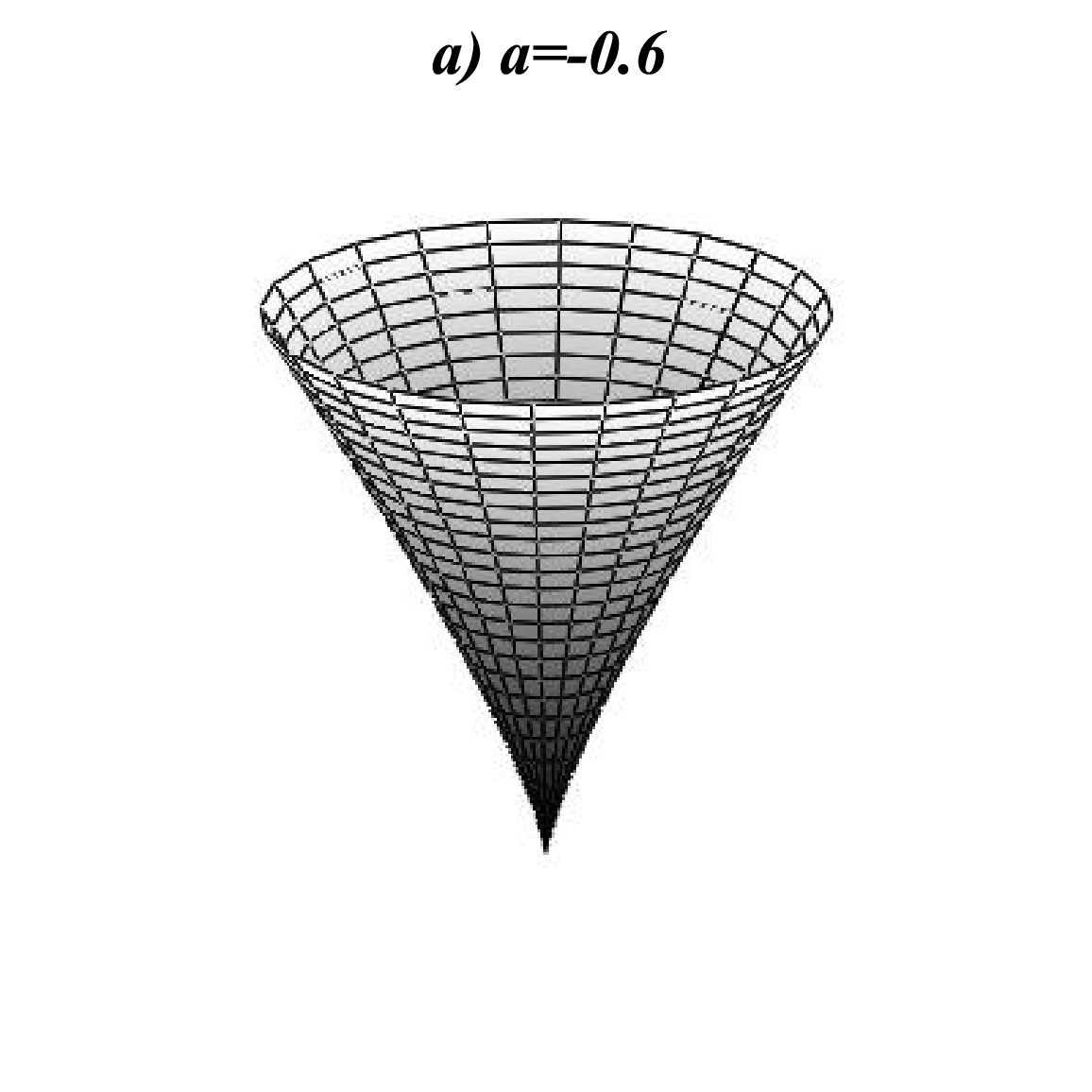}
}
\resizebox{0.32\textwidth}{!}{%
  \includegraphics{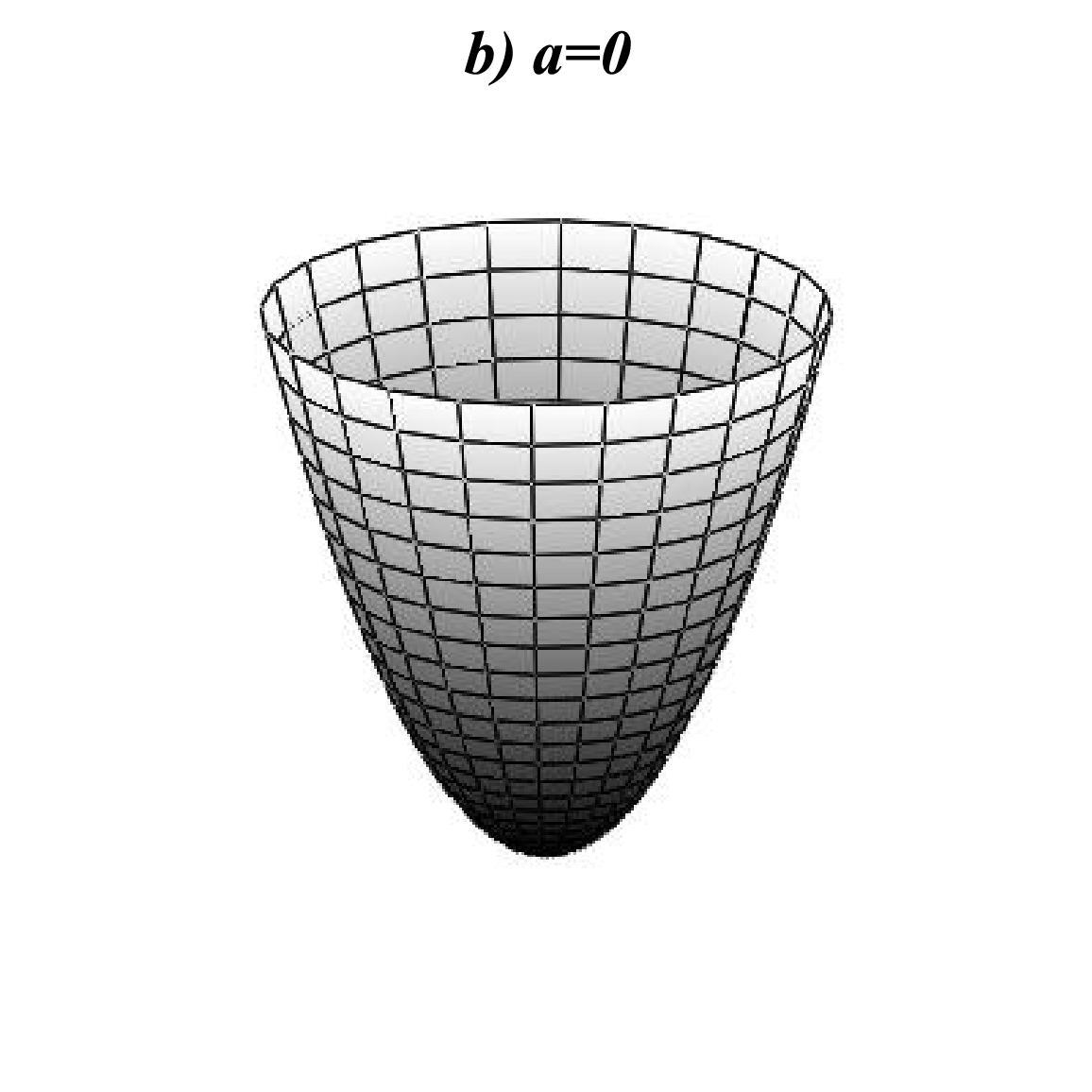}
}
\resizebox{0.32\textwidth}{!}{%
  \includegraphics{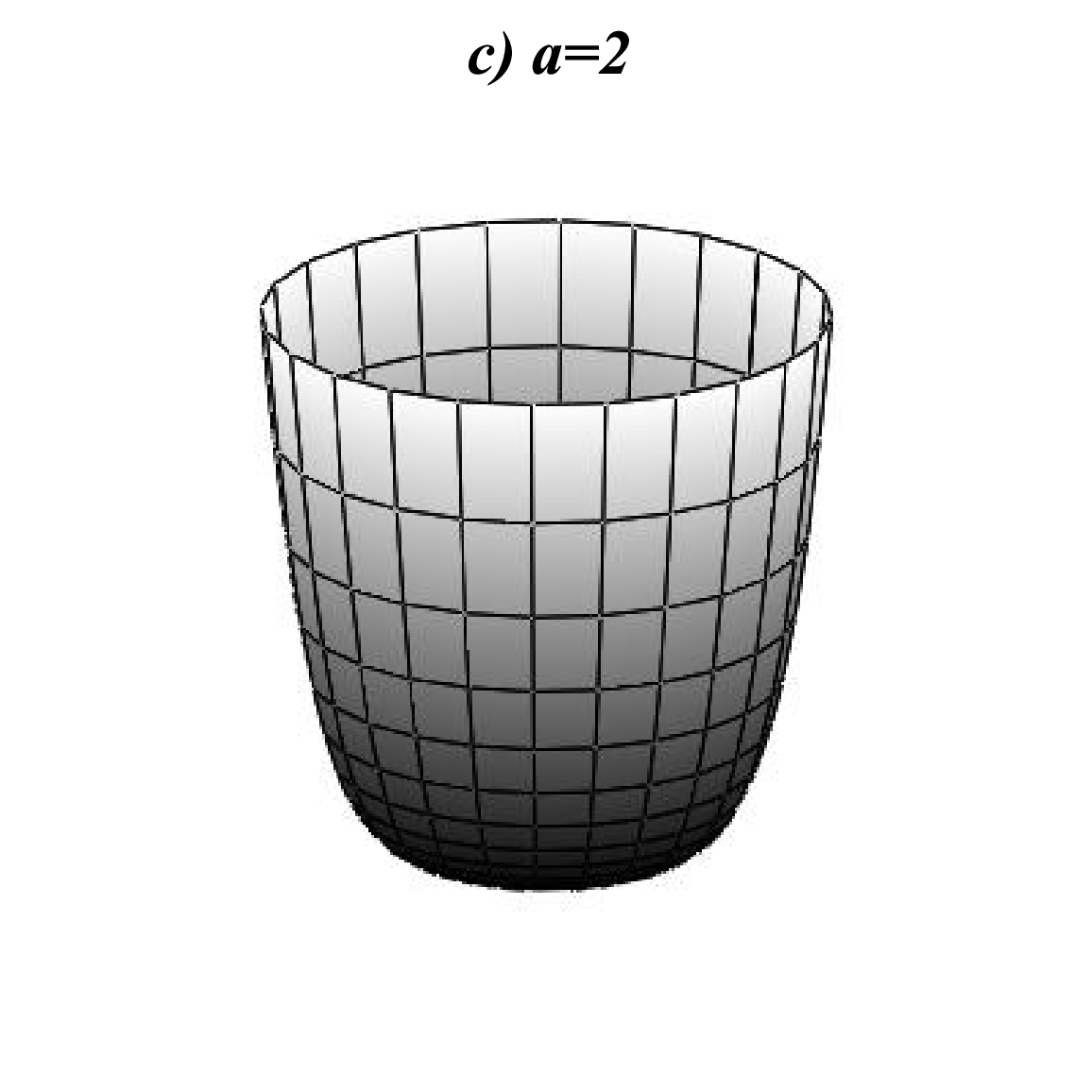}
}
\end{center}
\caption{The circular harmonic oscillator potential~(\ref{p-uc-osc}) with varying mass $M\left( \rho \right)$~(\ref{m-rho}) for values of the deformation parameter $a=-0.6;$ $0;$ $2.0$ ($m_0=\omega=\hbar=1$).} 
\label{fig.1}
\end{figure}

One observes that the above expression of varying mass changes the circular harmonic oscillator potential (\ref{p-uc-osc}) in such a way that the potential exhibits both triangular- and well-shaped behavior for different values of the parameter $a$. This is demonstrated in Fig.~1. 
Such a behavior is achieved thanks to the appearance of the parameter $a$ in the definition of the varying mass (\ref{m-rho}) and its restriction between $-1$ and $+\infty$. 
The restriction $a>-1$ is necessary, because the value $a=-1$ simply converts the circular oscillator potential (\ref{p-uc-osc}) to the constant $\frac 12 \hbar \omega$. 
When $a \to +\infty$ the potential gets infinitely high walls of almost rectangular behavior.

Substitution of (\ref{m-rho}) in Eq.~(\ref{scheq-rho-01}) yields:
\be
\label{scheq-rho-02}
\left[ {\rho ^2 \frac{{\partial ^2 }}{{\partial \rho ^2 }} + \left( {1 - 2a} \right)\rho \frac{\partial }{{\partial \rho }} + \kappa _{\rho ,\varphi } ^2 \lambda _0 ^{2a} \rho^{2a + 2}  - \lambda _0 ^{4a + 4} \rho^{4a + 4}  - m^2  + \frac{3}{4}a^2 } \right]{\rm P}\left( \rho  \right) = 0.
\ee
Here,
\be
\label{kappa-rho-phi}
\kappa _{\rho ,\varphi }  = \sqrt {\frac{{2m_0 E_{\rho ,\varphi } }}{{\hbar ^2 }}} .
\ee
Equation~(\ref{scheq-rho-02}) can be slightly modified as a result of its division by $\rho^2$:
\be
\label{scheq-rho-03}
\frac{{\partial ^2 {\rm P}}}{{\partial \rho ^2 }} + \left( {1 - 2a} \right)\frac{1}{\rho }\frac{{\partial {\rm P}}}{{\partial \rho }} + \left[ {\kappa _{\rho ,\varphi } ^2 \lambda _0 ^{2a} \rho ^{2a}  - \lambda _0 ^{4a + 4} \rho ^{4a + 2}  - \frac{{m^2  - \frac{3}{4}a^2 }}{{\rho^2 }}} \right]{\rm P} = 0.
\ee
The contact point transformation
\be
\label{cpt}
\xi  = \frac{{\lambda _0 ^{a + 1} }}{{\sqrt {a + 1} }}\rho ^{a + 1} ,
\ee
applied to Eq.~(\ref{scheq-rho-03}) changes it after straightforward calculations as follows:
\be
\label{scheq-rho-04}
{\rm P}'' + \frac{{1 - a}}{{\left( {a + 1} \right)\xi }}{\rm P}' + \left[ {\frac{{\kappa _{\rho ,\varphi } ^2 \lambda _0 ^{ - 2} }}{{a + 1}} - \xi ^2  - \frac{{m^2  - \frac{3}{4}a^2 }}{{\left( {a + 1} \right)^2 \xi ^2 }}} \right]{\rm P} = 0.
\ee
This second-order differential equation can be rewritten in the following compact form:
\be
\label{scheq-rho-05}
{\rm P}'' + \frac{{1 - a}}{{\left( {a + 1} \right)\xi }}{\rm P}' + \frac{{\kappa _{\rho ,\varphi } ^2 \lambda _0 ^{ - 2} \left( {a + 1} \right)\xi ^2  - \left( {a + 1} \right)^2 \xi ^4  - m^2  + \frac{3}{4}a^2 }}{{\left( {a + 1} \right)^2 \xi ^2 }}{\rm P} = 0.
\ee

We look for solutions ${\rm P}$ of Eq.~(\ref{scheq-rho-05}) as a product of two functions $f\left( \xi  \right)$ and $\Pi \left( \xi  \right)$:
\be
\label{p-01}
{\rm P} = f\left( \xi  \right)\Pi \left( \xi  \right).
\ee
Here, the function $f\left( \xi  \right)$ is responsible for the boundary behavior of the quantum system under consideration, whereas the function $\Pi \left( \xi  \right)$ defines its polynomial behavior following the potential (\ref{p-uc-osc}). 
Substitution of (\ref{p-01}) in Eq.~(\ref{scheq-rho-05}) yields:
\be
\label{scheq-rho-06}
\begin{split}
f\Pi '' &+ \left[ {2f' + \frac{{1 - a}}{{\left( {a + 1} \right)\xi }}f} \right]\Pi ' \\
&+ \left[ {f'' + \frac{{1 - a}}{{\left( {a + 1} \right)\xi }}f' + \frac{{\kappa _{\rho ,\varphi } ^2 \lambda _0 ^{ - 2} \left( {a + 1} \right)\xi ^2  - \left( {a + 1} \right)^2 \xi ^4  - m^2  + \frac{3}{4}a^2 }}{{\left( {a + 1} \right)^2 \xi ^2 }}f} \right]\Pi  = 0.
\end{split}
\ee
The boundary conditions for the radial wavefunction ${\rm P}\left( \xi  \right)$ require that 
\be
\label{bc-wf}
\begin{cases}
f\left( 0 \right) = 0, \\ 
 f\left( \infty  \right) = 0.
\end{cases}
\ee
Then, one introduces the following general expression of $f\left( \xi  \right)$ that will satisfy (\ref{bc-wf}):
\be
\label{f-01}
f\left( \xi  \right) = \xi ^\alpha  e^{\beta \xi ^2 } ,
\ee
where $\alpha$ is positive due to the boundary condition $f\left( 0 \right) = 0$, and $\beta$ is negative due to the boundary condition $f\left( \infty  \right) = 0$. 
Substitution of (\ref{f-01}) in Eq.~(\ref{scheq-rho-06}) yields:
\be
\label{alphabeta}
\begin{cases}
\alpha  = \frac{{a + \sqrt {m^2  + a^2 /4} }}{{a + 1}}, \\ 
\beta  =  - \frac{1}{2}.
\end{cases}
\ee
Hence,
\be
\label{scheq-rho-07}
\Pi '' + \left( {\frac{{a + 1 + \sqrt {a^2  + 4m^2 } }}{{a + 1}}\xi ^{ - 1}  - 2\xi } \right)\Pi ' + \frac{{\kappa _{\rho ,\varphi } ^2 \lambda _0 ^{ - 2}  - 2\left( {a + 1} \right) - \sqrt {a^2  + 4m^2 } }}{{a + 1}}\Pi  = 0.
\ee
Comparison with the following second-order differential equation for the generalized Laguerre polynomials
\[
y''\left( {\zeta ^2 } \right) + \left[ {\left( {2\gamma  + 1} \right)\zeta ^{ - 1}  - 2\zeta } \right]y'\left( {\zeta ^2 } \right) + 4ny\left( \zeta  \right) = 0,\quad y\left( {\zeta ^2 } \right) = L_n^{\left( \gamma  \right)} \left( {\zeta ^2 } \right),
\]
allows us to obtain the expression for $\Pi \left( \xi  \right)$ in terms of Laguerre polynomials:
\be
\label{pi-01}
\Pi  = L_n^{\left( {\sqrt {m^2  + a^2 /4} /\left( {a + 1} \right)} \right)} \left( {\xi ^2 } \right).
\ee

The energy spectrum can also be obtained from this comparison:
\be
\label{kappa-01}
\kappa _{\rho ,\varphi } ^2 \lambda _0 ^{ - 2}  - 2\left( {a + 1} \right) - \sqrt {a^2  + 4m^2 }  = 4n\left( {a + 1} \right).
\ee
Easy computations via substitutions (\ref{m-rho}) and (\ref{kappa-rho-phi}) in (\ref{kappa-01}) yield:
\be
\label{e-rho-phi}
E_{\rho ,\varphi }  = \hbar \omega \left( {a + 1} \right)\left[ {2n + 1 + \frac{{\sqrt {m^2  + a^2 /4} }}{{a + 1}}} \right],\quad n = 0,1,2, \ldots .
\ee
The total energy of the quantum system under consideration is the following:
\be
\label{e-tot}
E = E_{\rho ,\varphi }  + E_z  = \hbar \omega \left( {a + 1} \right)\left[ {2n + 1 + \frac{{\sqrt {m^2  + a^2 /4} }}{{a + 1}}} \right] + \frac{{\hbar ^2 \kappa _z ^2 }}{{2m_0 }}.
\ee
Substitution of both (\ref{f-01}) and (\ref{pi-01}) in (\ref{p-01}) yields:
\be
\label{p-02}
{\rm P} = \xi ^{\frac{{a + \sqrt {m^2  + a^2 /4} }}{{a + 1}}} e^{ - \frac{1}{2}\xi ^2 } L_n^{\left( {\sqrt {m^2  + a^2 /4} /\left( {a + 1} \right)} \right)} \left( {\xi ^2 } \right).
\ee
It can be orthonormalized thanks to the following well-known orthogonality relation for the generalized Laguerre polynomials:
\be
\label{lp-or}
\int\limits_0^\infty  {e^{ - \zeta} \zeta^{ d} L_m^{\left( d \right)} \left( \zeta \right)L_n^{\left( d \right)} \left( \zeta \right)\textrm{d} \zeta}  = \frac{{\Gamma \left( {n + d + 1} \right)}}{{n!}}\delta _{mn} ,\quad d >  - 1.
\ee
We drop here the straightforward computations and immediately write down the analytical expression of the radial wavefunction (\ref{p-02}):
\be
\label{p-03}
{\rm P}\left( \rho  \right) \equiv {\rm P}_{nm} \left( \rho  \right) = C_{nm}  \cdot \rho ^{a + \sqrt {m^2  + a^2 /4} } e^{ - \frac{{\lambda _0 ^{2\left( {a + 1} \right)} }}{{2\left( {a + 1} \right)}}\rho ^{2\left( {a + 1} \right)} } L_n^{\left( {\sqrt {m^2  + a^2 /4} /\left( {a + 1} \right)} \right)} \left( {\frac{{\lambda _0 ^{2\left( {a + 1} \right)} }}{{a + 1}}\rho ^{2\left( {a + 1} \right)} } \right).
\ee
Here, $C_{nm}$ is the normalization coefficient, the exact expression of which can be easily written down from the orthogonality relation (\ref{lp-or}). 
It has the following analytical expression:
\[
C_{nm}  = \sqrt 2 \lambda _0 ^{a + \sqrt {m^2  + a^2 /4}  + 1} \left( {a + 1} \right)^{ - \frac{{\sqrt {m^2  + a^2 /4} }}{{2\left( {a + 1} \right)}}} \sqrt {\frac{{n!}}{{\Gamma \left( {n + \frac{{\sqrt {m^2  + a^2 /4} }}{{a + 1}} + 1} \right)}}} .
\]

It is clear that the $\left( {\rho,\varphi} \right)$-dependent part of the wavefunction has the following analytical expression:
\be
\label{psi-rhophi-01}
\Psi \left( {\rho ,\varphi } \right) \equiv \Psi _{nm} \left( {\rho ,\varphi } \right) = \frac{1}{{\sqrt {2\pi } }}C_{nm}  \cdot e^{im\varphi } \rho ^{a + \sqrt {m^2  + a^2 /4} } e^{ - \frac{{\lambda _0 ^{2\left( {a + 1} \right)} }}{{2\left( {a + 1} \right)}}\rho ^{2\left( {a + 1} \right)} } L_n^{\left( {\sqrt {m^2  + a^2 /4} /\left( {a + 1} \right)} \right)} \left( {\frac{{\lambda _0 ^{2\left( {a + 1} \right)} }}{{a + 1}}\rho ^{2\left( {a + 1} \right)} } \right).
\ee
Hence, one obtains the following exact expression of the total wavefunction $\psi \left( {x,y,z} \right)$:
\be
\label{total-wf}
\begin{split}
\psi \left( {x,y,z} \right) &\equiv \Psi _{n,m,\kappa _z } \left( {\rho,\varphi,z} \right) \\
&= \frac{1}{{2\pi  }}C_{nm}  \cdot e^{i\left( {m\varphi  + \kappa _z z} \right)} \rho ^{a + \sqrt {m^2  + a^2 /4} } e^{ - \frac{{\lambda _0 ^{2\left( {a + 1} \right)} }}{{2\left( {a + 1} \right)}}\rho ^{2\left( {a + 1} \right)} } L_n^{\left( {\sqrt {m^2  + a^2 /4} /\left( {a + 1} \right)} \right)} \left( {\frac{{\lambda _0 ^{2\left( {a + 1} \right)} }}{{a + 1}}\rho ^{2\left( {a + 1} \right)} } \right).
\end{split}
\ee

The total wavefunctions $\psi \left( {x,y,z} \right)$ defined in (\ref{total-wf}) satisfy the following general orthogonality relation:

\[
\int\limits_{ - \infty }^{ + \infty } {\int\limits_0^{2\pi } {\int\limits_0^\infty  {\Psi _{n,m,\kappa _z } ^* \left( {\rho ,\varphi ,z} \right) \cdot \Psi _{n',m',\kappa _z '} \left( {\rho ,\varphi ,z} \right) \cdot r\textrm{d}r} }  \cdot \textrm{d}\varphi }  \cdot \textrm{d}z = \delta _{n,n'}  \cdot \delta _{m,m'}  \cdot \delta _{\kappa _z ,\kappa _z '} .
\]

\section{Generalization of the solution to the non-canonical case}

Now, we generalize the problem to the non-canonical case.
In this setting, one substitutes in Eq.~(\ref{scheq-03}) the following analytical expressions of the $\left(x;y\right)$ components of the momentum operator in the position representation:
\be
\label{p-ncan}
\hat p_x  \to \hat p_x^p  = \hat p_x  + i\hbar \frac{{\gamma  - 1/2}}{x}\hat R,\quad \hat p_y  \to \hat p_y^p  = \hat p_y  + i\hbar \frac{{\gamma  - 1/2}}{y}\hat R,
\ee
where $\hat R$ acts as a parity or reflection operator in $x$ and $y$~\cite{jafarov2025}: $\hat R f(x,y) = f(-x,-y)$. 
The origin of the parameter $\gamma$ appearing in the components of the momentum operators above is interesting. 
The parameter has a value $\gamma \geq 1/2$, and was introduced by Wigner in his paper~\cite{wigner1950}.
Wigner solved analytically the Heisenberg-Lie equations for the non-relativistic quantum harmonic oscillator problem, and found that an additional parameter $\gamma$ appears as a  deformation parameter of the oscillator Heisenberg-Weyl algebra. It appears as a multiplication parameter: the ground state oscillator energy of the single photon is $\gamma \hbar \omega$ under the non-canonical commutation rules.

Both components of the momentum operator in the position representation introduced in Eq.~(\ref{p-ncan}), commute with the corresponding position operators $\hat x$ and $\hat y$ via the following non-canonical commutation relations in the position representation:
\be
\label{nccr}
\left[ {\hat p_x ,\hat x} \right] = - i\hbar \left[ {1 + \left( {2\gamma  - 1} \right)\hat R} \right],\quad \left[ {\hat p_y ,\hat y} \right] = - i\hbar \left[ {1 + \left( {2\gamma  - 1} \right)\hat R} \right].
\ee

Although both components of the momentum operator introduced in Eq.~(\ref{p-ncan}) exhibit a singularity at $x=0$ (or at $y=0$), 
making the Schr\"odinger equation look singular, it is well-known that this is not a real problem.
Indeed, the admissible wavefunctions are well-known and vanish at the origin due to the appearance of a factor $x^{\gamma-1/2}$ in the wavefunctions~\cite{mukunda1980}. Hence, the singularity is neutralized when the operators act on wavefunctions. 

Of course, everything written up to here is true if one deals with the effective electron mass that is homogeneous. We are going to present below the computations of the Hamiltonian for the case when the effective electron mass varying with position replaces the homogeneous one.
 
In order to keep the solution simpler, we preserved the canonical behavior of the $z$-component momentum operator. 
However, the case of the non-canonical behavior of the momentum operator in the $z$ direction is also exactly solvable in terms of the Bessel functions. 
Some similar examples of such a solution appear in~\cite{landau1977}, where the radial part of the Schr\"odinger equation of the free motion of a particle in a centrally symmetric field is solved analytically in terms of the spherical Bessel functions. 

Again, one preserves the simpler exact solubility property of Eq.~(\ref{scheq-03}) and assumes that the mass $M$ is still of central symmetric nature. Now, straightforward computations yield:
\be
\label{scheq-04-m}
\begin{split}
- \frac{1}{2}\hbar ^2 \frac{1}{{M^{\frac{1}{4}} }}&\left[ {\left( {\frac{\partial }{{\partial x}} - \frac{{\gamma  - 1/2}}{x}\hat R} \right)\frac{1}{{M^{\frac{1}{2}} }}\left( {\frac{\partial }{{\partial x}} - \frac{{\gamma  - 1/2}}{x}\hat R} \right) + \left( {\frac{\partial }{{\partial y}} - \frac{{\gamma  - 1/2}}{y}\hat R} \right)\frac{1}{{M^{\frac{1}{2}} }}\left( {\frac{\partial }{{\partial y}} - \frac{{\gamma  - 1/2}}{y}\hat R} \right)} \right]\frac{1}{{M^{\frac{1}{4}} }} \\
&=- \frac{{\hbar ^2 }}{{2M}}\left[ {\frac{{\partial ^2 }}{{\partial \rho ^2 }} + \left( {\frac{1}{\rho } - \frac{{M'}}{M}} \right)\frac{\partial }{{\partial \rho }} - \frac{1}{4}\frac{{M''}}{M} + \frac{7}{{16}}\left( {\frac{{M'}}{M}} \right)^2  - \left( {\frac{1}{4} - \left( {\gamma  - 1/2} \right)\hat R} \right)\frac{1}{\rho }\frac{{M'}}{M} } \right. \\
&\left. { + \frac{1}{{\rho ^2 }}\frac{{\partial ^2 }}{{\partial \varphi ^2 }} - \frac{{\left( {\gamma  - 1/2} \right)\left( {\gamma  - 1/2 - \hat R} \right)}}{{\rho ^2 }}\left( {\frac{1}{{\cos ^2 \varphi }} + \frac{1}{{\sin ^2 \varphi }}} \right)} \right].
\end{split}
\ee
Substitution in Eq.~(\ref{scheq-03}) leads to the following two second-order differential equations:
\begin{align}
\label{scheq-05-m}
\begin{split}
&\left\{ { - \frac{{\hbar ^2 }}{{2M}}\left[ {\frac{{\partial ^2 }}{{\partial \rho ^2 }} + \left( {\frac{1}{\rho } - \frac{{M'}}{M}} \right)\frac{\partial }{{\partial \rho }} - \frac{1}{4}\frac{{M''}}{M} + \frac{7}{{16}}\left( {\frac{{M'}}{M}} \right)^2  - \left( {\frac{1}{4} - \left( {\gamma  - 1/2} \right)\hat R} \right)\frac{1}{\rho }\frac{{M'}}{M} } \right.} \right. \\
&\left. {\left. { + \frac{1}{{\rho ^2 }}\frac{{\partial ^2 }}{{\partial \varphi ^2 }} - \frac{{\left( {\gamma  - 1/2} \right)\left( {\gamma  - 1/2 - \hat R} \right)}}{{\rho ^2 }}\left( {\frac{1}{{\cos ^2 \varphi }} + \frac{1}{{\sin ^2 \varphi }}} \right)} \right] + \frac{{M\omega ^2 }}{2}\rho ^2 } \right\}\Psi \left( {\rho ,\varphi } \right) = E_{\rho ,\varphi } \Psi \left( {\rho ,\varphi } \right), 
\end{split}
\\ 
\label{scheq-06-m}
&  - \frac{{\hbar ^2 }}{{2m_0 }}\frac{{\partial ^2 }}{{\partial z^2 }}Z\left( z \right) = E_z Z\left( z \right).
\end{align}

Due to the conserved behavior of the $z$-component of the momentum operator, eq.~(\ref{scheq-06-m}) completely overlaps with eq.~(\ref{scheq-05-02}). 
Therefore, the solution of Eq.~(\ref{scheq-06-m}) is simply given by Eq.~(\ref{z-wf}).
Next, Eq.~(\ref{scheq-05-m}) can also be rewritten in terms of eigenfunctions $\Psi \left( {\rho ,\varphi } \right) \equiv {\rm P}\left( \rho  \right)\Phi \left( \varphi  \right)$ and eigenvalues $E_{\rho ,\varphi }$ as follows:
\be
\label{scheq-07-m}
\begin{split}
&\left[ {\rho ^2 \frac{{\partial ^2 }}{{\partial \rho ^2 }} + \left( {\rho  - \rho ^2 \frac{{M'}}{M}} \right)\frac{\partial }{{\partial \rho }} - \frac{1}{4}\frac{{M''}}{M}\rho ^2  + \frac{7}{{16}}\left( {\frac{{M'}}{M}} \right)^2 \rho ^2  - \left( {\frac{1}{4} - \left( {\gamma  - 1/2} \right)\hat R} \right)\frac{{M'}}{M}\rho  + \frac{{2M\rho ^2 E_{\rho ,\varphi } }}{{\hbar ^2 }} - \frac{{M^2 \omega ^2 }}{{\hbar ^2 }}\rho ^4  + } \right. \\
&\left. { + \frac{{\partial ^2 }}{{\partial \varphi ^2 }} - \left( {\gamma  - 1/2} \right)\left( {\gamma  - 1/2 - \hat R} \right)\left( {\frac{1}{{\cos ^2 \varphi }} + \frac{1}{{\sin ^2 \varphi }}} \right)} \right]{\rm P}\left( \rho  \right)\Phi \left( \varphi  \right) = 0. 
\end{split}
\ee
Equation~(\ref{scheq-07-m}) is the non-canonical generalization of Eq.~(\ref{scheq-05-03}).
The non-canonical generalization has an extra parameter $\gamma$ and an extra operator $\hat R$ in its mathematical expression. 
The reflection operator $\hat R$ has eigenvalue $+1$ (resp.\ $-1$) when acting on even (resp.\ odd) functions.
This implies the splitting of the above equation in the following two second-order differential equations.
For the even case, one has
\begin{align}
\label{scheq-08-m}
&\left[ {\rho ^2 \frac{{\partial ^2 }}{{\partial \rho ^2 }} + \left( {\rho  - \rho ^2 \frac{{M'}}{M}} \right)\frac{\partial }{{\partial \rho }} - \frac{1}{4}\frac{{M''}}{M}\rho ^2  + \frac{7}{{16}}\left( {\frac{{M'}}{M}} \right)^2 \rho ^2  + \left( {\gamma  - \frac{3}{4}} \right)\frac{{M'}}{M}\rho  + \frac{{2M\rho ^2 E_{\rho ,\varphi } }}{{\hbar ^2 }} - m^{\left( e \right) ^2}  - \frac{{M^2 \omega ^2 }}{{\hbar ^2 }}\rho ^4 } \right]{\rm P}^{\left( e \right)} \left( \rho  \right) = 0,\\ 
\label{scheq-09-m}
&\left[ {\frac{{\partial ^2 }}{{\partial \varphi ^2 }} - \left( {\gamma  - 1/2} \right)\left( {\gamma  - 3/2} \right)\left( {\frac{1}{{\cos ^2 \varphi }} + \frac{1}{{\sin ^2 \varphi }}} \right)} \right]\Phi ^{\left( e \right)} \left( \varphi  \right) =  - m^{\left( e \right) ^2} \Phi ^{\left( e \right)} \left( \varphi  \right),
\end{align}
and for the odd case:
\begin{align}
\label{scheq-10-m}
&\left[ {\rho ^2 \frac{{\partial ^2 }}{{\partial \rho ^2 }} + \left( {\rho  - \rho ^2 \frac{{M'}}{M}} \right)\frac{\partial }{{\partial \rho }} - \frac{1}{4}\frac{{M''}}{M}\rho ^2  + \frac{7}{{16}}\left( {\frac{{M'}}{M}} \right)^2 \rho ^2  - \left( {\gamma  - \frac{1}{4}} \right)\frac{{M'}}{M}\rho  + \frac{{2M\rho ^2 E_{\rho ,\varphi } }}{{\hbar ^2 }} - m^{\left( o \right) ^2}  - \frac{{M^2 \omega ^2 }}{{\hbar ^2 }}\rho ^4 } \right]{\rm P}^{\left( o \right)} \left( \rho  \right) = 0,\\ 
\label{scheq-11-m}
&\left[ {\frac{{\partial ^2 }}{{\partial \varphi ^2 }} - \left( {\gamma  - 1/2} \right)\left( {\gamma  + 1/2} \right)\left( {\frac{1}{{\cos ^2 \varphi }} + \frac{1}{{\sin ^2 \varphi }}} \right)} \right]\Phi ^{\left( o \right)} \left( \varphi  \right) =  - m^{\left( o \right) ^2} \Phi ^{\left( o \right)} \left( \varphi  \right).
\end{align}

Observe that Eqs.~(\ref{scheq-09-m}) and~(\ref{scheq-11-m}) are essentially the same up to a shift of $\left( {\gamma  - 1/2} \right)$ to $\left( {\gamma  + 1/2} \right)$. 
So it is sufficient to solve only one of the equations. 
Both equations coincide with the second order differential equation for the non-relativistic P\"oschl-Teller potential problem~\cite{flugge2012}. 
We shall solve Eq.~(\ref{scheq-11-m}) for the region $0 < \varphi  < \pi /2$ and then extend the solution appropriately to the region $\pi /2 < \varphi  < 2\pi$.
Introducing the new variable $\eta  = \cos \left( {2\varphi } \right)$, one obtains
\be
\label{scheq-12-m}
4\left( {1 - \eta ^2 } \right){\Phi ^{\left( o \right)}} '' - 4\eta {\Phi ^{\left( o \right)}} ' + \left( {m^{\left( o \right) ^2}  - \frac{{4\gamma ^2  - 1}}{{1 - \eta ^2 }}} \right)\Phi ^{\left( o \right)}  = 0.
\ee

We look for solutions of Eq.~(\ref{scheq-12-m}) of the following form:
\be
\label{phichixi-1}
\Phi ^{\left( o \right)}  = \chi \left( \eta  \right) \cdot \xi \left( \eta  \right).
\ee
Its substitution in Eq.~(\ref{scheq-12-m}) transforms this equation to the following second-order differential equation:
\be
\label{scheq-13-m}
4\left( {1 - \eta ^2 } \right)\chi  \cdot \xi '' + 4\left[ {2\left( {1 - \eta ^2 } \right)\chi ' - \eta \chi } \right] \cdot \xi ' + \left[ {4\left( {1 - \eta ^2 } \right)\chi '' - 4\eta \chi ' + \left( {m^{\left( o \right) ^2}  - \frac{{4\gamma ^2  - 1}}{{1 - \eta ^2 }}} \right)\chi } \right] \cdot \xi  = 0.
\ee
We assume that the function $\chi$ is responsible for the boundary behavior of the wave function $\Phi ^{\left( o \right)}$ at singular values $0$ and $\pi / 2$ of $\varphi$, whereas the other function $\xi$ is of polynomial nature.
Then, one needs to substitute
\be
\label{chi1}
\chi  = \left( {1 - \eta ^2 } \right)^A ,
\ee
where our next task is to find analytical expressions of $A$. 
Straightforward computations lead to the following quadratic equation for the parameter $A$:
\[
4A^2  - 2A - \left( {\gamma ^2  - 1/4} \right) = 0,
\]
with solution
\[
A = \frac{1}{4}\left( {1 + 2\gamma } \right).
\]
Substitution of (\ref{chi1}), with this $A$-value, in Eq.~(\ref{scheq-13-m}) yields:
\be
\label{scheq-14-m}
\left( {1 - \eta ^2 } \right)\xi '' - \left( {2\gamma  + 2} \right)\eta \xi ' + \left[ {\left( {m^{\left( o \right)} /2} \right)^2  - \left( {\gamma  + 1/2} \right)^2 } \right]\xi  = 0.
\ee

From comparison of the above second-order differential equation with the following equation for the Gegenbauer polynomials~\cite{koekoek2010}
\begin{align}
\label{scheq-15-m}
&\left( {1 - \eta ^2 } \right)\xi '' - \left( {2\bar \lambda  + 1} \right)\eta \xi ' + m\left( {m + 2\bar \lambda } \right)\xi  = 0,\quad m = 0,1,2, \cdots , \\
\label{geg-p}
&\xi  = C_m^{\left( {\bar \lambda } \right)} \left( \eta  \right) = \frac{{\left( {2\bar \lambda } \right)_m }}{{m!}}\,_2 F_1 \left( {\begin{array}{*{20}c}
   { - m,m + 2\bar \lambda }  \\
   {\bar \lambda  + 1/2}  \\
\end{array};\frac{{1 - \eta }}{2}} \right),
\end{align}
one obtains that
\begin{align}
\label{wf-odd-f}
\Phi ^{\left( o \right)}  \equiv \Phi _m^{\left( o \right)} \left( {\varphi } \right)  &= C_m^o  \cdot \left[ {1 - \cos ^2 \left( {2\varphi } \right)} \right]^{\frac{1}{2}\left( {\gamma  + 1/2} \right)}  \cdot C_m^{\left( {\gamma  + 1/2} \right)} \left( {\cos \left( {2\varphi } \right)} \right), \\
\label{m-odd}
&m^{\left( o \right)}  =  \pm \left[ {2\left( {\gamma  + m} \right) + 1} \right],
\end{align}
where the normalization coefficient $C_m^o$ can be easily deduced from well known orthogonality relation for the Gegenbauer polynomials.
Careful analysis allows the extension from $0 < \varphi  < \pi / 2$ to $0 < \varphi  < 2 \pi$ as follows:
\be
\label{cm-odd}
C_m^o  = 2^{\gamma-1}   \cdot \varepsilon  \cdot \Gamma \left( {\gamma  + 1/2} \right)\sqrt {\frac{{\left( {m + \gamma  + 1/2} \right)m!}}{{\pi \Gamma \left( {m + 2\gamma  + 1} \right)}}} ,
\ee
with the coefficient $\varepsilon$ defining the phase behavior of the angular distance wavefunction in its singularity points:
\be
\label{epsil}
\varepsilon  = \begin{cases}
 1,& 0 < \varphi  < \pi /2, \\ 
 \left( { - 1} \right)^m ,& \pi /2 < \varphi  < \pi , \\ 
 1 ,& \pi  < \varphi  < 3\pi /2, \\ 
 \left( { - 1} \right)^m,& 3\pi /2 < \varphi  < 2\pi . 
 \end{cases}
\ee

Now, the exact solution to Eq.~(\ref{scheq-09-m}) can be easily obtained by the action $\gamma  \to \gamma  - 1$:
\begin{align}
\label{wf-even-f}
\Phi ^{\left( e \right)}  \equiv \Phi _m^{\left( e \right)} \left( {\varphi } \right) &= C_m^e  \cdot \left[ {1 - \cos ^2 \left( {2\varphi } \right)} \right]^{\frac{1}{2}\left( {\gamma  - 1/2} \right)}  \cdot C_m^{\left( {\gamma  - 1/2} \right)} \left( {\cos \left( {2\varphi } \right)} \right), \\
\label{m-even}
&m^{\left( e \right)}  =  \pm \left[ {2\left( {\gamma  + m} \right) - 1} \right],
\end{align}
where
\be
\label{cm-even}
C_m^e  = 2^{\gamma  - 2} \cdot \varepsilon  \cdot \Gamma \left( {\gamma  - 1/2} \right)\sqrt {\frac{{\left( {m + \gamma  - 1/2} \right)m!}}{{\pi \Gamma \left( {m + 2\gamma  - 1} \right)}}} .
\ee

One observes that both wavefunctions $\Phi _m^{\left( e \right)} \left( {\varphi } \right)$ and $\Phi _m^{\left( o \right)} \left( {\varphi } \right)$ correctly satisfy the condition $\Phi \left( {\varphi + 2 \pi } \right)=\Phi \left( {\varphi } \right)$. Therefore, our solution to the angular part of the non-canonical wavefunction is complete.

Let us also analyze the possibility of obtaining exact solutions to the radial equations (\ref{scheq-08-m}) and (\ref{scheq-10-m}). 
Substitution of (\ref{m-rho}) in the even states equation (\ref{scheq-08-m}) and further division by $\rho^2$ yields:
\be
\label{scheq-16-m}
\frac{{\partial ^2 {\rm P}^{\left( e \right)} }}{{\partial \rho ^2 }} + \left( {1 - 2a} \right)\frac{1}{\rho }\frac{{\partial {\rm P}^{\left( e \right)} }}{{\partial \rho }} + \left[ {\kappa _{\left( {\rho ,\varphi } \right)} ^2 \lambda _0 ^{2a} \rho ^{2a}  - \lambda _0 ^{4a + 4} \rho ^{4a + 2}  - \frac{{m^{\left( e \right) ^2}  - \frac{3}{4}a^2  - \left( {2\gamma  - 1} \right)a}}{{\rho ^2 }}} \right]{\rm P}^{\left( e \right)}  = 0.
\ee
Similarly, substitution of (\ref{m-rho}) in the odd states equation (\ref{scheq-10-m}) yields:
\be
\label{scheq-17-m}
\frac{{\partial ^2 {\rm P}^{\left( o \right)} }}{{\partial \rho ^2 }} + \left( {1 - 2a} \right)\frac{1}{\rho }\frac{{\partial {\rm P}^{\left( o \right)} }}{{\partial \rho }} + \left[ {\kappa _{\left( {\rho ,\varphi } \right)} ^2 \lambda _0 ^{2a} \rho ^{2a}  - \lambda _0 ^{4a + 4} \rho ^{4a + 2}  - \frac{{m^{\left( o \right) ^2}  - \frac{3}{4}a^2  + \left( {2\gamma  - 1} \right)a}}{{\rho ^2 }}} \right]{\rm P}^{\left( o \right)}  = 0.
\ee
The contact point transformation (\ref{cpt}) changes them as follows:
\begin{align}
\label{r-even-01}
& {\rm P}^{\left( e \right)''} + \frac{{1 - a}}{{\left( {a + 1} \right)\xi }} \cdot {\rm P}^{\left( e \right)'}  + \left[ {\frac{{\kappa _{\left( {\rho ,\varphi } \right)} ^2 \lambda _0 ^{ - 2} }}{{a + 1}} - \xi ^2  - \frac{{m^{\left( e \right) ^2}  - \frac{3}{4}a^2  - \left( {2\gamma  - 1} \right)a}}{{\left( {a + 1} \right)^2 \xi ^2 }}} \right]{\rm P}^{\left( e \right)}  = 0, \\
\label{r-odd-01}
& {\rm P}^{\left( o \right)''} + \frac{{1 - a}}{{\left( {a + 1} \right)\xi }} \cdot {\rm P}^{\left( o \right)'}  + \left[ {\frac{{\kappa _{\left( {\rho ,\varphi } \right)} ^2 \lambda _0 ^{ - 2} }}{{a + 1}} - \xi ^2  - \frac{{m^{\left( o \right) ^2}  - \frac{3}{4}a^2  + \left( {2\gamma  - 1} \right)a}}{{\left( {a + 1} \right)^2 \xi ^2 }}} \right]{\rm P}^{\left( o \right)}  = 0.
\end{align}
Both equations also can be rewritten in the following compact form similar to Eq.~(\ref{scheq-rho-05}):
\begin{align}
\label{r-even-02}
& {\rm P}^{\left( e \right) ''} + \frac{{1 - a}}{{\left( {a + 1} \right)\xi }} \cdot {\rm P}^{\left( e \right) '} + \frac{{\kappa _{\left( {\rho ,\varphi } \right)} ^2 \lambda _0 ^{ - 2} \left( {a + 1} \right)\xi ^2  - \left( {a + 1} \right)^2 \xi ^4  - m^{\left( e \right) ^2}  + \frac{3}{4}a^2  + \left( {2\gamma  - 1} \right)a}}{{\left( {a + 1} \right)^2 \xi ^2 }}{\rm P}^{\left( e \right)}  = 0, \\
\label{r-odd-02}
& {\rm P}^{\left( o \right) ''} + \frac{{1 - a}}{{\left( {a + 1} \right)\xi }} \cdot {\rm P}^{\left( o \right) '} + \frac{{\kappa _{\left( {\rho ,\varphi } \right)} ^2 \lambda _0 ^{ - 2} \left( {a + 1} \right)\xi ^2  - \left( {a + 1} \right)^2 \xi ^4  - m^{\left( o \right) ^2}  + \frac{3}{4}a^2  - \left( {2\gamma  - 1} \right)a}}{{\left( {a + 1} \right)^2 \xi ^2 }}{\rm P}^{\left( o \right)}  = 0. 
\end{align}
Then, the exact solution of both of them can easily be obtained in a similar way as that of Eq.~(\ref{scheq-rho-05}). 
We drop these long computations and immediately write down the solution:
\begin{align}
\label{r-even-03}
& {\rm P}^{\left( e \right)} \left( \rho  \right) \equiv {\rm P}_{nm}^{\left( e \right)} \left( \rho  \right) = C_{nm}^{\left( e \right)}  \cdot \rho ^{a + \sqrt {m^{\left( e \right) ^2}  + a^2 /4 - \left( {2\gamma  - 1} \right)a} } e^{ - \frac{{\lambda _0 ^{2\left( {a + 1} \right)} }}{{2\left( {a + 1} \right)}}\rho ^{2\left( {a + 1} \right)} } L_n^{\left( {\sqrt {m^{\left( e \right) ^2}  + a^2 /4 - \left( {2\gamma  - 1} \right)a} /\left( {a + 1} \right)} \right)} \left( {\frac{{\lambda _0 ^{2\left( {a + 1} \right)} }}{{a + 1}}\rho ^{2\left( {a + 1} \right)} } \right), \\
\label{r-odd-03}
& {\rm P}^{\left( o \right)} \left( \rho  \right) \equiv {\rm P}_{nm}^{\left( o \right)} \left( \rho  \right) = C_{nm}^{\left( o \right)}  \cdot \rho ^{a + \sqrt {m^{\left( o \right) ^2}  + a^2 /4 + \left( {2\gamma  - 1} \right)a} } e^{ - \frac{{\lambda _0 ^{2\left( {a + 1} \right)} }}{{2\left( {a + 1} \right)}}\rho ^{2\left( {a + 1} \right)} } L_n^{\left( {\sqrt {m^{\left( o \right) ^2}  + a^2 /4 + \left( {2\gamma  - 1} \right)a} /\left( {a + 1} \right)} \right)} \left( {\frac{{\lambda _0 ^{2\left( {a + 1} \right)} }}{{a + 1}}\rho ^{2\left( {a + 1} \right)} } \right), 
\end{align}
where the exact expressions of the normalization coefficients of even and odd cases are as follows:
\begin{align}
\label{c-even}
&  C_{nm}^{\left( e \right)}  = \lambda _0 ^{a + \sqrt {m^{\left( e \right) ^2}  + a^2 /4 - \left( {2\gamma  - 1} \right)a}  + 1} \left( {a + 1} \right)^{ - \frac{{\sqrt {m^{\left( e \right) ^2}  + a^2 /4 - \left( {2\gamma  - 1} \right)a} }}{{2\left( {a + 1} \right)}}} \sqrt {\frac{{2 \cdot n!}}{{\Gamma \left( {n + \frac{{\sqrt {m^{\left( e \right) ^2}  + a^2 /4 - \left( {2\gamma  - 1} \right)a} }}{{a + 1}} + 1} \right)}}} , \\
\label{c-odd}
& C_{nm}^{\left( o \right)}  = \lambda _0 ^{a + \sqrt {m^{\left( o \right) ^2}  + a^2 /4 + \left( {2\gamma  - 1} \right)a}  + 1} \left( {a + 1} \right)^{ - \frac{{\sqrt {m^{\left( o \right) ^2}  + a^2 /4 + \left( {2\gamma  - 1} \right)a} }}{{2\left( {a + 1} \right)}}} \sqrt {\frac{{2 \cdot n!}}{{\Gamma \left( {n + \frac{{\sqrt {m^{\left( o \right) ^2}  + a^2 /4 + \left( {2\gamma  - 1} \right)a} }}{{a + 1}} + 1} \right)}}} . 
\end{align}

The energy spectrum of both states also can be written down easily by applying a method similar to Eq.~(\ref{e-rho-phi}). It yields:
\begin{align}
\label{e-even}
& E^{\left( e \right)}  = E_{\rho ,\varphi }^{\left( e \right)}  + E_z  = \hbar \omega \left( {a + 1} \right)\left[ {2n + 1 + \frac{{\sqrt {m^{\left( e \right) ^2}  + a^2 /4 - \left( {2\gamma  - 1} \right)a} }}{{a + 1}}} \right] + \frac{{\hbar ^2 \kappa _z ^2 }}{{2m_0 }}, \\ 
\label{e-odd}
& E^{\left( o \right)}  = E_{\rho ,\varphi }^{\left( o \right)}  + E_z  = \hbar \omega \left( {a + 1} \right)\left[ {2n + 1 + \frac{{\sqrt {m^{\left( o \right) ^2}  + a^2 /4 + \left( {2\gamma  - 1} \right)a} }}{{a + 1}}} \right] + \frac{{\hbar ^2 \kappa _z ^2 }}{{2m_0 }}.
\end{align}

One also writes down the total wavefunctions of the stationary states as follows:

\begin{align}
\label{psit-even-01}
& \Psi _{n,m,\kappa _z } ^{\left( e \right)} \left( {\rho ,\varphi ,z} \right)={\rm P}_{nm}^{\left( e \right)} \left( \rho  \right) \cdot \Phi _m^{\left( e \right)} \left( {\varphi } \right) \cdot Z_{\kappa _z}\left( z \right) , \\
\label{psit-odd-01}
& \Psi _{n,m,\kappa _z } ^{\left( o \right)} \left( {\rho ,\varphi ,z} \right)={\rm P}_{nm}^{\left( o \right)} \left( \rho  \right) \cdot \Phi _m^{\left( o \right)} \left( {\varphi } \right) \cdot Z_{\kappa _z}\left( z \right). 
\end{align}

Both the even and odd total wavefunctions satisfy the same orthogonality relation as in the canonical case:

\[
\int\limits_{ - \infty }^{ + \infty } {\int\limits_0^{2\pi } {\int\limits_0^\infty  {\Psi _{n,m,\kappa _z } ^* \left( {\rho ,\varphi ,z} \right) \cdot \Psi _{n',m',\kappa _z '} \left( {\rho ,\varphi ,z} \right) \cdot r\textrm{d}r} }  \cdot \textrm{d}\varphi }  \cdot \textrm{d}z = \delta _{n,n'}  \cdot \delta _{m,m'}  \cdot \delta _{\kappa _z ,\kappa _z '} .
\]

The main difference from the canonical case is that now the computations rely on the orthogonality relation for Gegenbauer polynomials:

\[
\int\limits_{ - 1}^1 {\left( {1 - \eta^2 } \right)^{\bar \lambda  - \frac{1}{2}} C_m^{\left( \bar \lambda  \right)} \left( \eta \right)C_{m'}^{\left( \bar \lambda  \right)} \left( \eta \right)\textrm{d}\eta}  = \frac{{\pi \Gamma \left( {n + 2\bar \lambda } \right)2^{1 - 2\bar \lambda } }}{{\left\{ {\Gamma \left( \bar \lambda  \right)} \right\}^2 \left( {n + \bar \lambda } \right)n!}}\delta _{m,m'} ,\;\bar \lambda  >  - \frac{1}{2},\;\bar \lambda  \ne 0.
\]

\section{Discussions}

\begin{figure}
\centering
\subfloat[$a=-0.6$, $\left| {\Psi _{00 } } \right|^2$]{\label{4figs-a} \includegraphics[width=0.24\textwidth]{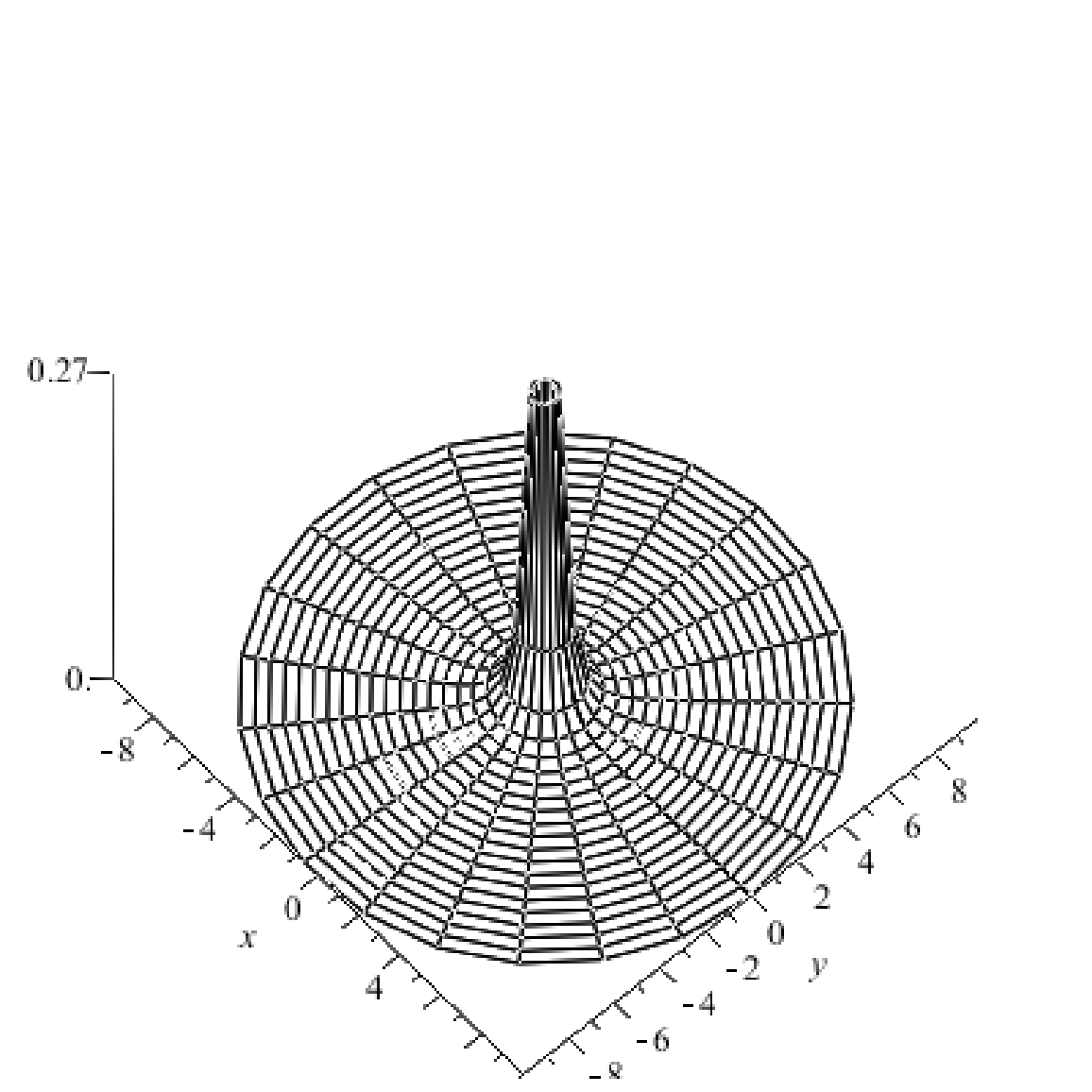}}
\hfill
\subfloat[$a=-0.6$, $\left| {\Psi _{10 } } \right|^2$]{\label{4figs-b} \includegraphics[width=0.24\textwidth]{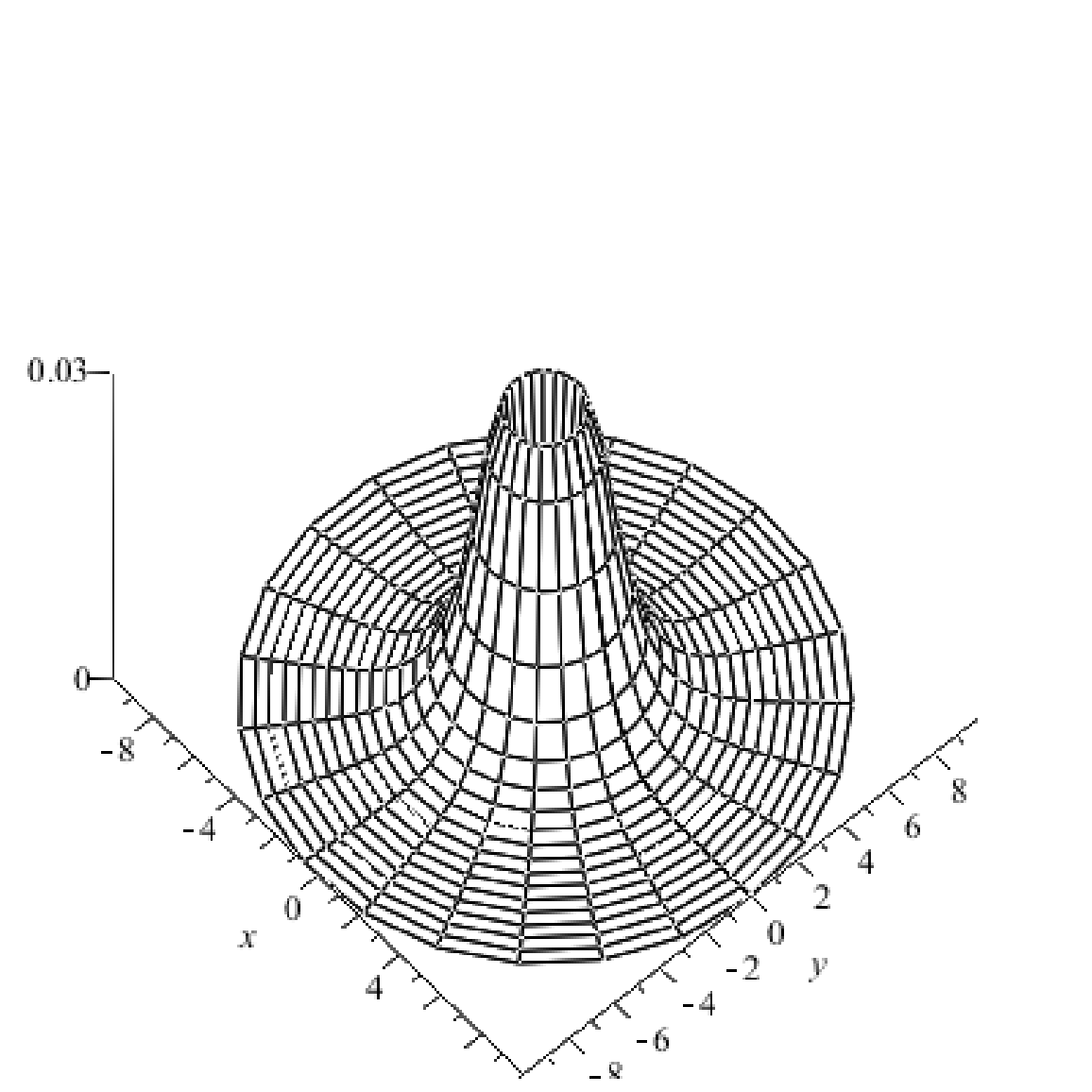}}%
\hfill
\subfloat[$a=-0.6$, $\left| {\Psi _{01 } } \right|^2$]{\label{4figs-c} \includegraphics[width=0.24\textwidth]{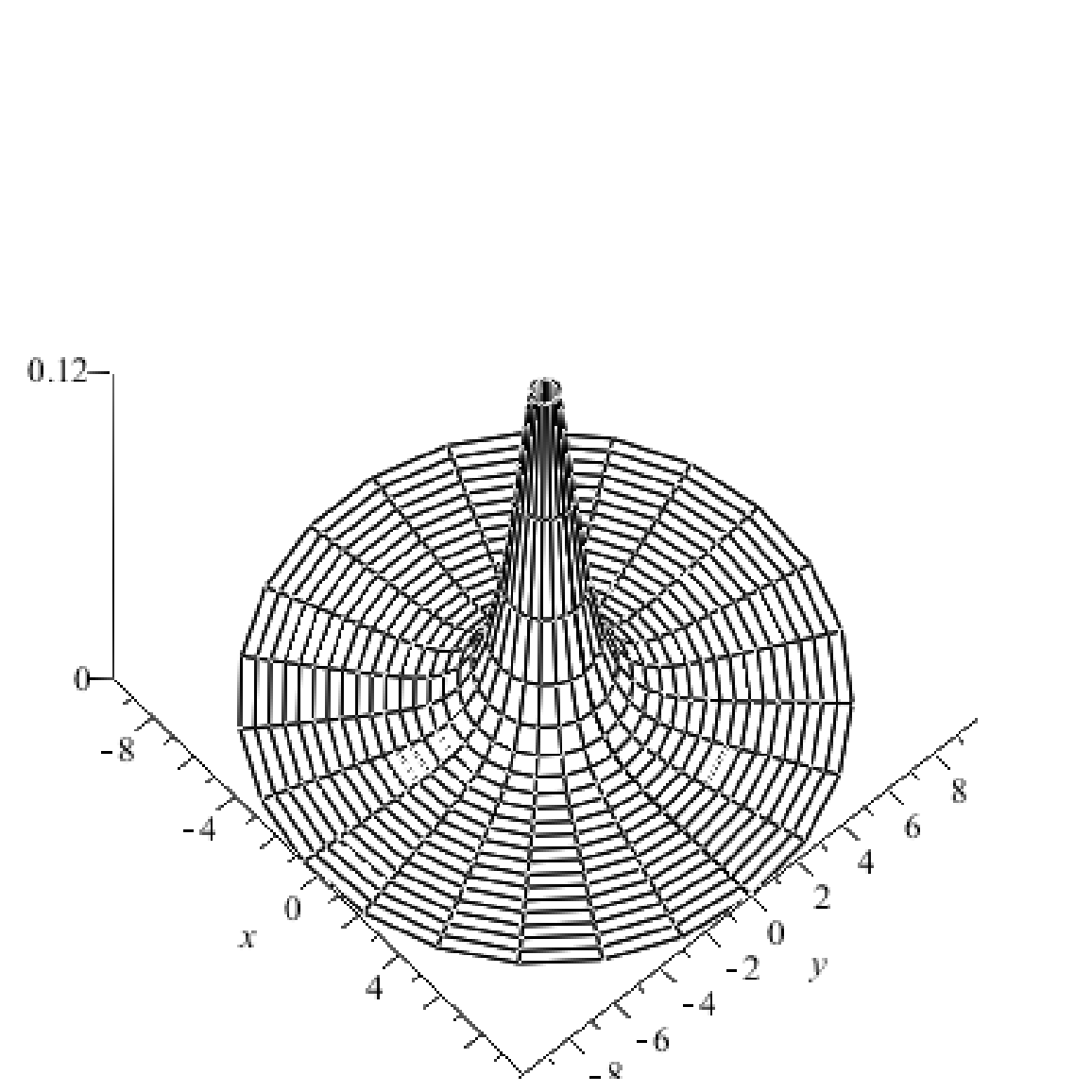}}%
\hfill
\subfloat[$a=-0.6$, $\left| {\Psi _{11 } } \right|^2$]{\label{4figs-d} \includegraphics[width=0.24\textwidth]{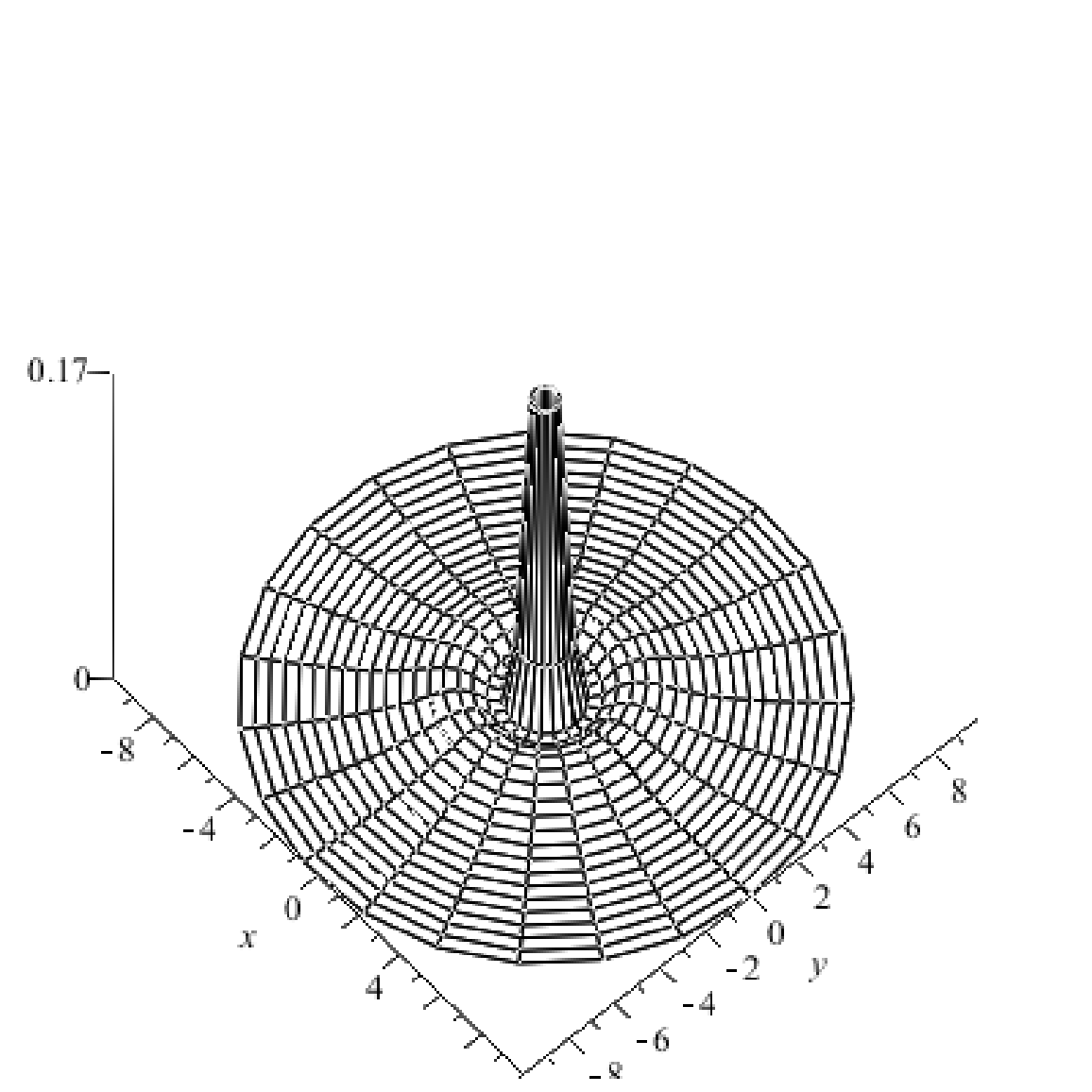}}%
\hfill
\subfloat[$a=0$, $\left| {\Psi _{00 } } \right|^2$]{\label{4figs-e} \includegraphics[width=0.24\textwidth]{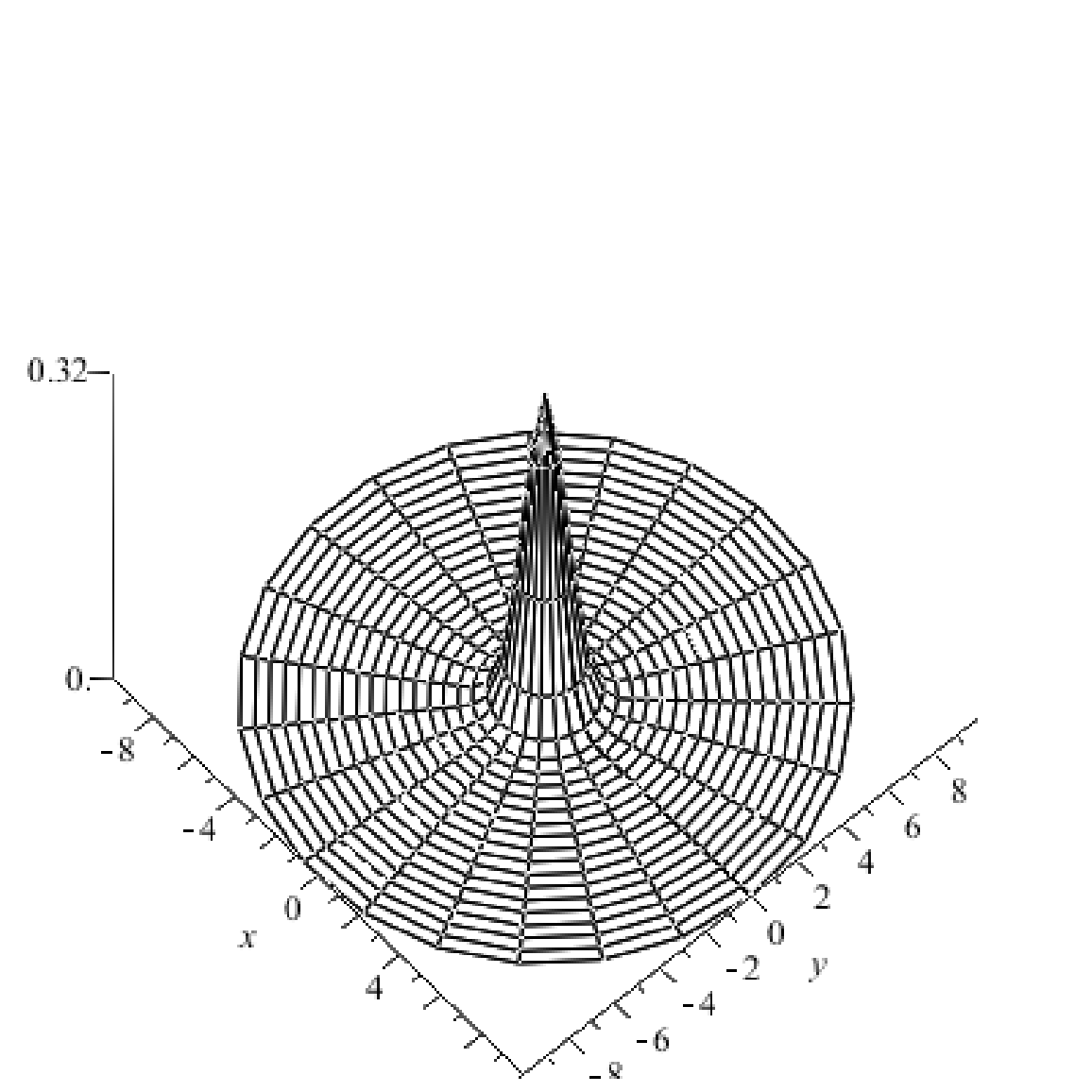}}
\hfill
\subfloat[$a=0$, $\left| {\Psi _{10 } } \right|^2$]{\label{4figs-f} \includegraphics[width=0.24\textwidth]{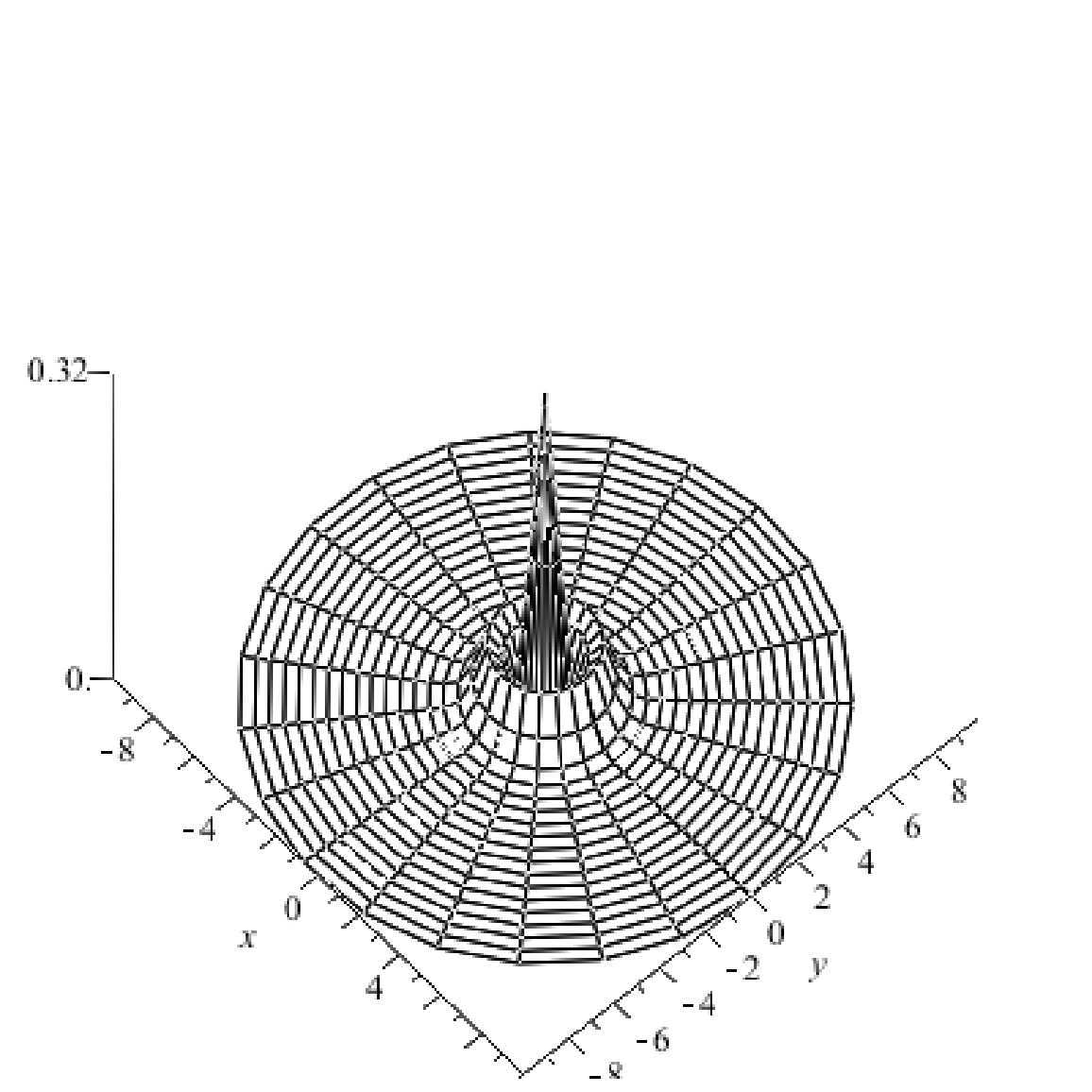}}%
\hfill
\subfloat[$a=0$, $\left| {\Psi _{01 } } \right|^2$]{\label{4figs-g} \includegraphics[width=0.24\textwidth]{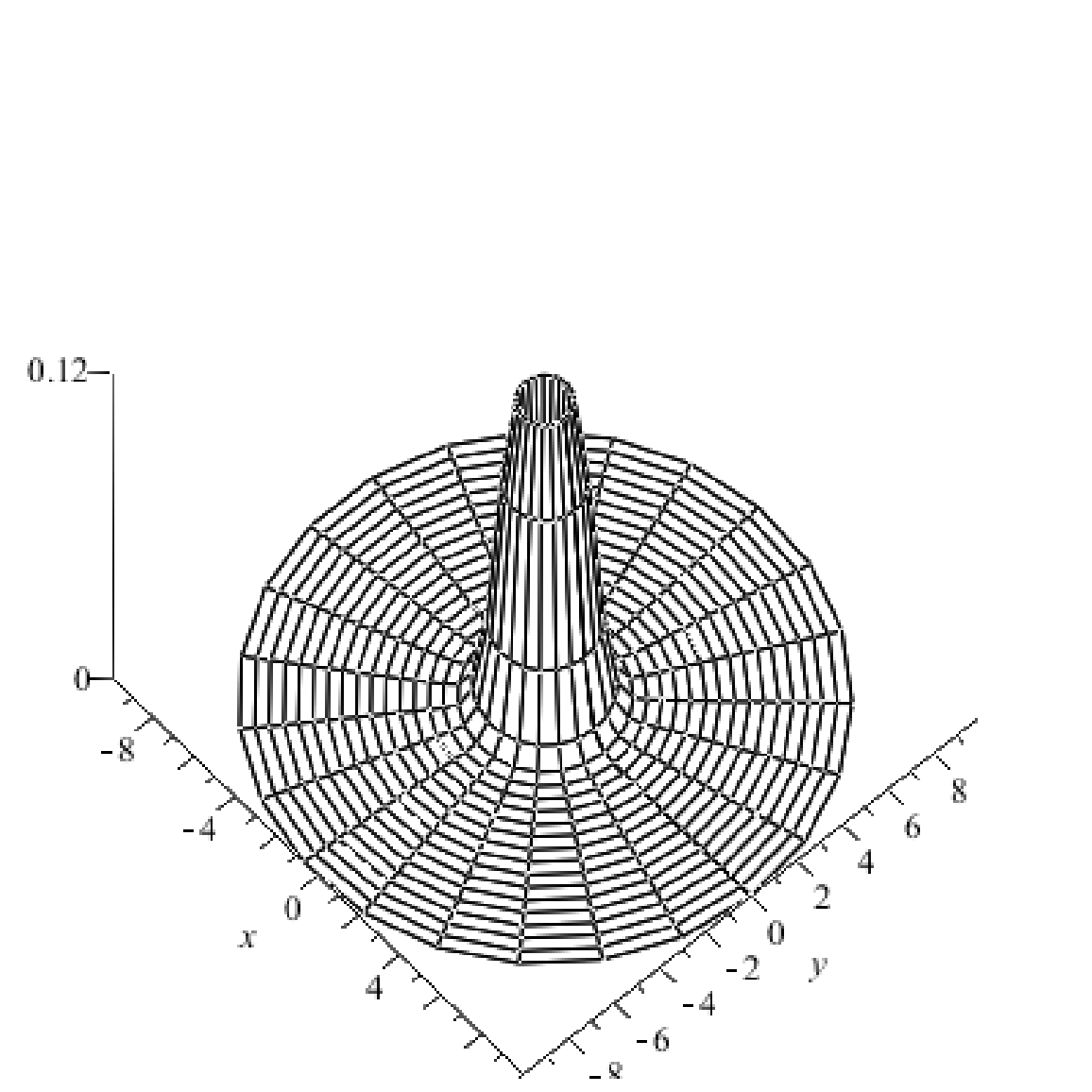}}%
\hfill
\subfloat[$a=0$, $\left| {\Psi _{11 } } \right|^2$]{\label{4figs-h} \includegraphics[width=0.24\textwidth]{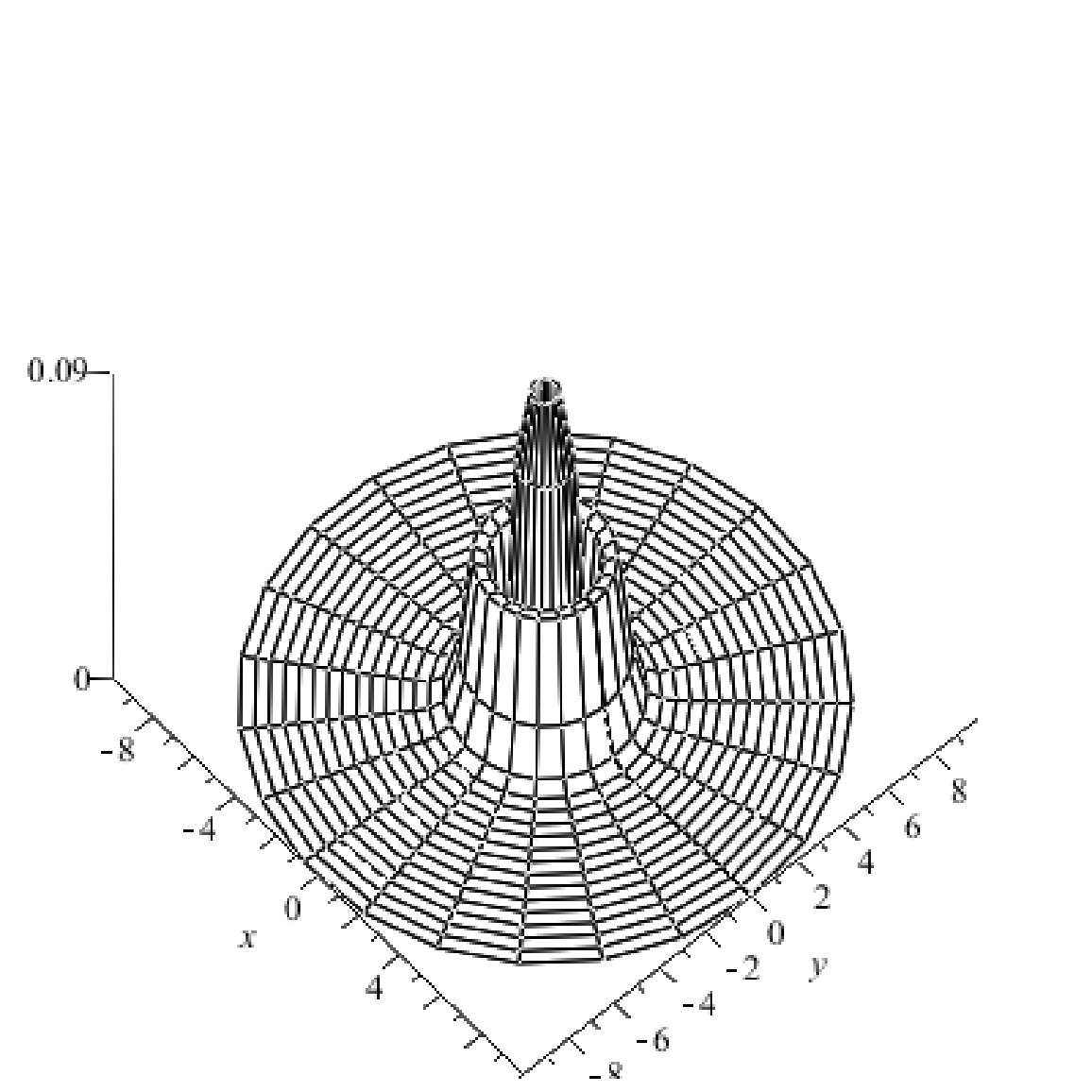}}%
\hfill
\subfloat[$a=2.0$, $\left| {\Psi _{00 } } \right|^2$]{\label{4figs-i} \includegraphics[width=0.24\textwidth]{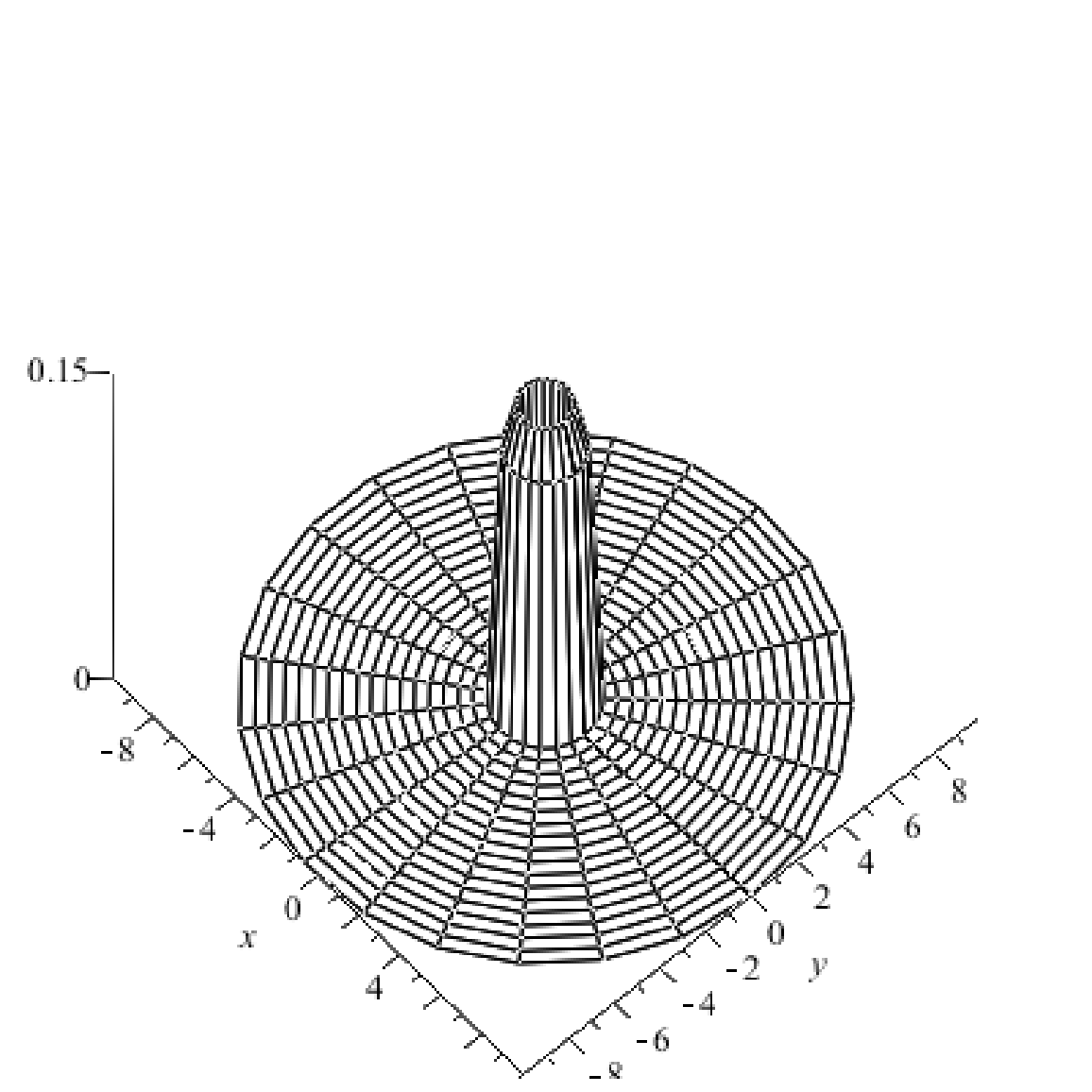}}
\hfill
\subfloat[$a=2.0$, $\left| {\Psi _{10 } } \right|^2$]{\label{4figs-j} \includegraphics[width=0.24\textwidth]{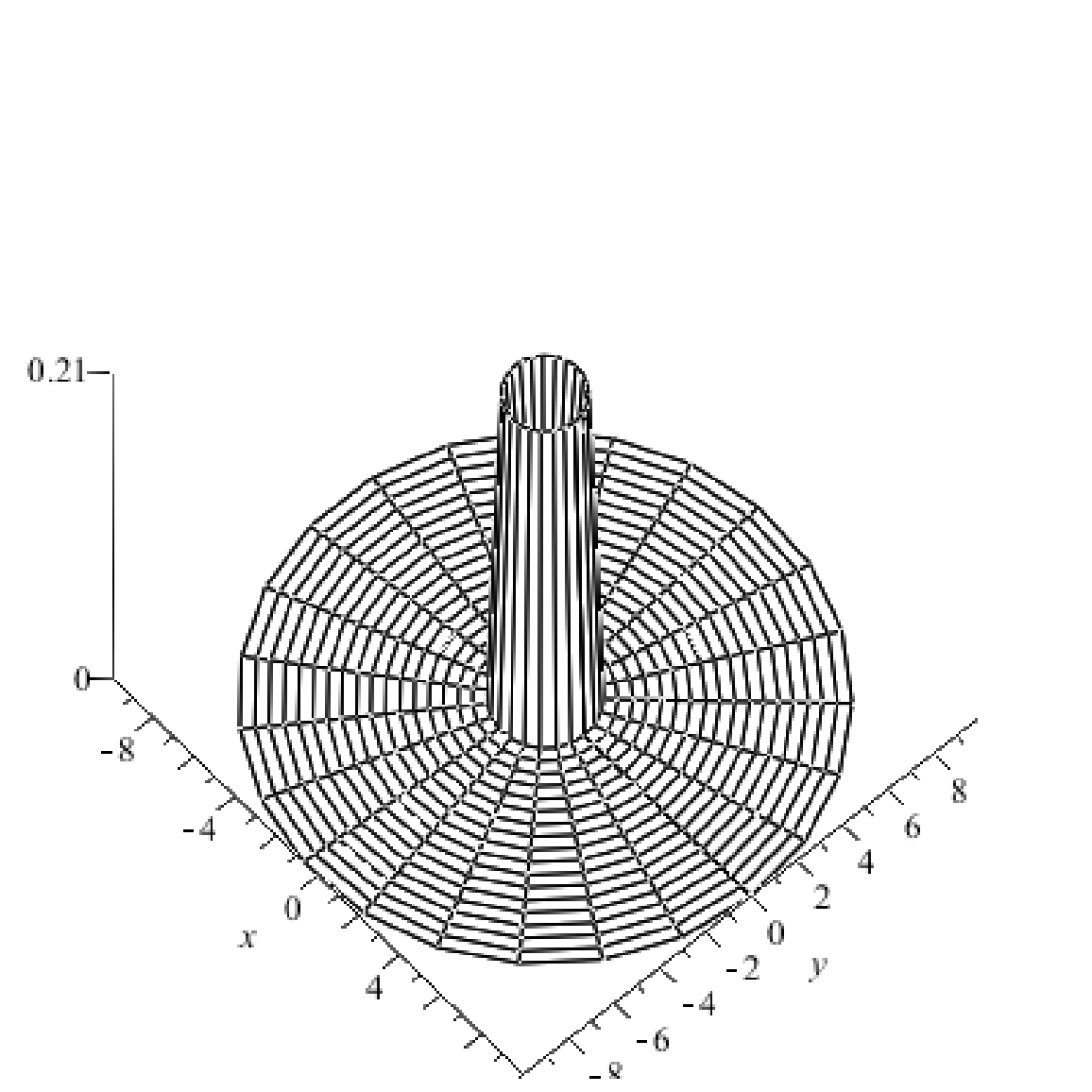}}%
\hfill
\subfloat[$a=2.0$, $\left| {\Psi _{01 } } \right|^2$]{\label{4figs-k} \includegraphics[width=0.24\textwidth]{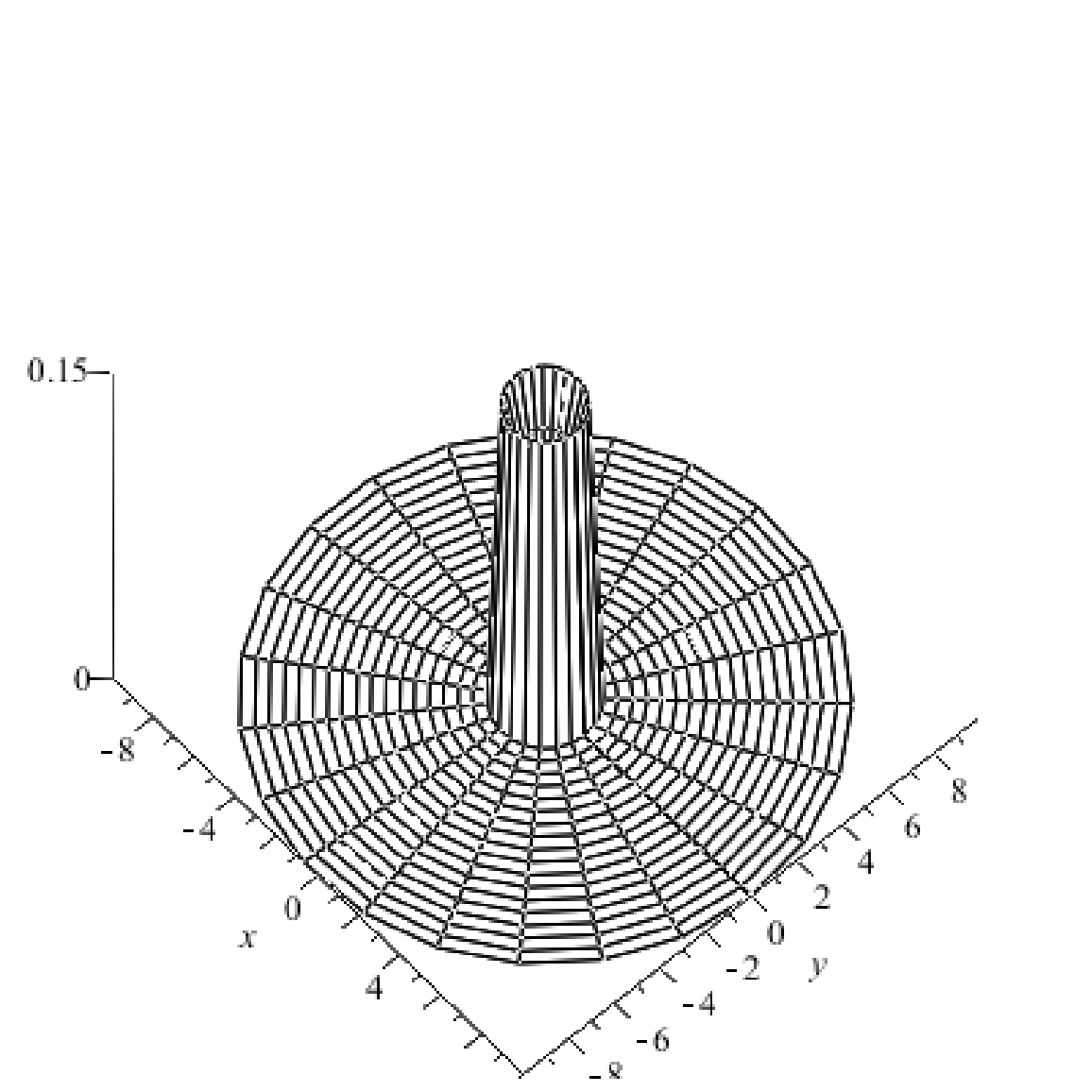}}%
\hfill
\subfloat[$a=2.0$, $\left| {\Psi _{11 } } \right|^2$]{\label{4figs-l} \includegraphics[width=0.24\textwidth]{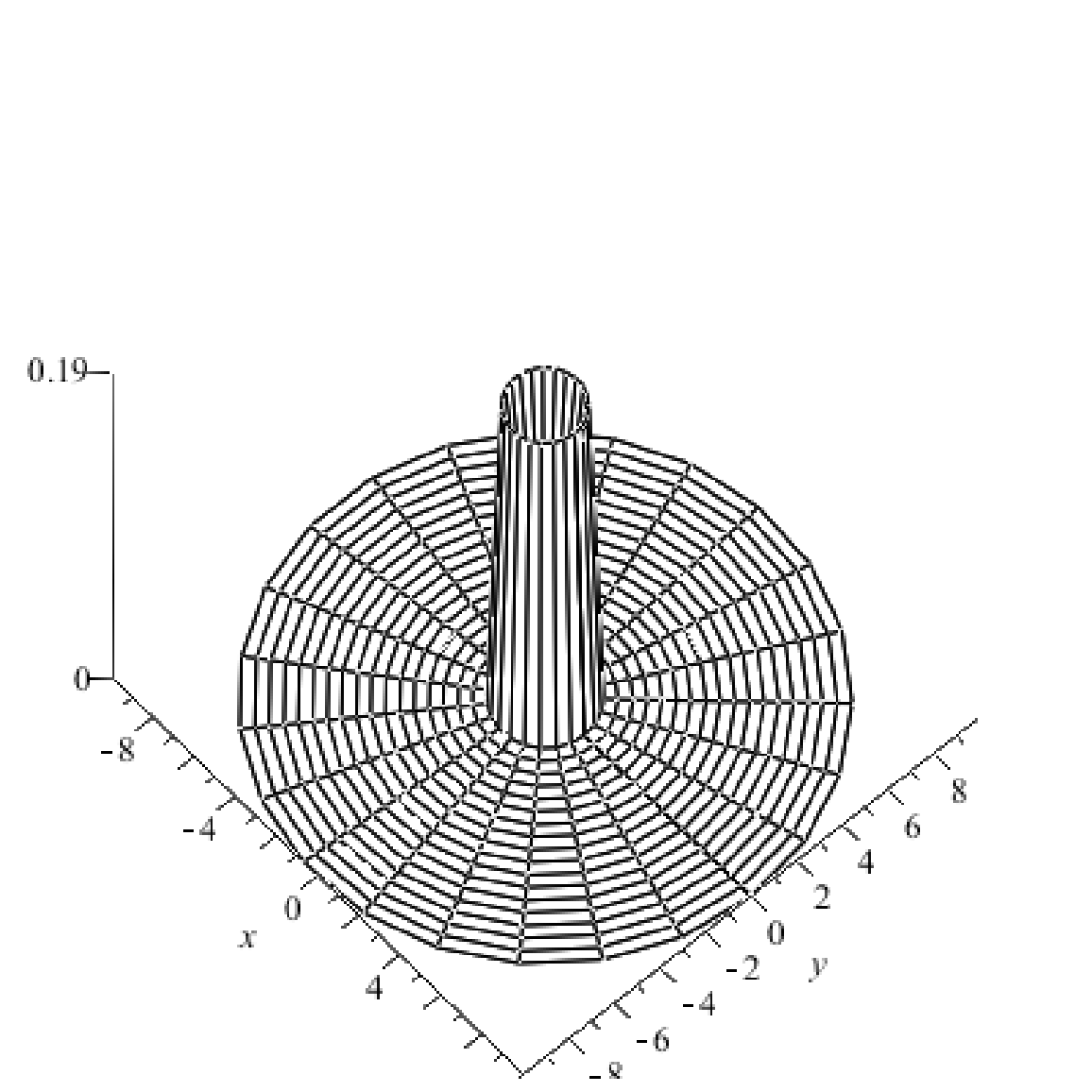}}%
\caption{Dependence of the probabilistic densities $\left| {\Psi _{nm } } \right|^2$ generated via (\ref{psi-rhophi-01}) for values of $a=\left(-0.6;0;2.0\right)$ as well as $n=0,1$ and $m=0,1$. Aiming to achieve the best informative observability of the central symmetric density, all the plots are generated for values $\varphi=0...2\pi$ ($m_0=\omega=\hbar=1$). All plots correspond to the canonical case, which means that $\gamma=0.5$.}
\label{4figs}
\end{figure}

As we highlighted in the Introduction, our main goal was to present an exact solution to the ideal one-dimensional electron gas confined with an anisotropic quantum wire of a non-rectangular profile. 
We achieved the initial step of our main goal because we managed to obtain exact expressions of both the energy spectrum (\ref{e-tot}) and wavefunctions of the stationary states (\ref{total-wf}) in the canonical case. 
The probabilistic densities $\left| {\Psi _{nm } } \right|^2$ generated via (\ref{psi-rhophi-01}) are presented in Fig.~\ref{4figs}. 
First of all, note that the energy spectrum (\ref{e-tot}) preserves its linear dependence property on the radial quantum number $n$, but loses this property for the quantum number $m$. 
The linear dependence on $m$ can be recovered only if $a = 0$. 
This case also defines the limit under which the confinement effect disappears and all expressions convert into the well-known exact solution of the Schr\"odinger equation for the circular harmonic oscillator. 
This can be observed through the following limit relation for the energy spectrum (\ref{e-tot}):
\be
\label{e-tot-lim}
\mathop {\lim }\limits_{a \to 0} E = \hbar \omega \left( {2n + \left| m \right| + 1} \right) + \frac{{\hbar ^2 \kappa _z ^2 }}{{2m_0 }}.
\ee
The same is true for the wavefunctions of the stationary states (\ref{total-wf}):
\be
\label{total-wf-lim}
\mathop {\lim }\limits_{a \to 0} \psi \left( {x,y,z} \right) 
= \frac{1}{{2\pi  }}C_{nm}  \cdot e^{i\left( {m\varphi  + \kappa _z z} \right)} \rho ^{\left| m \right| } e^{ - \frac{{\lambda _0 ^{2} }}{{2}}\rho ^{2} } L_n^{\left( {\left| m \right|} \right)} \left( {\lambda _0 ^{2} \rho ^{2} } \right).
\ee
Here,
\[
C_{nm}  = \sqrt 2 \lambda _0 ^{m  + 1} \sqrt {\frac{{n!}}{{\Gamma \left( {n + m + 1} \right)}}} .
\]

For the value of $a=-0.6$ the circular oscillator potential (\ref{p-uc-osc}) transforms to the two-dimensional analogue of the one-dimensional triangular-shaped potential (it can be considered as a cone-shaped potential, see Fig.~\ref{fig.1}).
Then, one observes from the plots (a)--(d) of Fig.~\ref{4figs} that for this value the wavefunctions change from ordinary harmonic oscillator-like to a very specific behavior that can be studied in depth. 
We restricted the plots of higher excited states to values of $n=0,1$ and $m=0,1$. 
The plots (e)--(h) of Fig.~\ref{4figs} are for the limit case $a=0$, corresponding to constant mass and no confinement.
The plots (i)--(l) of Fig.~\ref{4figs} correspond to the case where the circular oscillator potential (\ref{p-uc-osc}) behaves itself as a quantum wire with doubled confinement effect.

\begin{figure}
\centering
\subfloat[$a=-0.6$, $\left| {\Psi _{00 } } \right|^2$]{\label{4figs2-a} \includegraphics[width=0.24\textwidth]{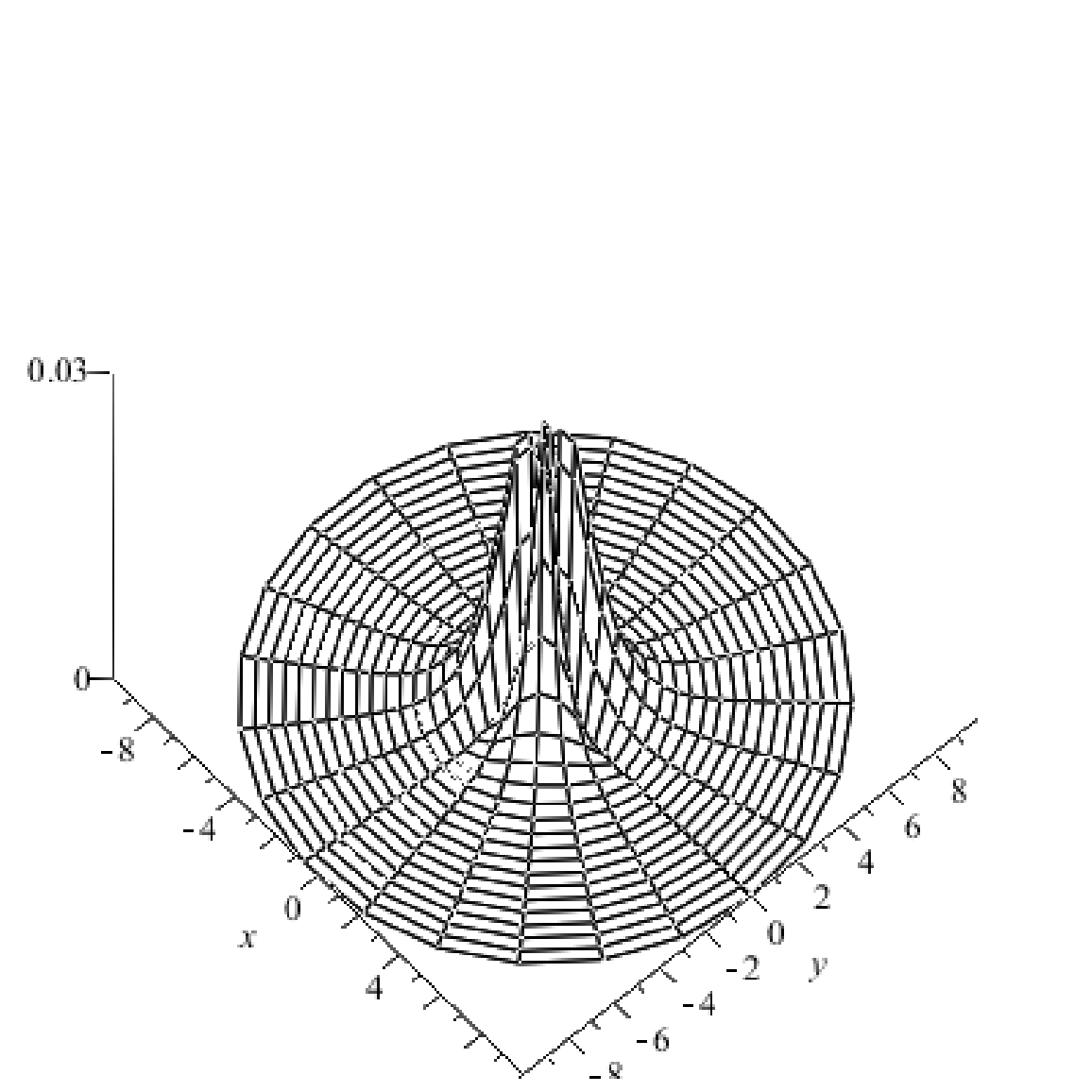}}
\hfill
\subfloat[$a=-0.6$, $\left| {\Psi _{10 } } \right|^2$]{\label{4figs2-b} \includegraphics[width=0.24\textwidth]{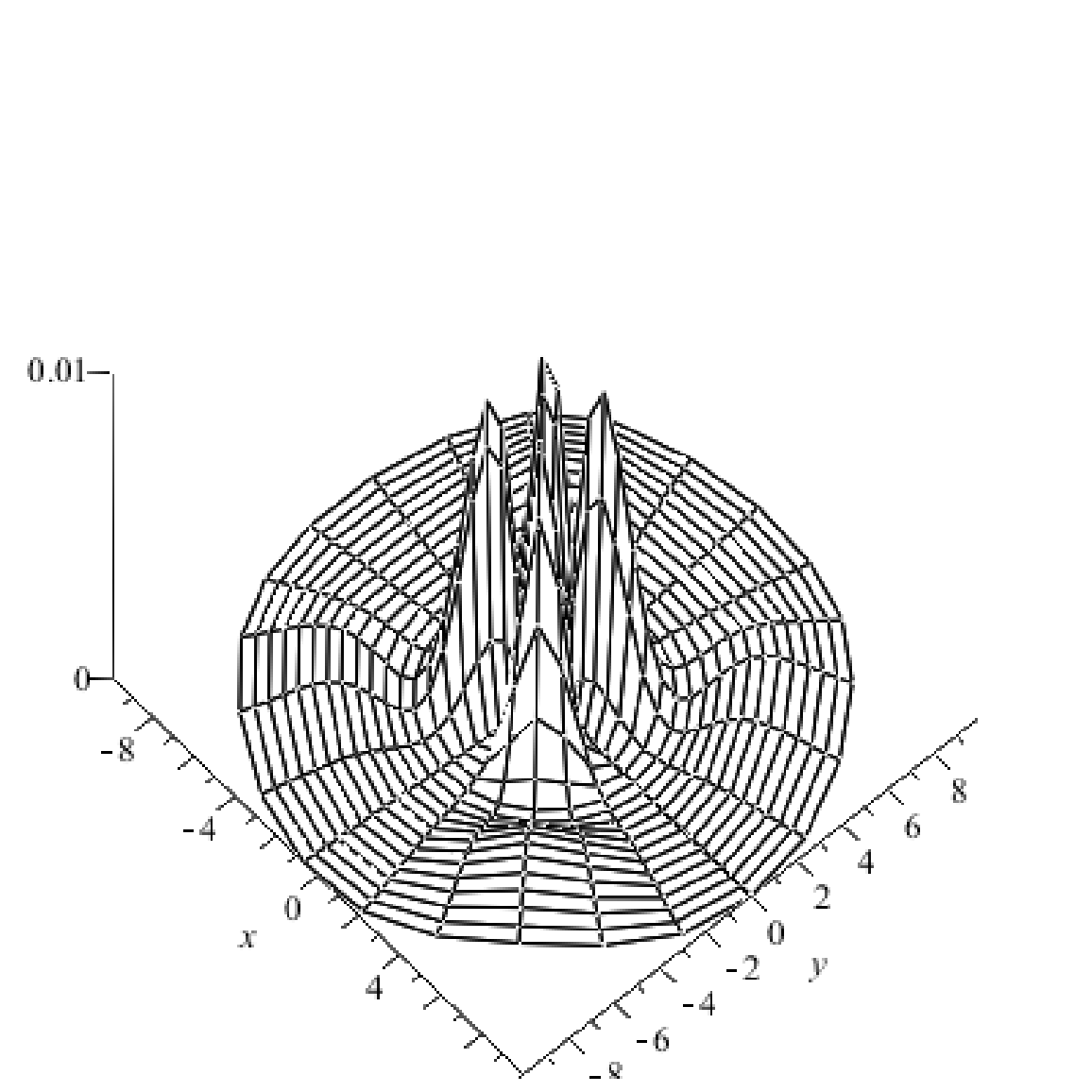}}%
\hfill
\subfloat[$a=-0.6$, $\left| {\Psi _{01 } } \right|^2$]{\label{4figs2-c} \includegraphics[width=0.24\textwidth]{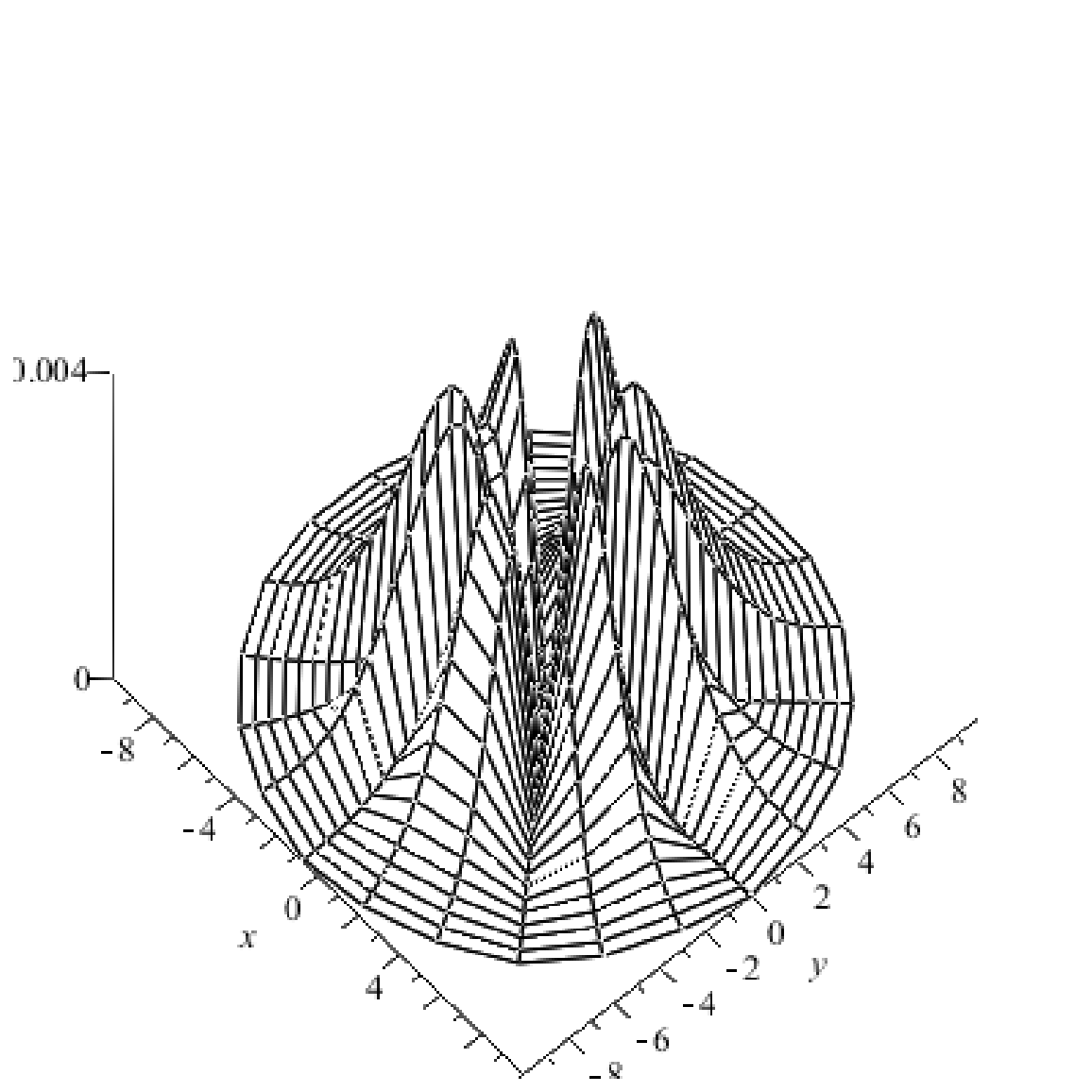}}%
\hfill
\subfloat[$a=-0.6$, $\left| {\Psi _{11 } } \right|^2$]{\label{4figs2-d} \includegraphics[width=0.24\textwidth]{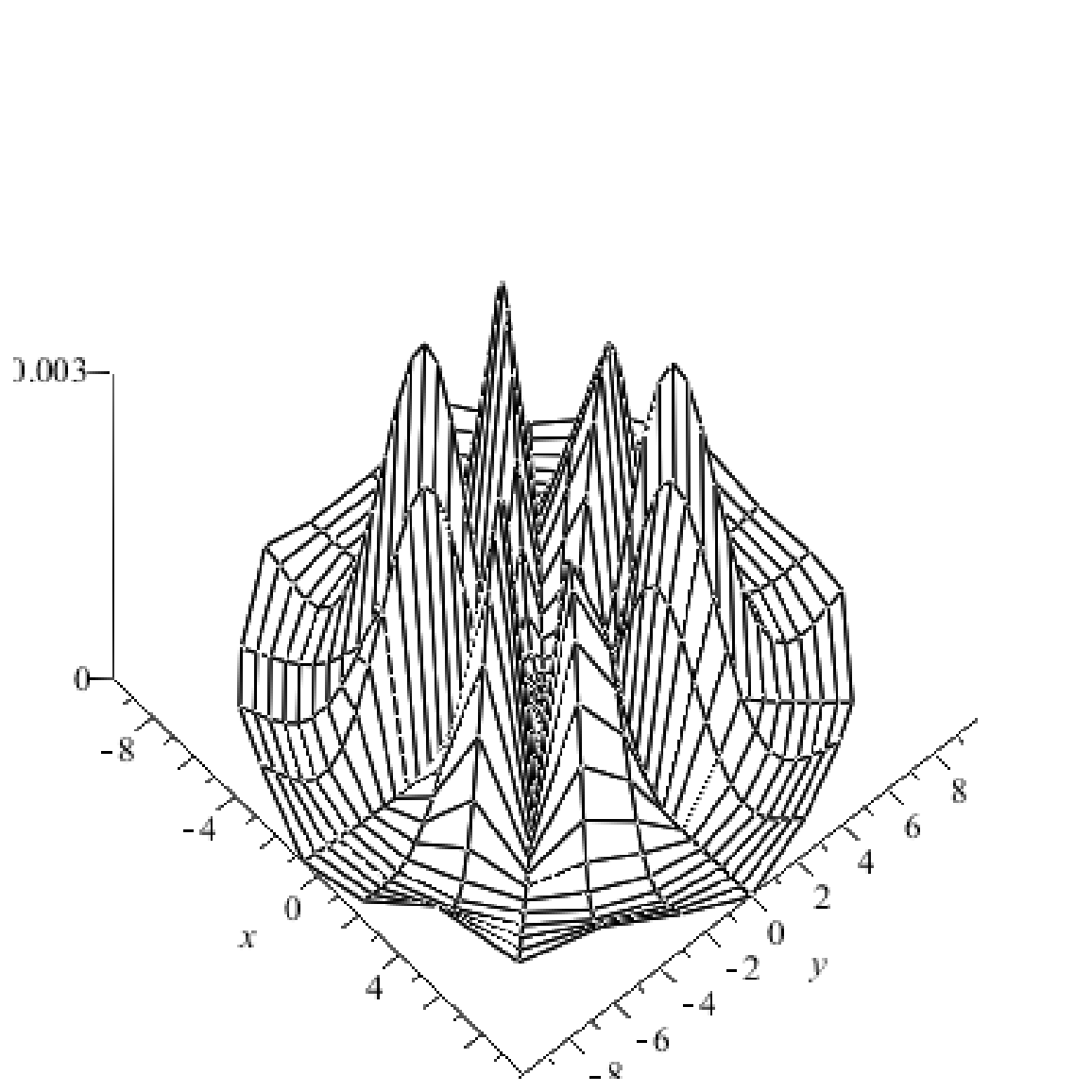}}%
\hfill
\subfloat[$a=0$, $\left| {\Psi _{00 } } \right|^2$]{\label{4figs2-e} \includegraphics[width=0.24\textwidth]{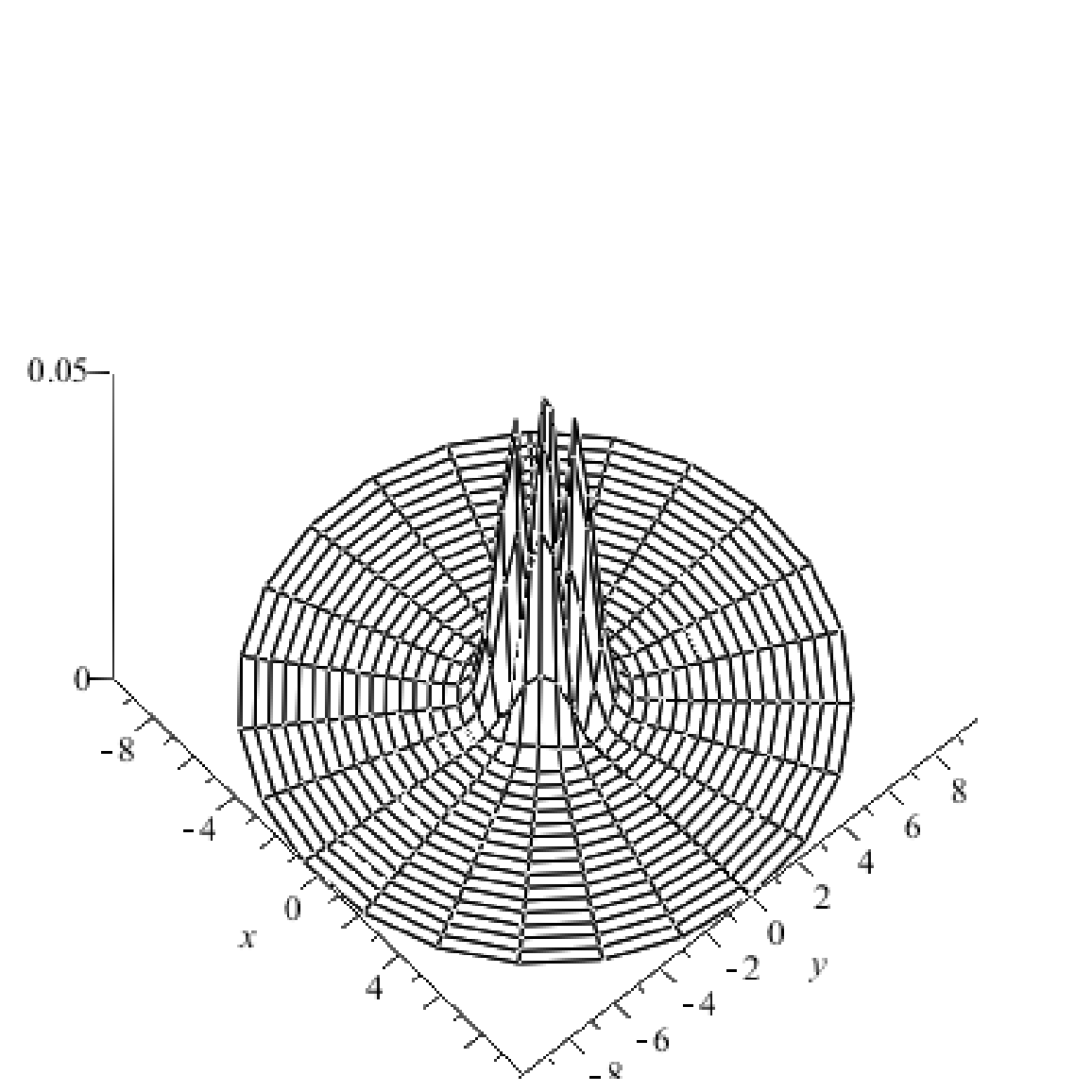}}
\hfill
\subfloat[$a=0$, $\left| {\Psi _{10 } } \right|^2$]{\label{4figs2-f} \includegraphics[width=0.24\textwidth]{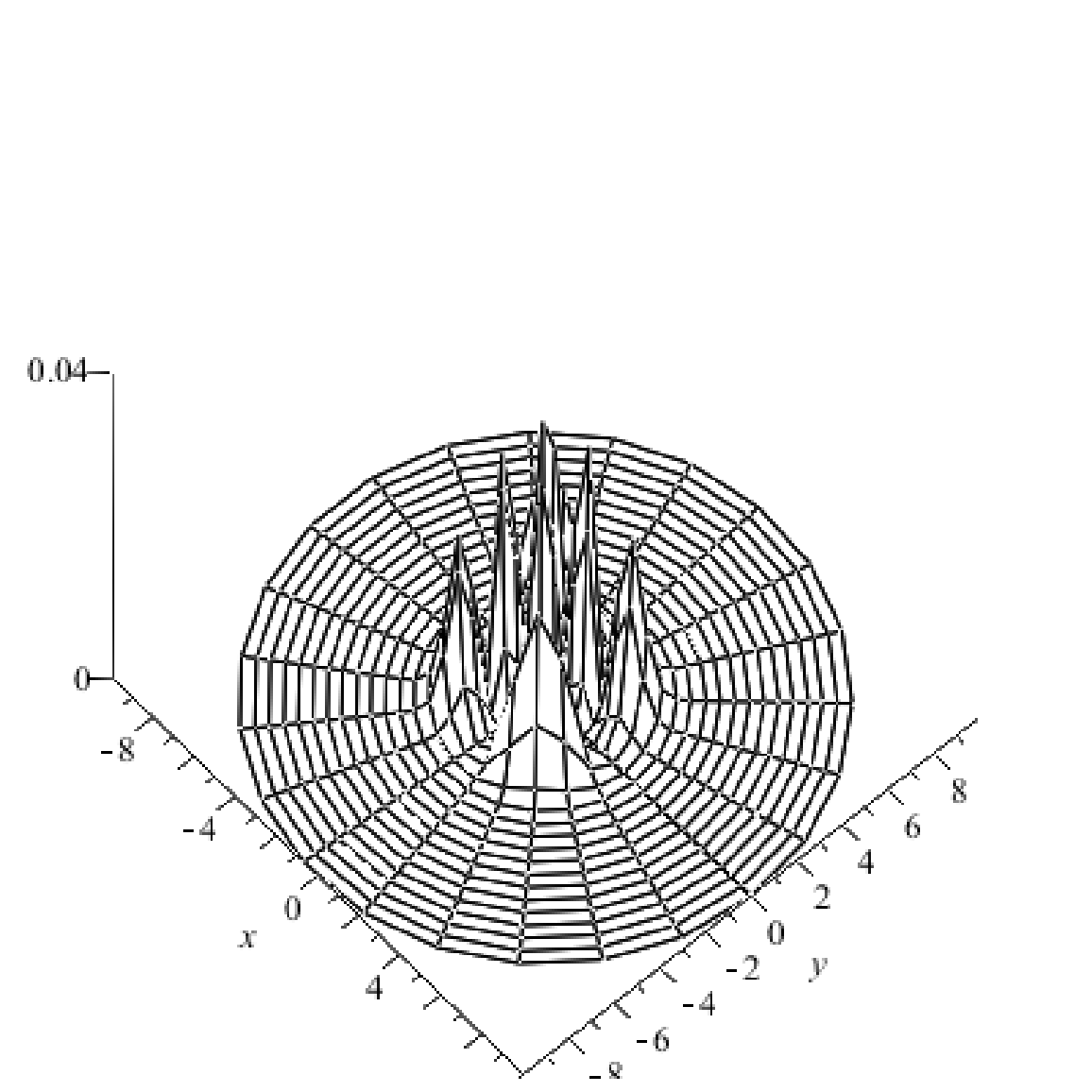}}%
\hfill
\subfloat[$a=0$, $\left| {\Psi _{01 } } \right|^2$]{\label{4figs2-g} \includegraphics[width=0.24\textwidth]{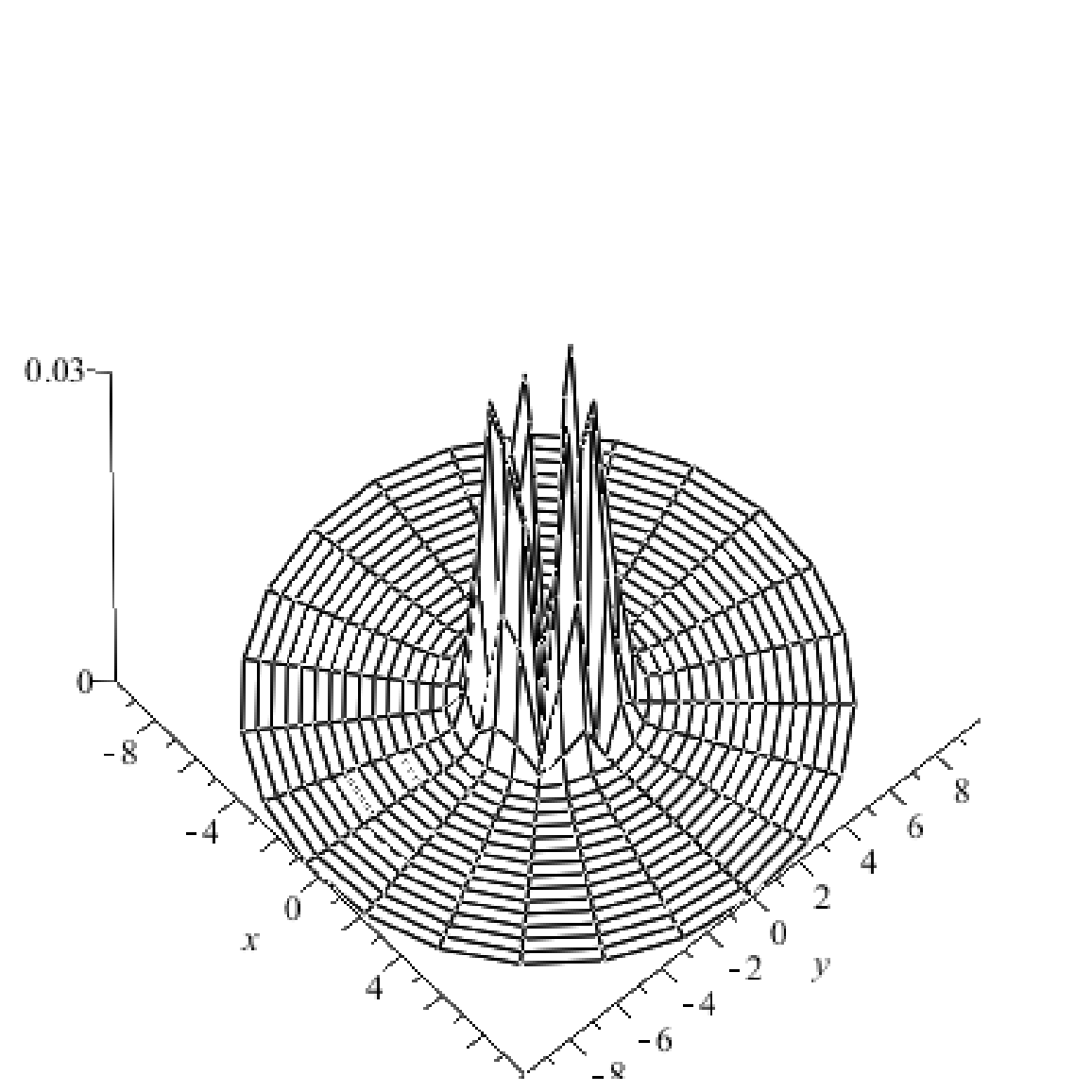}}%
\hfill
\subfloat[$a=0$, $\left| {\Psi _{11 } } \right|^2$]{\label{4figs2-h} \includegraphics[width=0.24\textwidth]{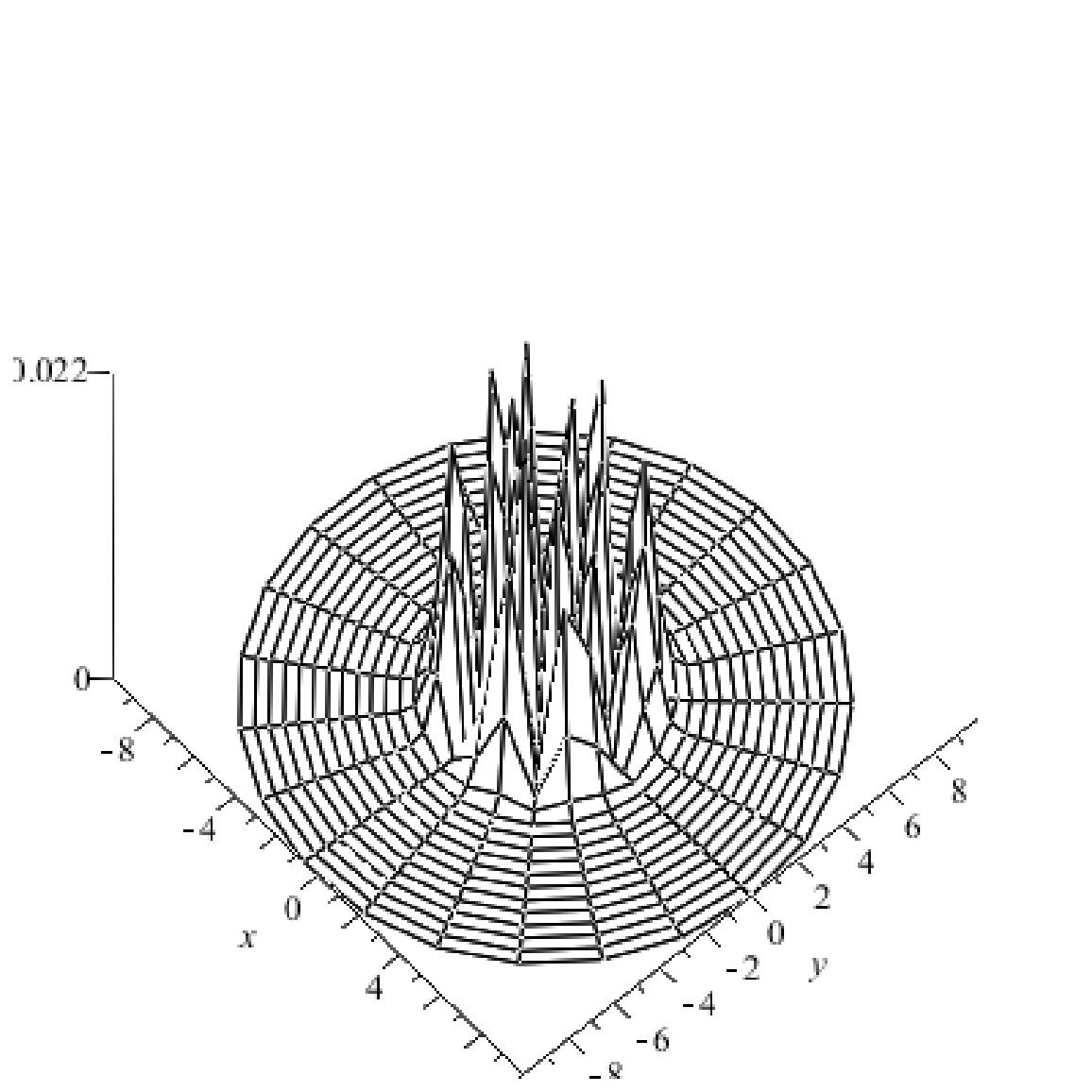}}%
\hfill
\subfloat[$a=2.0$, $\left| {\Psi _{00 } } \right|^2$]{\label{4figs2-i} \includegraphics[width=0.24\textwidth]{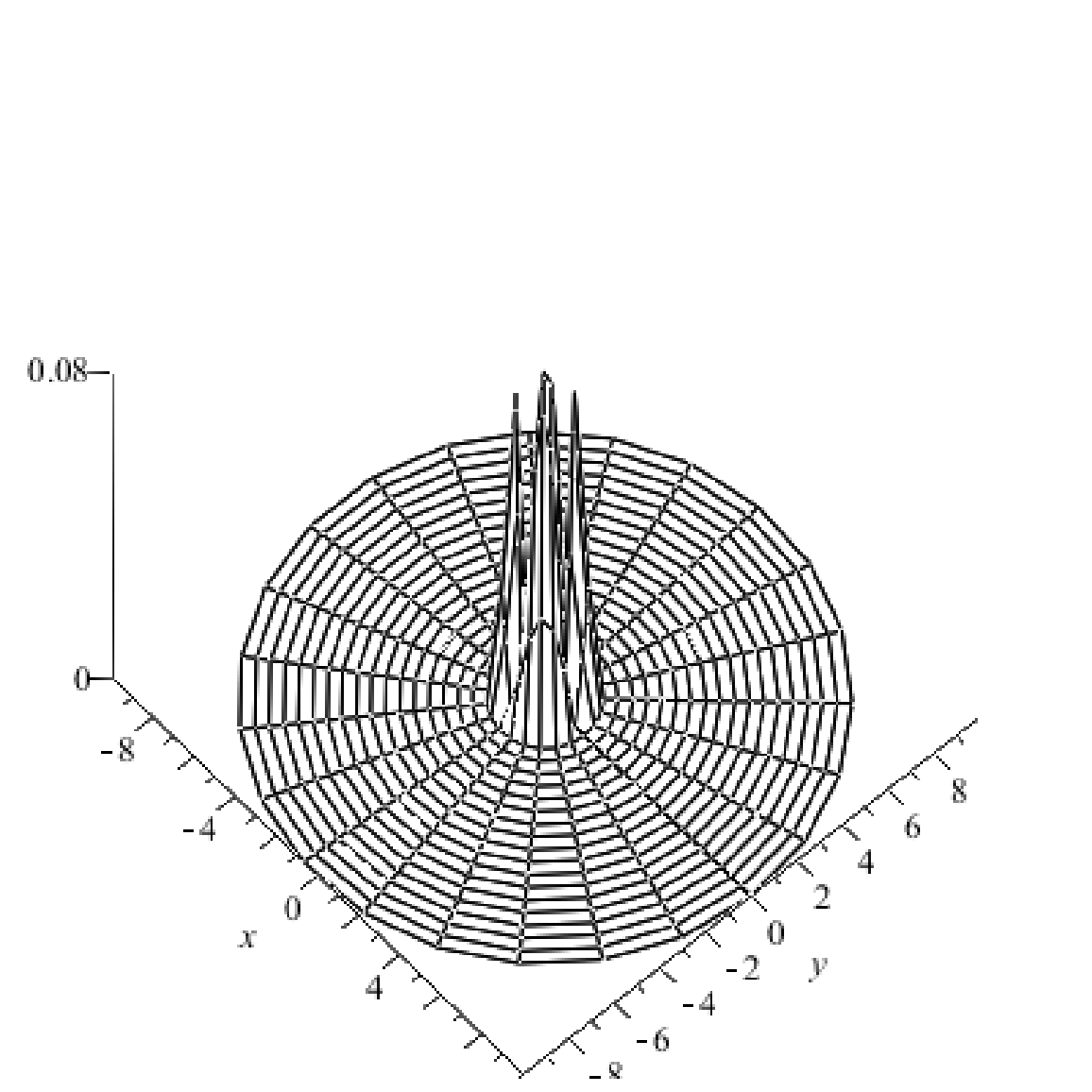}}
\hfill
\subfloat[$a=2.0$, $\left| {\Psi _{10 } } \right|^2$]{\label{4figs2-j} \includegraphics[width=0.24\textwidth]{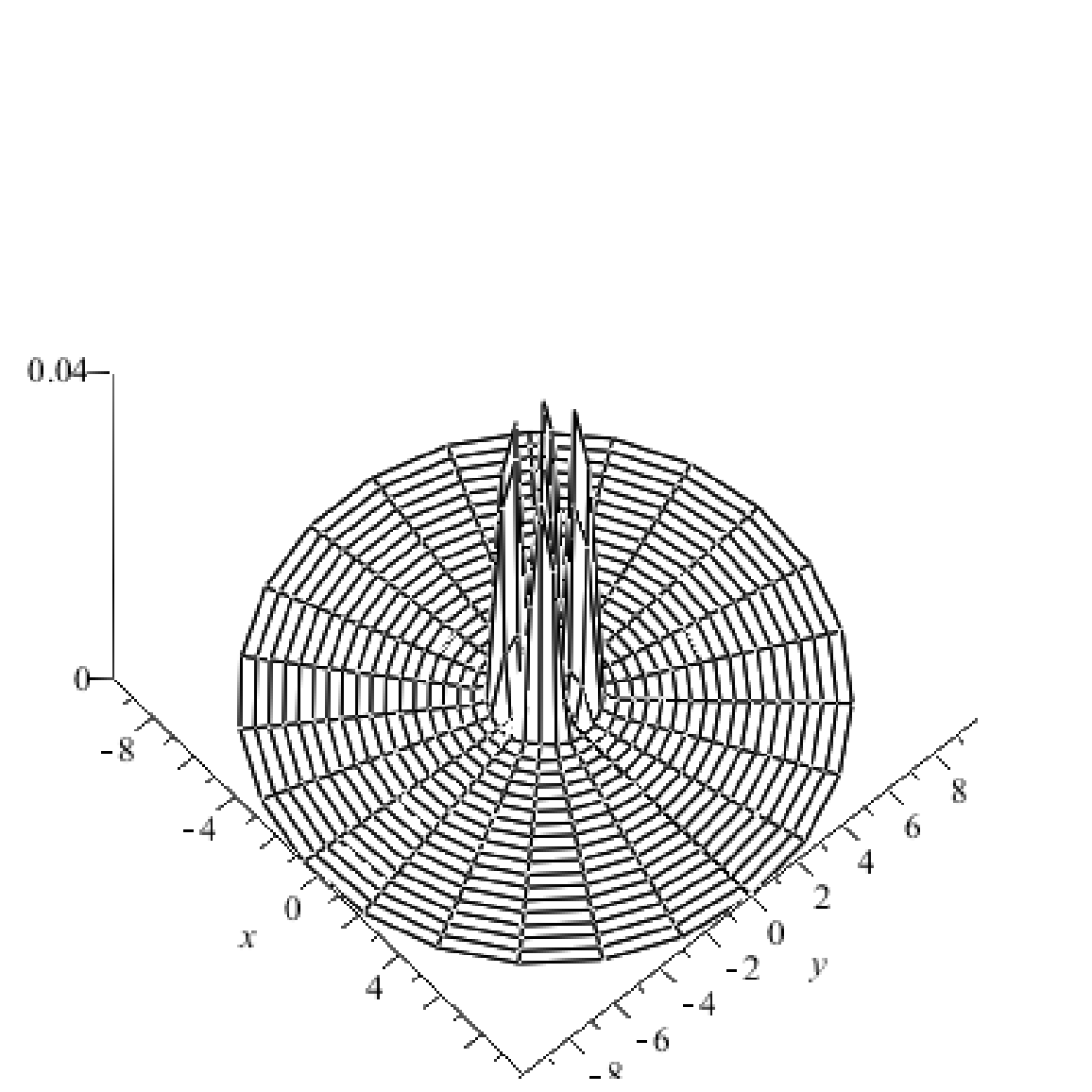}}%
\hfill
\subfloat[$a=2.0$, $\left| {\Psi _{01 } } \right|^2$]{\label{4figs2-k} \includegraphics[width=0.24\textwidth]{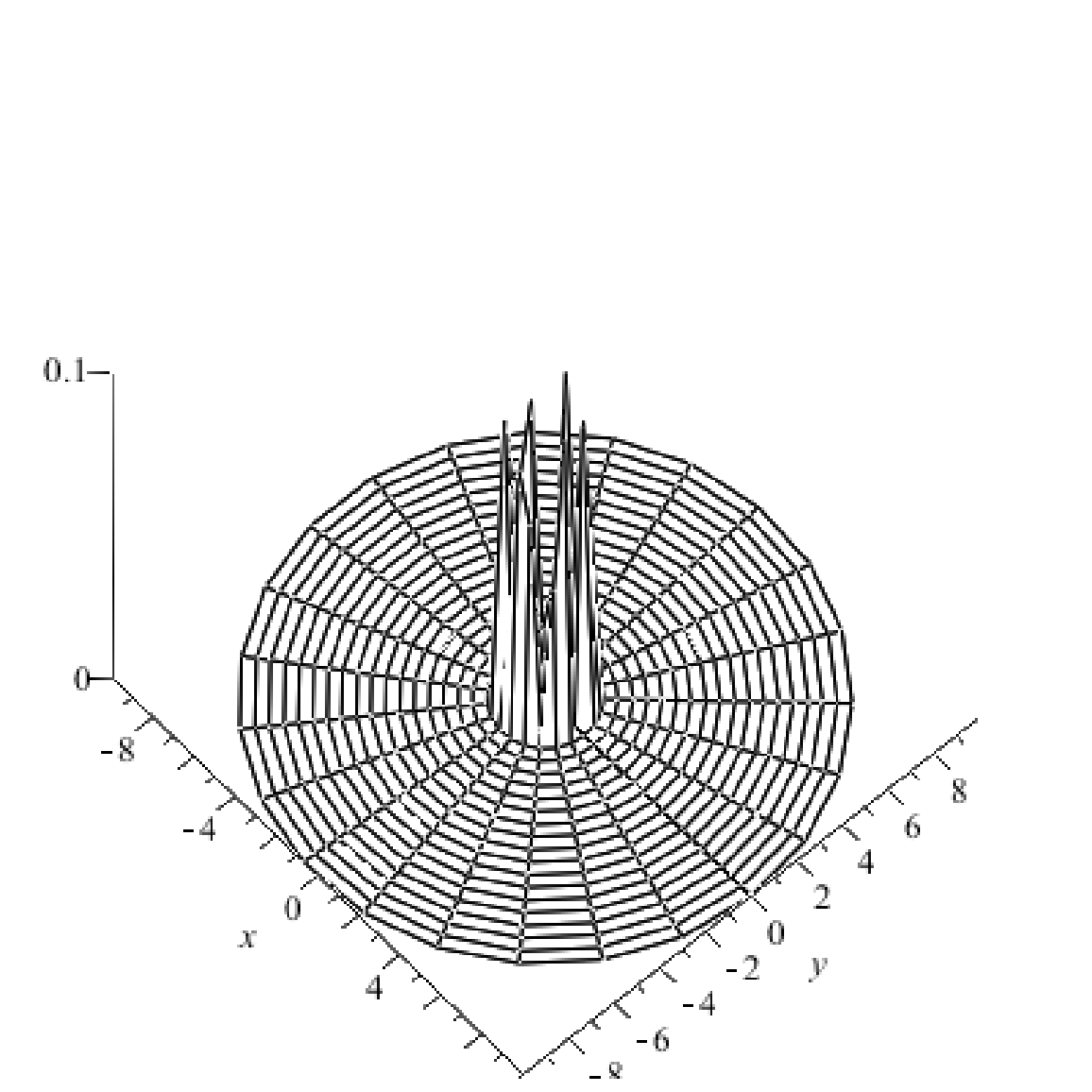}}%
\hfill
\subfloat[$a=2.0$, $\left| {\Psi _{11 } } \right|^2$]{\label{4figs2-l} \includegraphics[width=0.24\textwidth]{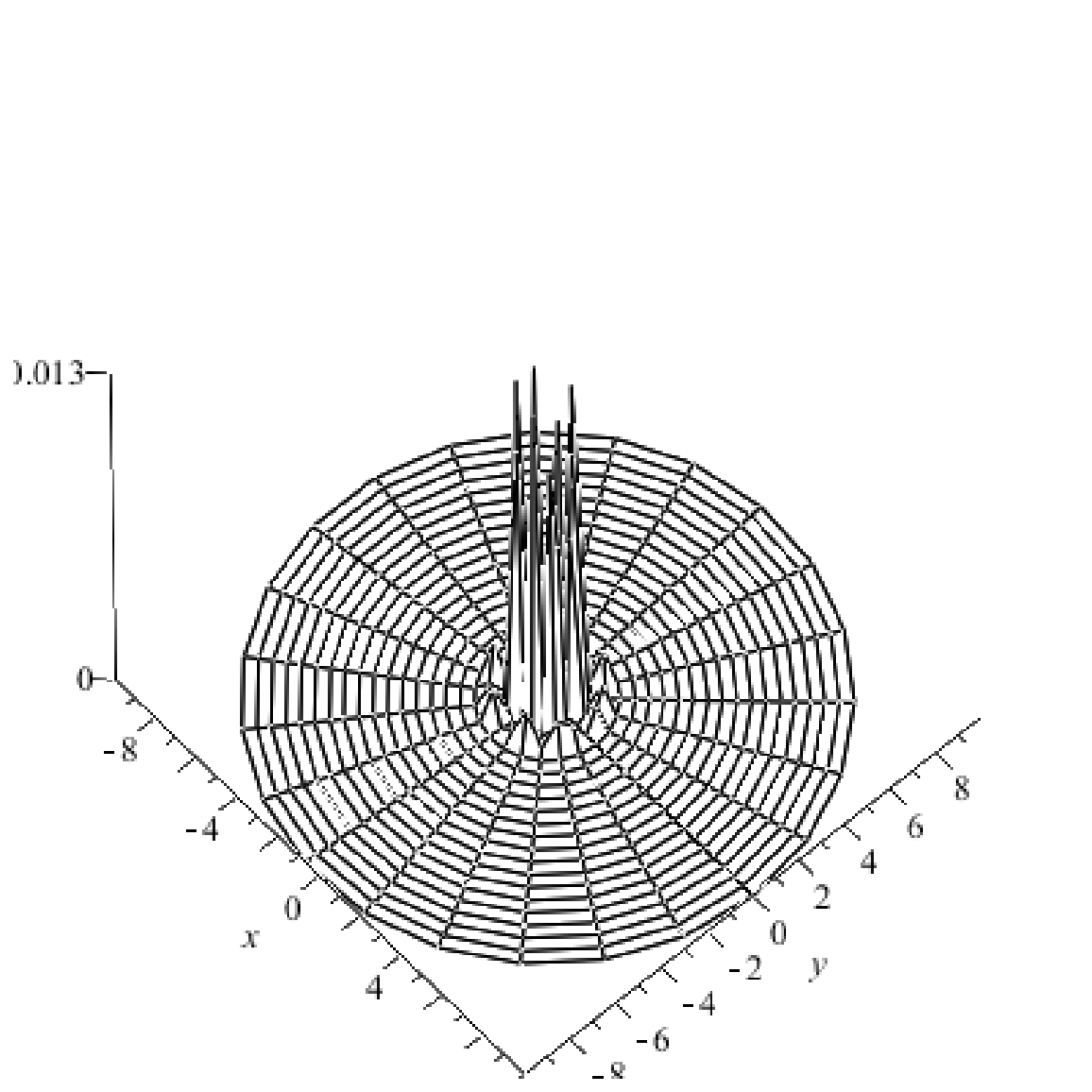}}%
\caption{Generalization of the plots of Figure \ref{4figs} to the non-canonical case $\gamma=1.0$.}
\label{4figs2}
\end{figure}

However, our main goal was to generalize the ideal one-dimensional electron gas problem confined with an anisotropic quantum wire to the case when non-canonical commutation relations between the momentum and position operators hold. 
It is clear from the definition of the momentum operator components~(\ref{p-ncan}) that the additional parameter $\gamma$ plays a role and that the reflection operator $\hat R$ has an impact. 
Both have an effect on the behavior of the energy spectrum and of the wavefunctions. 
The influence of the parameter $\gamma$ on the behavior of the wavefunctions is visible in Fig.~\ref{4figs2}, where the non-canonical case $\gamma=1.0$ generalizes the plots of Fig.~\ref{4figs}. 
Let us visually analyze Eqs.~(\ref{scheq-16-m}) and~(\ref{scheq-17-m}). 
The case $a=0$, corresponding to the limit $M\left(\rho\right) \to m_0$, just reduces both equations to the same differential equation in which the quantum number $m$ is denoted by $m^{\left( {e,o} \right)}$:
\be
\label{can-scheq}
\frac{{\partial ^2 {\rm P}^{\left( {e,o} \right)} }}{{\partial \rho ^2 }} + \frac{1}{\rho }\frac{{\partial {\rm P}^{\left( {e,o} \right)} }}{{\partial \rho }} +\left[ {\kappa _{\left( {\rho ,\varphi } \right)} ^2  - \lambda _0 ^4 \rho ^2  - \frac{{m^{\left( {e,o} \right) ^2} }}{{\rho ^2 }}} \right]{\rm P}^{\left( {e,o} \right)}  = 0.
\ee
Note that the operator $\hat R$ appears in two terms of eq.~(\ref{scheq-07-m}), once in combination with $M(\rho)$ and once without.
Since $M(\rho)$ is the source of the appearance of the parameter $a$, the limit $a\rightarrow 0$ reduces $M(\rho)$ to the constant mass $m_0$, and the splitting of the radial part of the Schr\"odinger equation in even and odd cases is elevated.
However, the angular part equations are preserved.
One observes that both Eqs. (\ref{scheq-09-m}) and (\ref{scheq-11-m}) simply do not depend on the parameter $a$. 
This is due to our assumption that the mass $M$ is central symmetric and only depends on the radial distance $\rho$. 
It is possible to introduce more generalized definitions for the mass $M$, depending on both $\rho$ and on the angular position $\varphi$, and still the radial and angular equations of the problem under consideration will be exactly solvable in terms of orthogonal polynomials. 
Such more general results will be studied in a forthcoming paper; but at present we tried to keep the problem as simple as possible.

\begin{figure}
\centering
\subfloat[$a=-0.6$, $\left| {\Psi _{00 } } \right|^2$]{\label{4figs3-a} \includegraphics[width=0.24\textwidth]{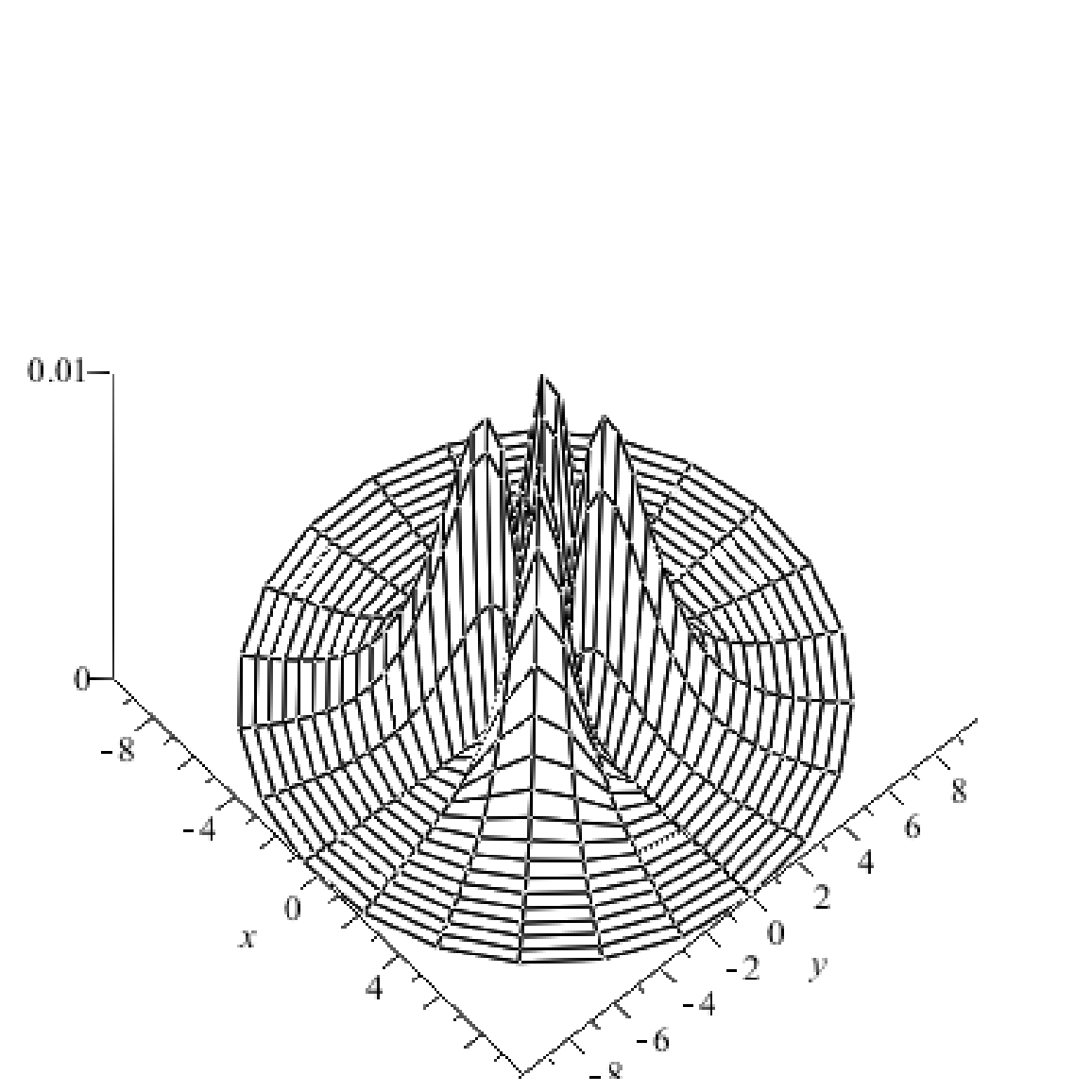}}
\hfill
\subfloat[$a=-0.6$, $\left| {\Psi _{10 } } \right|^2$]{\label{4figs3-b} \includegraphics[width=0.24\textwidth]{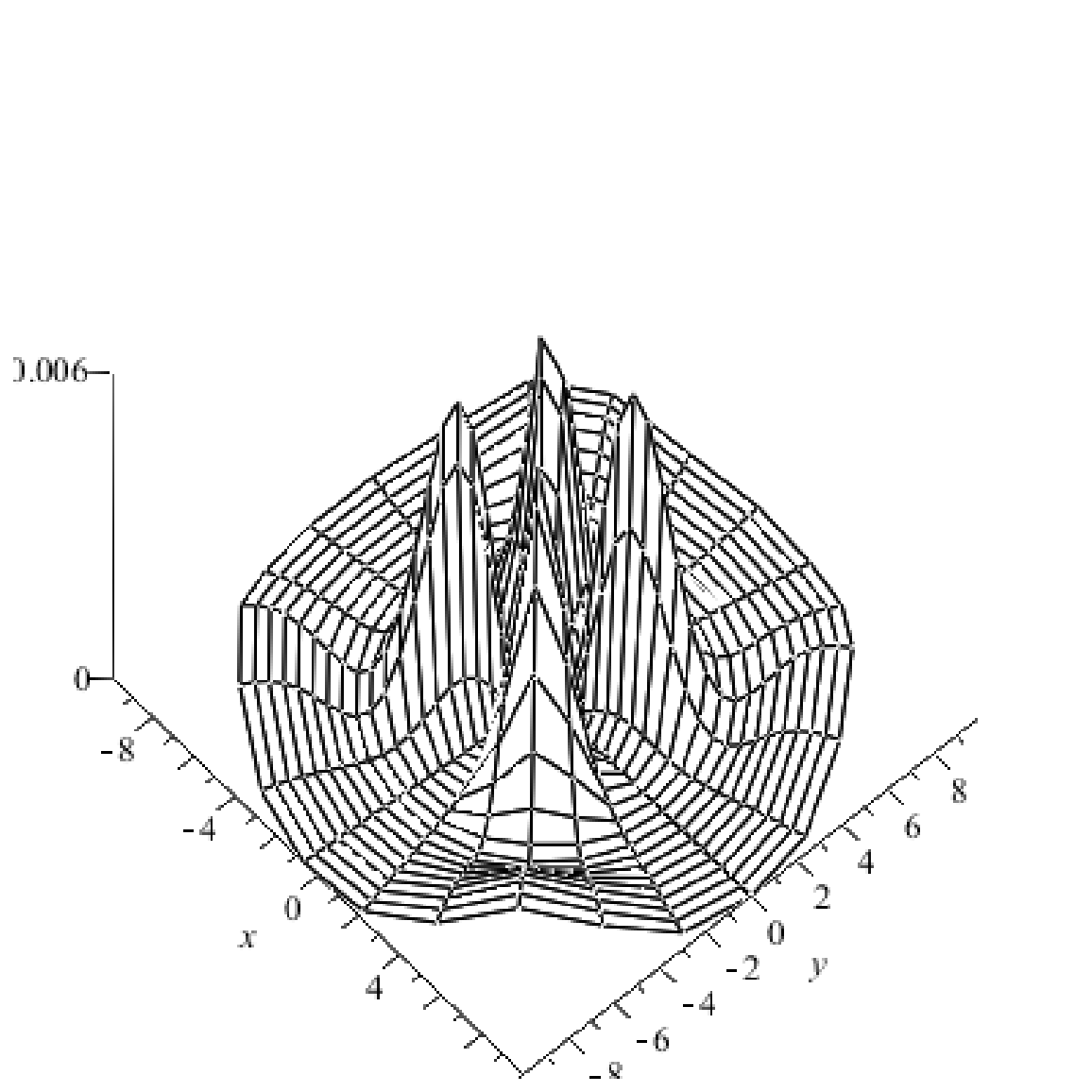}}%
\hfill
\subfloat[$a=-0.6$, $\left| {\Psi _{01 } } \right|^2$]{\label{4figs3-c} \includegraphics[width=0.24\textwidth]{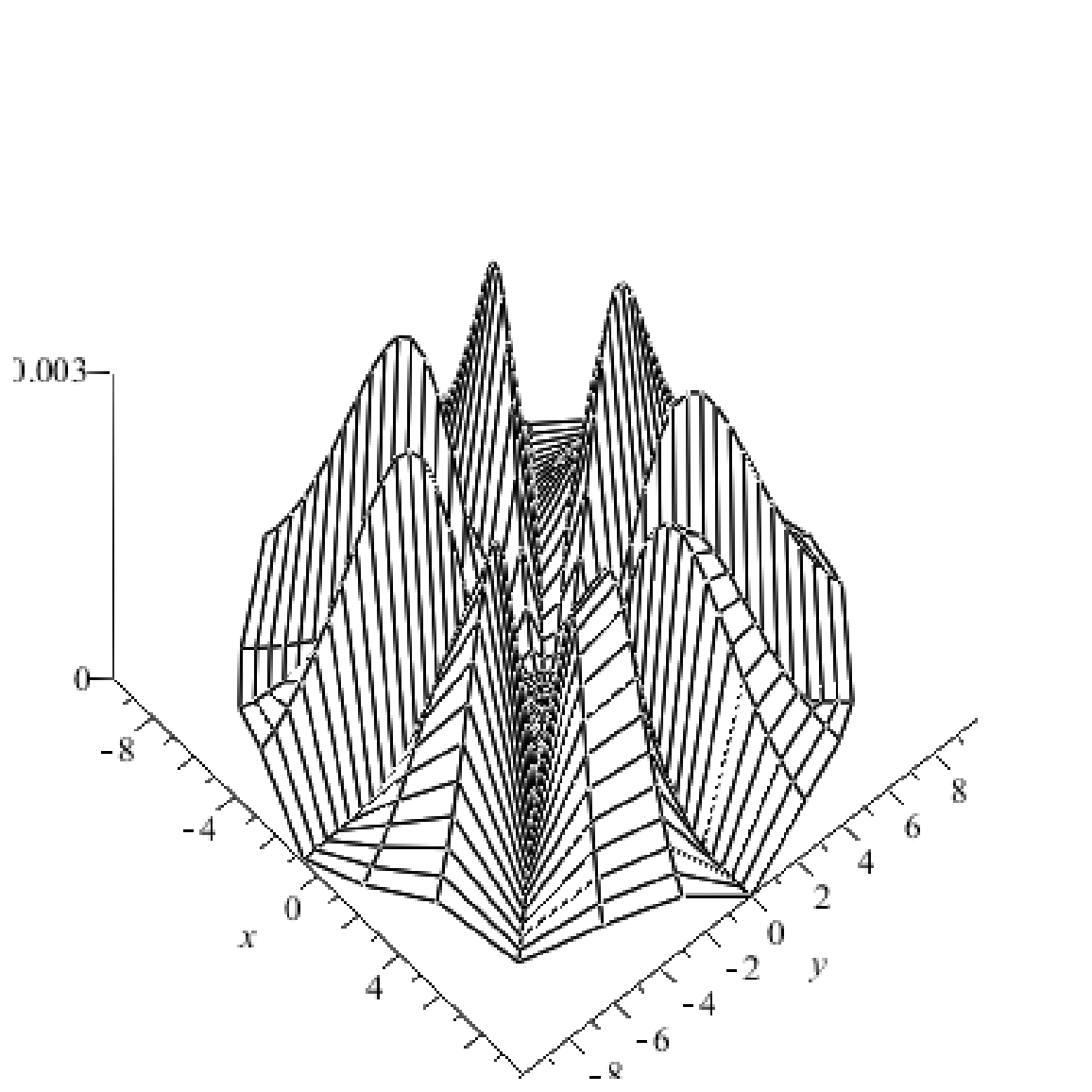}}%
\hfill
\subfloat[$a=-0.6$, $\left| {\Psi _{11 } } \right|^2$]{\label{4figs3-d} \includegraphics[width=0.24\textwidth]{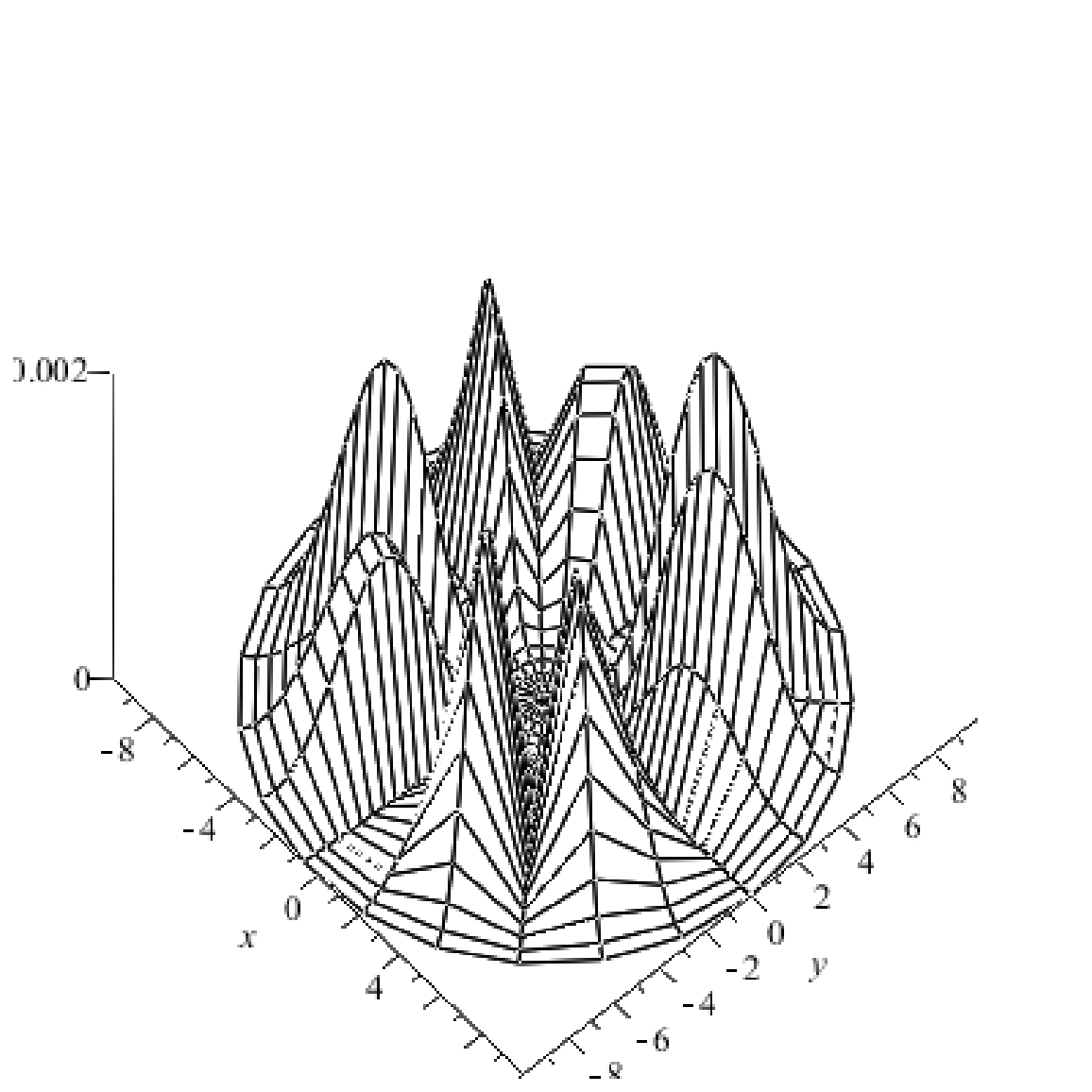}}%
\hfill
\subfloat[$a=0$, $\left| {\Psi _{00 } } \right|^2$]{\label{4figs3-e} \includegraphics[width=0.24\textwidth]{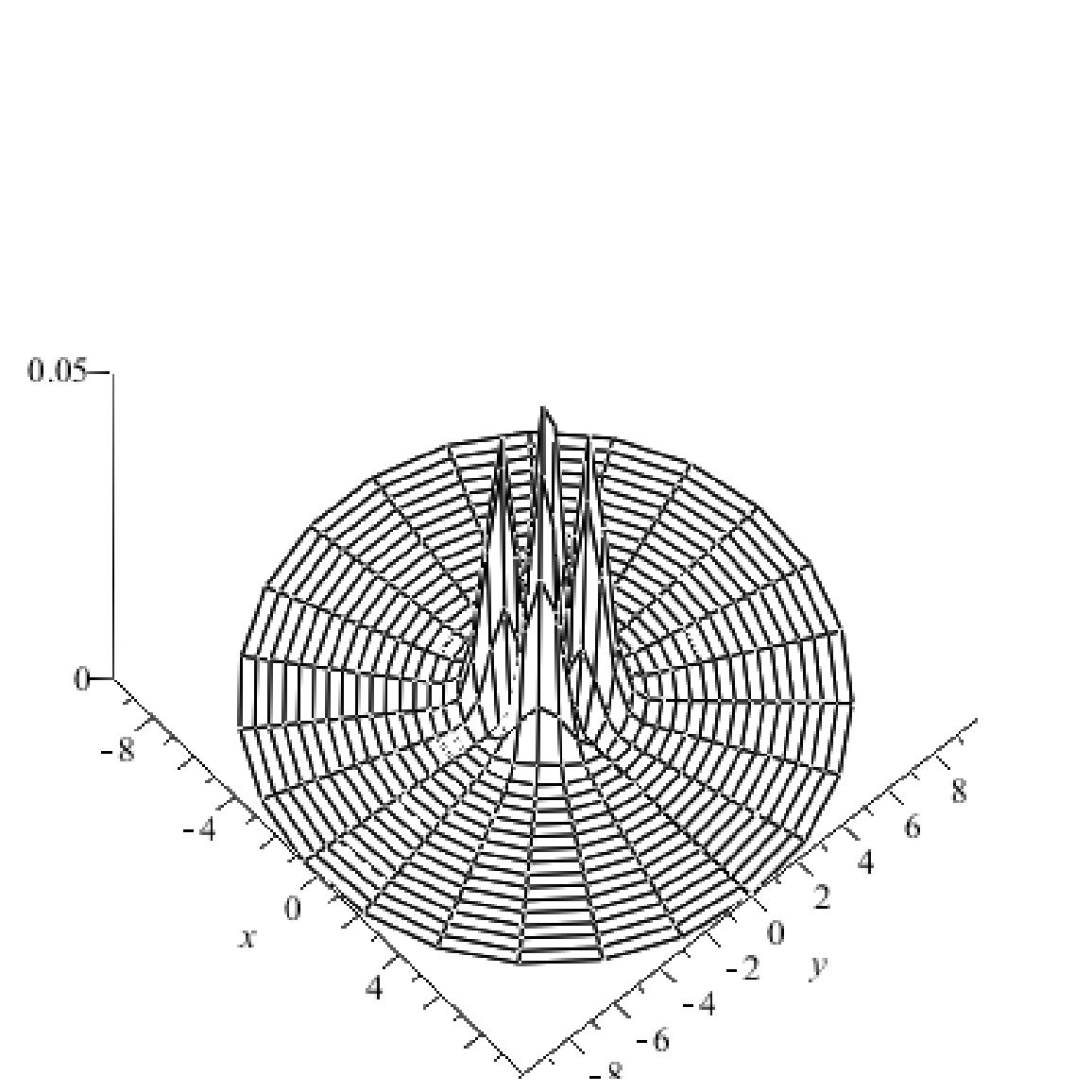}}
\hfill
\subfloat[$a=0$, $\left| {\Psi _{10 } } \right|^2$]{\label{4figs3-f} \includegraphics[width=0.24\textwidth]{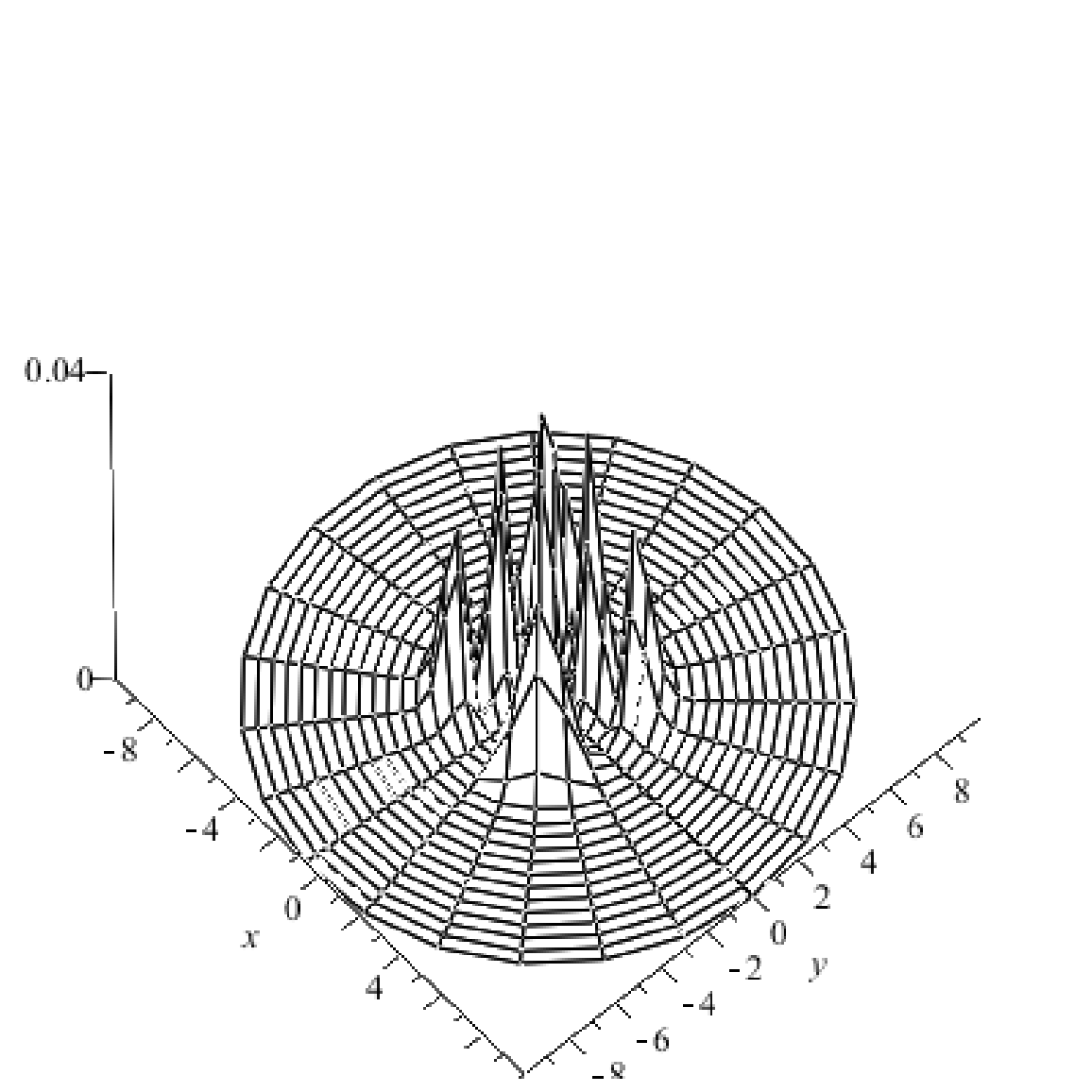}}%
\hfill
\subfloat[$a=0$, $\left| {\Psi _{01 } } \right|^2$]{\label{4figs3-g} \includegraphics[width=0.24\textwidth]{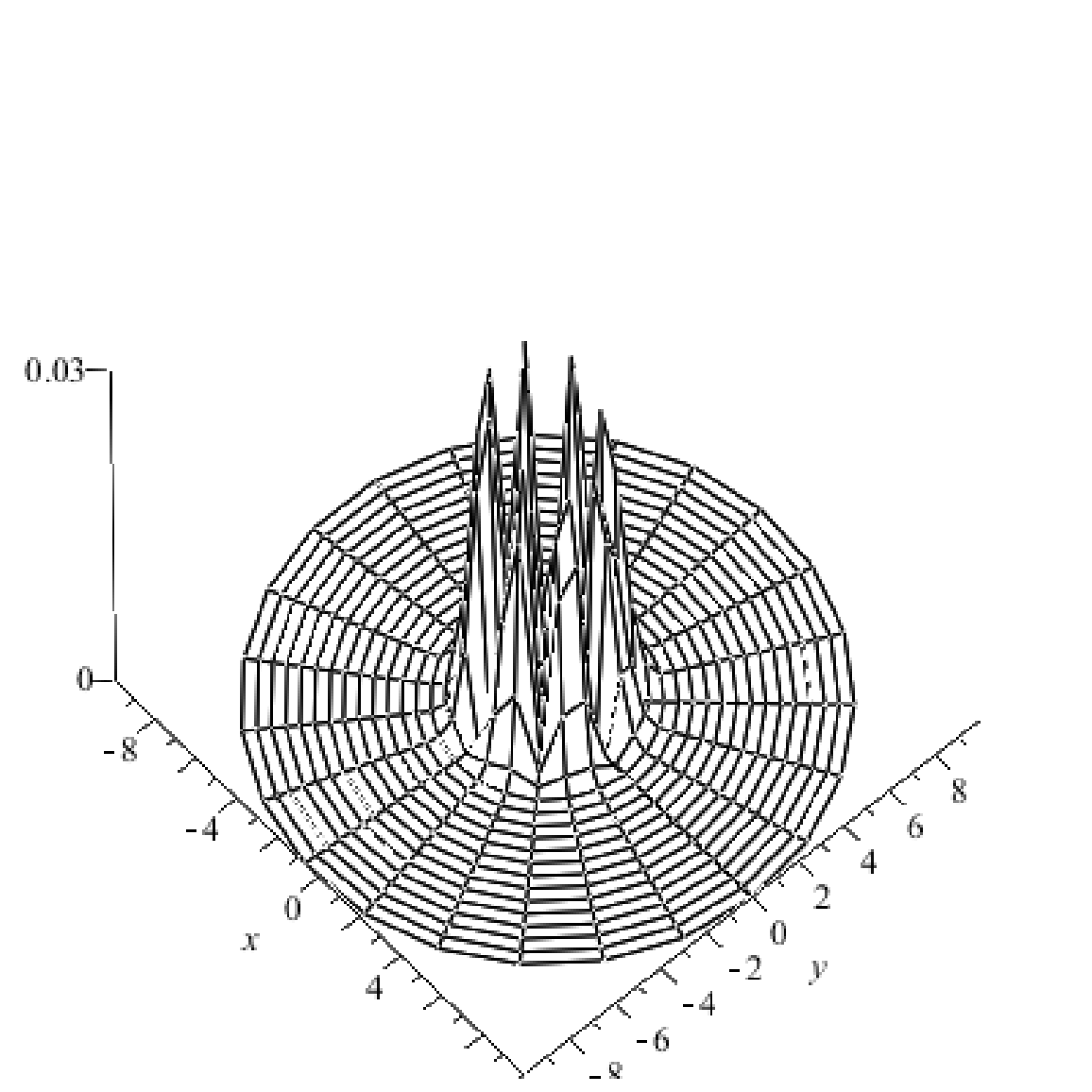}}%
\hfill
\subfloat[$a=0$, $\left| {\Psi _{11 } } \right|^2$]{\label{4figs3-h} \includegraphics[width=0.24\textwidth]{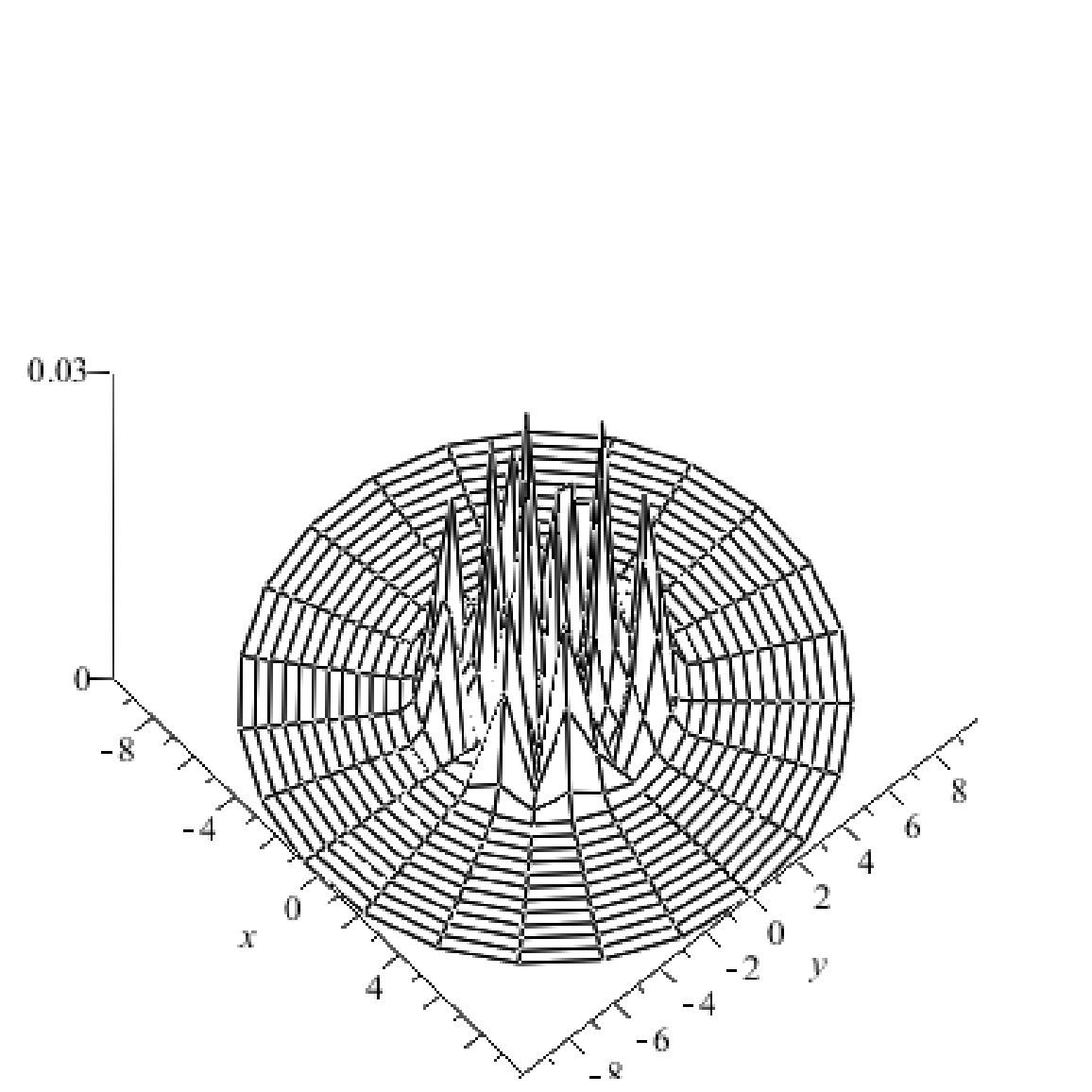}}%
\hfill
\subfloat[$a=2.0$, $\left| {\Psi _{00 } } \right|^2$]{\label{4figs3-i} \includegraphics[width=0.24\textwidth]{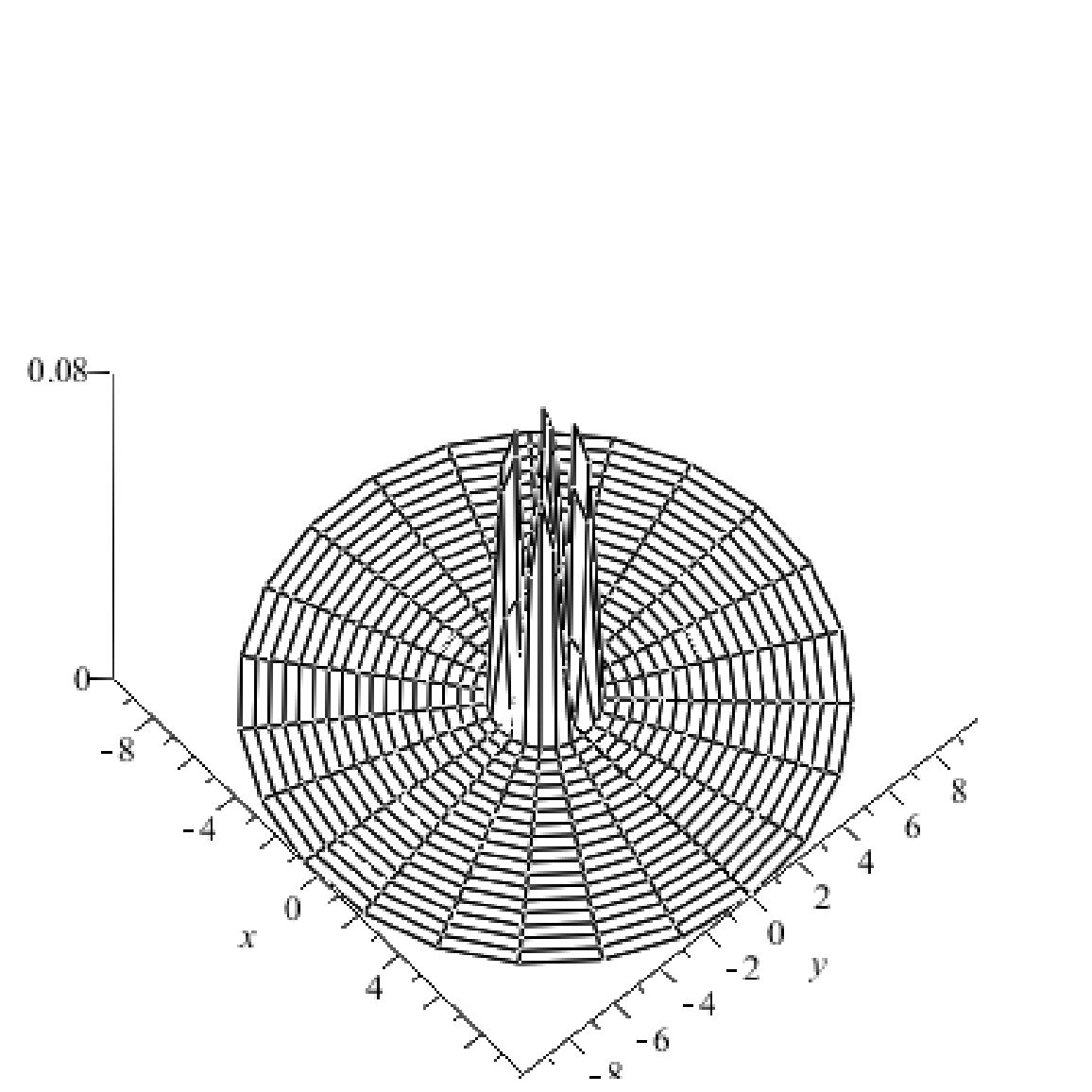}}
\hfill
\subfloat[$a=2.0$, $\left| {\Psi _{10 } } \right|^2$]{\label{4figs3-j} \includegraphics[width=0.24\textwidth]{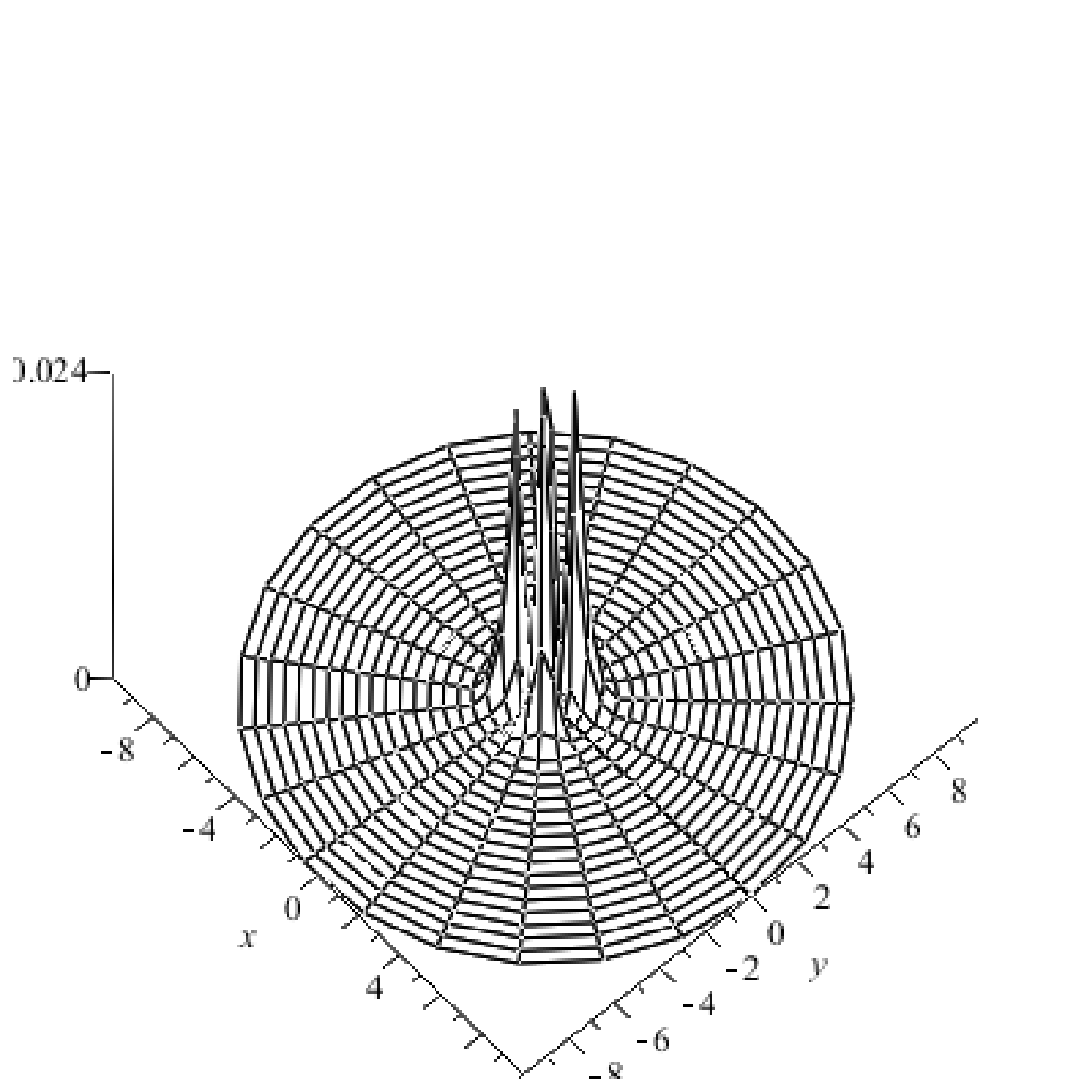}}%
\hfill
\subfloat[$a=2.0$, $\left| {\Psi _{01 } } \right|^2$]{\label{4figs3-k} \includegraphics[width=0.24\textwidth]{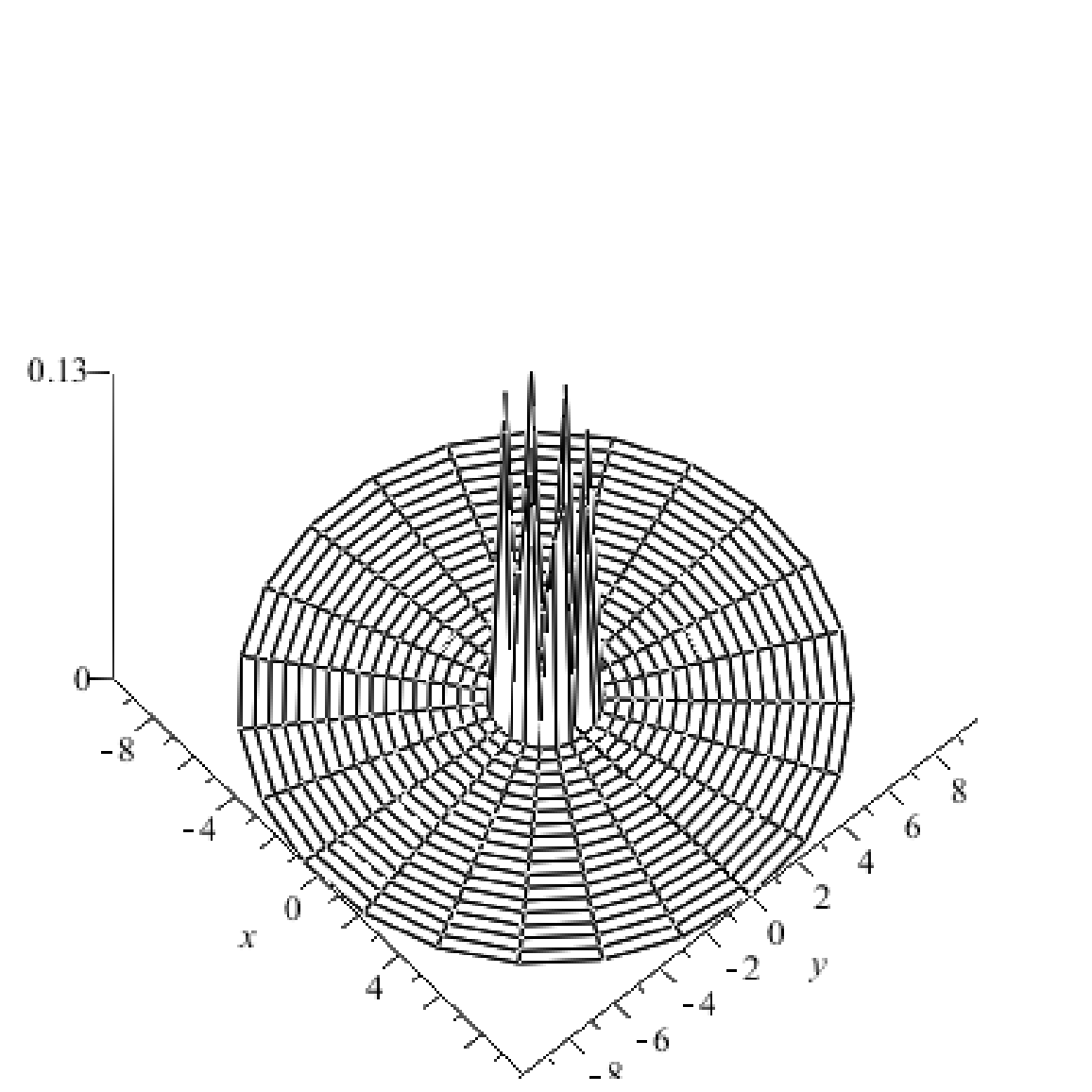}}%
\hfill
\subfloat[$a=2.0$, $\left| {\Psi _{11 } } \right|^2$]{\label{4figs3-l} \includegraphics[width=0.24\textwidth]{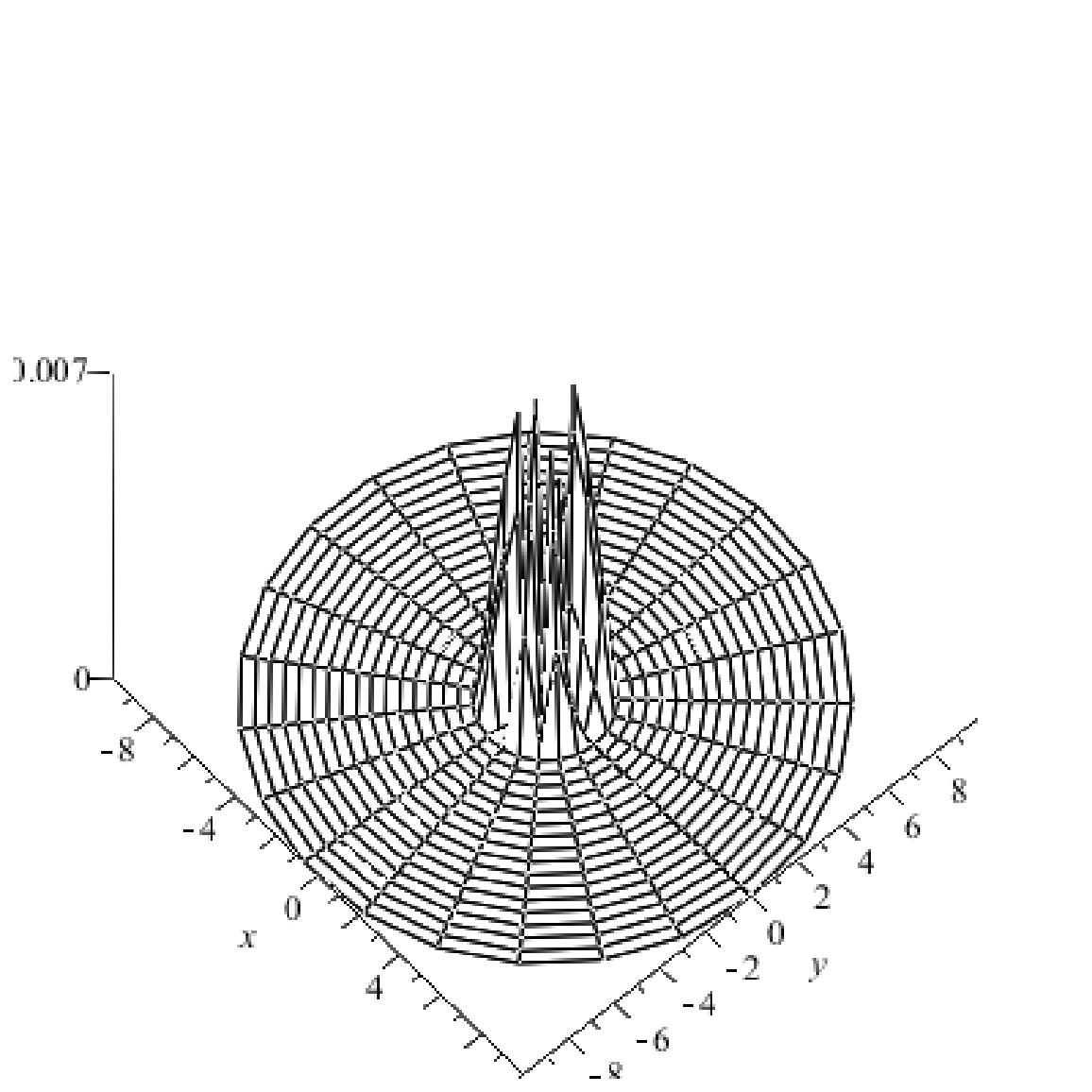}}%
\caption{Generalization of the plots from Figure \ref{4figs} to the non-canonical case $\gamma=1.5$.}
\label{4figs3}
\end{figure}

Figure \ref{4figs3} generalizes all plots from Fig.~\ref{4figs} to the non-canonical case with $\gamma=1.5$. 
We briefly discuss the plots of figs. \ref{4figs2} and \ref{4figs3} together, 
since they both belong to the non-canonical approach but with different values of the parameter $\gamma$. 
One needs to emphasize that in these plots in the non-canonical approach the pure Gaussian-like solution of the radial part splits into at least four different peaks due to existence of the confinement walls $\varphi \to \left(0;\;\pi/2;\;\pi;3\pi/2\right)$, appearing in the solution of Eqs. (\ref{scheq-09-m}) and (\ref{scheq-11-m}). 
This is clearly visible for the ground state.
Then, the number of these peaks increases for the excited states. 
Additionally, one observes an interesting but hidden phenomenon here with the increase of the value of parameter $a$. 
It simply reduces the width of all appearing peaks simultaneously, leading to a quantum foam-like behavior of the states. 
Such states can be very informative for the structure of physics at sub-Planck sizes~\cite{wheeler2010,zurek2001,jafarov2010}.

In this paper, we have presented an exactly solvable model of an ideal one-dimensional electron gas confined in an anisotropic quantum wire with an oscillator-type potential, including a position-dependent effective mass. The problem has been analyzed within both canonical and non-canonical frameworks. In both cases, analytical expressions for the stationary-state wavefunctions and the corresponding discrete energy spectra have been obtained. In particular, the solutions are expressed in terms of Laguerre polynomials for the radial part. In contrast, within the non-canonical approach, the angular part is described by even and odd states in terms of Gegenbauer polynomials. Several limiting cases have also been studied. All limits exhibit consistency with known results.

First of all, one needs to highlight that the introduction of a position-dependent effective mass in two directions allows one to model realistically inhomogeneous semiconductor nanowires, where the effective mass varies due to spatial changes in material composition. Additionally, the obtained analytical solutions for such systems allow qualitatively describe carrier confinement beyond the constant-mass approximation, commonly used in standard models. Therefore, the results obtained in the present paper can find successful applications for semiconductor heterostructures based on materials such as $GaAs/AlGaAs$ or $InAs/InP$, where spatial variations of the band structure naturally lead to an effective mass depending on position.

The non-canonical formulation further introduces an additional degree of freedom through the deformation of the commutation relations via the $\gamma$ parameter. This leads to the modified structures of the energy spectrum and wavefunctions, which do not exist within the conventional quantum mechanics framework. Such a feature may be particularly relevant for modeling systems with effective interactions or non-parabolic dispersion, where standard approaches become less accurate. In this sense, the proposed approach can be useful for the effective description of correlated or constrained motion in low-dimensional systems.

From the viewpoint of applications in nanotechnology, the availability of exact analytical expressions for both energy levels and wavefunctions is of direct importance for the evaluation of physical observables. For example, the appearance of the $\gamma$ deformation parameter in the generalized energy spectrum and wave functions will definitely affect the dipole transition matrix elements and transition energies of the nano-sized wire fabricated in the framework of the proposed quantum system~\cite{al2022,al2026}. Consequently, the optical response of the system is expected to differ from that of conventional quantum wires, suggesting the possibility of achieving tunable optical characteristics. We believe that such a difference will definitely play a vital role for possible applications in nanoscale optoelectronic and photonic devices, including quantum wire–based photodetectors and emitters.

Also, all computations performed above can be generalized under the Frenet--Serret coordinate system~\cite{frenet1852,serret1851}, which can be considered somehow as an extension of the cylindrical coordinate system used in the current paper. 
There, one keeps $x=\rho\cos\varphi$, and takes the $Z$-direction dependent on the angular position $\varphi$ as $z=h \varphi$ ($0< \varphi<2 \pi$). 
Here, $h$ defines the height of the helix that appears as a result of the torsional motion. 
The torsion can be right-handed if one keeps the definition of the $Y$-direction as $y=\rho \sin \varphi$, or becomes a left-handed helix with the torsional motion if one modifies the $Y$-direction as $y=-\rho \sin \varphi$. 
Further deeper study of this idea may extend the computations to the torsional motion with the helix of finite height, while still trying to preserve the exact solubility of the quantum system under consideration. 
Taking into account some examples of the successful application of this curvature in the quantum computations, as in~\cite{costa1981,vieira2012,alsing2024}, an attempt at the above-described generalization can also be successfully achieved in terms of the (quasi) exact solutions.

Summarizing, we want to highlight that the proposed exactly solvable model provides a consistent and flexible framework for describing quantum wires with position-dependent effective mass within both canonical and non-canonical approaches. The results establish a direct connection between (super)algebraic generalizations of quantum theory and physically relevant nanostructures, and they offer practical tools for the analysis and design of low-dimensional semiconductor devices. Future work may focus on extending the present approach to transport phenomena, nonlinear optical properties, and experimentally realizable nanowire systems, as well as generalization of the non-canonical approach to the exactly solvable anisotropic quantum dot models.

\section*{Data Availability Statement}

No data associated in the manuscript.

\bibliographystyle{unsrt}
\bibliography{qw-epjp}

\end{document}